\documentclass[proof]{pasj01}
\usepackage{url}
\usepackage{multirow}
\usepackage{lscape}
\usepackage{ulem}
\usepackage{natbib}
\usepackage{subfigure}
\usepackage[figuresright]{rotating}
\usepackage{longtable}
\usepackage{booktabs}
\Received{$\langle$reception date$\rangle$}
\Accepted{$\langle$acception date$\rangle$}
\Published{$\langle$publication date$\rangle$}

\begin{document}
	
\title{Multiple HC$_3$N line observations towards 19 Galactic massive star forming regions}
\author{Huanxue Feng$^{1,2 *}$\email{fhx@shao.ac.cn}, Junzhi Wang$^{1,2 *}$\email{jzwang@shao.ac.cn}, Shanghuo Li$^{2}$, Yong Shi$^{3}$, Fengyao Zhu$^{2}$, Minzhi Kong$^{4}$, Ripeng Gao$^{2,4}$, Fei Li$^{2}$}%
\altaffiltext{1}{Shanghai Astronomical Observatory, Chinese Academy of Sciences, No. 80 Nandan Road, Shanghai, 200030, China}
\altaffiltext{2}{Key Laboratory of Radio Astronomy, Chinese Academy of Sciences, No. 10 Yuanhua Road, Nanjing, JiangSu 210033, China}
\altaffiltext{3}{School of Astronomy and Space Science, Nanjing University, Nanjing, No. 163, Xianlin Avenue, Qixia District, Nanjing, 210093, China}
\altaffiltext{4}{Hebei Normal University, No.20 Road East. 2nd Ring South, Yuhua District, Shijiazhuang, Hebei, 050024, China}

\maketitle

\begin{abstract}
We performed observations of the HC$_3$N (24-23, 17-16, 11-10, 8-7) lines towards a sample consisting of 19 Galactic massive star-forming regions with the Arizona Radio Observatory 12-m and Caltech Submillimeter Observatory 10.4-m telescopes. HC$_3$N (24-23, 17-16, 11-10, 8-7) lines were detected in sources except for W44, where only HC$_3$N (17-16, 11-10) were
detected. Twelve of the nineteen sources showed probable line wing features. The excitation temperatures were estimated from the line ratio of HC$_3$N (24-23) to HC$_3$N (17-16) for 18 sources and are in the range 23-57\,K. The line widths of higher-$J$ transitions are larger than lower-$J$ ones for most sources. This indicates that the inner dense warm regions have more violent turbulence or other motions (such as rotation) than outer regions in these sources. A possible cutoff tendency was found around $L_{\rm IR}$ $\sim$ 10$^{6}$ $L_\odot$ in the relation between $L_{\rm IR}$ and full width at half maximum line
widths. 
\end{abstract}

\section{Introduction}

Molecular gases, especially dense ones, play a significant role in star-formation and as a tool to investigate the cause of star formation activity. In order to understand the evolution and regulation of current star formation in galaxies, dense gas properties should be precisely characterized \citep{2011AJ....142...32M}. Dense gas tracers can be used as powerful tools to study properties of molecular gas in star-forming regions, such as temperature, pressure, density and so on. 

In extended parts of molecular clouds, the low-$J$ transitions of CO are excellent tracers to probe the molecular gas. CO lines are usually optically thick and are used to probe relatively low-density gas \citep{2000ApJ...531..200M}. HC$_{2n+1}$N (n $\geq$ 1) \citep{2016MNRAS.460.2103F} are major components of the
carbon chain molecule series \citep{2008ApJ...681.1385H}. Among them, HC$_3$N is the simplest and the most abundant of the group of cyanopolyyne molecules. The majority of radio lines of the HC$_3$N molecule are usually optically thin and are used to probe dense gas \citep{1976ApJ...205...82M}. Since HC$_3$N has a relatively high electric dipole moment ($\mu$ = 3.7\,D), it can only be excited in dense gas regions \citep{2003A&AT...22...33P}. HC$_3$N is a highly linear molecule with a low rotational constant ($\sim$1/13 of CO and $B_o$ $\simeq$ 4.5 GHz),  which supplies a lot of lines dispersed from centimeter to (sub-)millimeter wavelengths \citep{1996ApJ...460..343B}. Multiple HC$_3 $N lines can be used to measure the gas temperature and barometric pressure in the interstellar medium \citep{2016MNRAS.460.2103F}. 

Using information on the line widths from different transitions (7-6, 5-4 and 2-1) of the CS molecule, \citet{2010ApJS..188..313W} suggested that turbulence injected from outflows, inflows, and expanding HII regions is more significant at smaller scales. The mean of the line width of the high-$J$ CS lines is larger than low-$J$ ones, which is inconsistent with Larso law and indicates high-$J$ CS emissions are from more turbulent regions than that of low-$J$ ones \citep{1981MNRAS.194..809L}. Kinematics, which can be obtained with line broadening, is important for studying star formation at different evolutionary stages. However, CS lines are normally optically thick, which can confuse determination of gas kinematics. Thus, multiple optically thin lines from one molecule, such as HC$_3$N, are necessary to obtain the gas kinematics from regions under different excitation conditions.

With the information of the HC$_3$N column density, the relation between the column density of HC$_3$N ($N_{\rm HC_3N}$) and the distance ($R$) from the center of a dense core of sources can be obtained in the form of $N_{\rm HC_3N}$ $\propto$ $R^{-\alpha}$  \citep{1991JKAS...24..217C}. Dissipation of turbulence and angular momentum is important during the star-forming process.  Such  turbulence and angular momentum can be reflected in line widths of optically thin lines, such as transitions of H$^{13}$CN, H$^{13}$CO$^+$, $^{13}$CS and HC$_3$N. The high-$J$ transitions of HC$_3$N can trace denser and warmer gas components compared with low-$J$ ones. Because the excitation conditions (e.g. density and temperature) of high-$J$ transitions are higher than low-$J$ ones, information on the turbulence and/or angular momentum of different regions can be derived from the line widths of multiple transitions of HC$_3$N observations, even if no spatially resolved information was provided. 

In this paper we present the survey results of HC$_3$N $J$=24-23, $J$=17-16, $J$=11-10, and $J$=8-7 lines of 19 star-formation regions with the Caltech Submillimeter Observatory (CSO) and the Arizona Radio Observatory (ARO) 12 m telescope. We introduce the observations and data reduction in Section \ref{ 2 sect: Obs and Data Reduction}. Detailed results are provided in Section \ref{ 3 sect :results }. In Section \ref{ 4 sect :discus }, we provide an analysis and discussion of the results. A brief summary is presented in Section \ref{ 5 sect :sum }.

\section{Observations and data reduction}
\label{ 2 sect: Obs and Data Reduction}

\subsection{Observation}
\label{ 2.1 sect :Obs }

HC$_3$N (17-16, 11-10, 8-7) observations were performed towards about 60 massive star-forming regions with the 12-m ARO radio telescope in 2008 January (Su et al. in preparation). The standard position-switching mode was used with an offset of 30$'$ in azimuth. HC$_3$N (17-16) was observed with the 2-mm receiver. The HC$_3$N (11-10) and HC$_3$N (8-7) lines were observed with the 3-mm-high and 3-mm-low receivers, respectively. The sample of  sources in our observation was selected from the large sample of dense clumps associated with water masers that had been surveyed with several CS transitions \citep{1992ApJS...78..505P, 1997ApJ...476..730P}. Most of these sources had been mapped with CS (5-4) \citep{2003ApJS..149..375S} and 350 $\mu m$ dust continuum emission \citep{2002ApJS..143..469M}. Table \ref{tabel 1} lists the source information including virial mass, infrared luminosity, angular extent of CS (5-4), excitation temperature of HC$_3$N, outflow detection, infall signature, and the evolutionary status of the HII region.

Based on the HC$_3$N (17-16, 11-10, 8-7) results, we selected 19 sources with HC$_3$N (17-16) peak emission higher than 0.5 K, promising a high detection rate, to perform further observations of HC$_3$N (24-23). HC$_3$N (24-23) was observed with the CSO 10.4-m radio telescope 230-receiver of a double-side-band (DSB) receiver in May 2013. The observation mode was position-switching with an offset of 30$'$ in azimuth. Pointing and focusing were checked approximately every two hours by measuring nearby strong millimeter-emitting sources. The Wideband fast Fourier transform spectrometer was used with 269\,kHz frequency spacing, which corresponds to 0.35\,km\,s$^{-1}$ at 230\,GHz. The bandwidth is 4\,GHz. The typical system temperature ranged from 200\,K to 400\,K. 

\subsection{Data reduction}
\label{ 2.2 sect :data reduction }

The data were reduced with the CLASS program of the GILDAS\footnote{http://iram.fr/IRAMFR/GILDAS/} package. We first visually checked each spectrum and discarded spectra with unstable baselines. Then spectra were averaged with time weighting. A first-order fitting was used to subtract baselines of averaged spectra.

In order to increase the signal-to-noise ratio (S/N), the spectra were smoothed to a velocity resolution of 0.3 - 0.4\,km\,s$^{-1}$. Peak intensity, velocity-integrated intensity, central velocity and line width were derived from the Gaussian fit of the spectra. We converted the antenna temperature ($T_{A}^{*}$) to the main beam brightness temperature ($T_{\rm mb}$) via $T_{\rm mb}$ = $T_{\rm A}^{*}$/$\eta_{\rm mb}$, where $\eta_{\rm mb}$ is the main beam efficiency (about 0.69 for the CSO telescope and 0.87 for the ARO telescope). The frequency, observational date, telescope name, beam size, root mean square (RMS) noise, and velocity resolution after smoothing of four transitions are listed in Table \ref{tabel 2}.

\begin{sidewaystable}[htb]
	\begin{minipage}{190mm}
		\caption{The 19 sources observed in HC$_3$N emissions. The columns are: (1) Source name. (2) and (3) Coordinates. (4) Source distance. (5) Virial mass. (6) Total infrared luminosity (8$-$1000 $\mu m$). (7) Excitation temperature. (8) Angular extent of CS (5-4) map at the half-power level. (9) Classification of outflow (Bi, bipolar outflow; MB, blue monopolar outflow). (10) Profile for $\Delta V$ \textless $-$0.25 judged from HCN (3-2) profile. (11) Classification of evolutionary status of HII region.}
		\label{tabel 1} 	
		\begin{tabular}{c c c c c c c c c c c }
			\hline
			\hline
			{Source}& {RA} & {DEC} &{Distance$^a$}&{$M_{\rm vir}$($R_{\rm CS\ 5-4}$)$^a$}&{log$L_{\rm IR}$$^c$}  & {$T_{\rm ex}$} &{$\theta$(CS$_{5-4}$,\ $A_{1/2}$)} $^a$ & Outflow$^b$& Infall?$^c$& HII?$^a$\\
			&(J2000)&(J2000)&(kpc)&($M_\odot$)&($L_\odot$)&(K)&(K)& & \\
			(1)&(2)&(3)&(4)&(5)&(6)&(7)&(8)&(9)&(10)&(11)\\
			\hline
			
			W3(OH)      & 02:27:04.69 & 61:52:25.5 & 2.4 &$1020\pm130$ & 5.09 & $33.1\pm1.3$ &$30.94\pm0.86$&...&Y&UCHII\\
			
			RCW142      & 17:50:15.13 & $-$28:54:31.5& 2.0 &$370\pm290$  & 4.82 & $34.1\pm0.8$ &$28.88\pm1.03$&...&Y&UCHII\\
			
			W28A2       & 18:00:30.42 & $-$24:03:58.5& 2.6 &$450\pm340$  & 5.45 & $45.2\pm0.5$ &$23.80\pm3.17$& Bi&...&UCHII\\
			
			M8E         & 18:04:53.25 & $-$24:26:42.4& 1.8 &$100\pm30$   & 4.26 & $26.2\pm1.2$ &$32.09\pm1.15$& Bi&...&UCHII\\
			
			G8.67-0.36  &18:06:18.87  & $-$21:37:37.8& 8.5 &$860\pm340$  & 5.12 & $40.9\pm1.5$ &$12.62\pm0.24$&...&Y&UCHII\\
			
			W31         & 18:08:38.32 & $-$19:51:49.7& 12.0&$7300\pm1670$& 6.22 & $50.6\pm1.3$ &$23.03\pm0.69$&...&Y&UCHII\\
			
			G10.6-0.4   & 18:10:28.70 & $-$19:55:48.6& 6.5 &$2750\pm460$ & 6.05 & $44.6\pm0.6$ &$26.02\pm0.32$&...&Y&CHII\\
			
			W33cont     & 18:14:13.67 & $-$17:55:25.2& 4.1 &$2950\pm520$ &$...$ & $32.0\pm0.3$ &$75.46\pm1.01$&...&...&UCHII\\
			
			G14.33-0.64 & 18:18:54.71 &$-$16:47:49.7 & 2.6 &$160\pm60$   & 4.33 & $25.1\pm1.0$ &$26.97\pm0.79$&...&...&UCHII\\
			
			W42         & 18:36:12.46 &$-$07:12:10.1 & 9.1 &$2160\pm480$ & $...$& $37.7\pm1.3$ &$22.21\pm0.91$&...&...&UCHII\\
			
			S76E        & 18:56:10.43 &07:53:14.1  & 2.1 &$240\pm20$   & 4.47 & $28.2\pm0.8$ &$39.29\pm0.98$&...&...&HII\\
			
			G35.20-0.74 & 18:58:12.73 &01:40:36.5  & 3.3 &$3200\pm1190$& 4.65 & $22.9\pm0.9$ &$37.50\pm1.25$& Bi&...&HII\\
			
			W51M        & 19:23:43.86 & 14:30:29.4 & 7.0 &$5930\pm980$ & 6.60 & $45.6\pm0.5$ &$29.47\pm0.29$&...&Y&CHII\\
			
			W75N        & 20:38:36.93 &42:37:37.4  & 3.0 &$700\pm180$  & 5.29 & $44.4\pm1.2$ &$37.13\pm0.69$& Bi&...&UCHII\\
			
			DR21S       & 20:39:00.80 & 42:19:29.7 & 3.0 &$990\pm220$  & 5.70 & $34.1\pm0.5$ &$37.13\pm0.69$& MB&Y&UCHII\\
			
			W75(OH)     & 20:39:01.00 & 42:22:49.8 & 3.0 &$1260\pm250$ & 4.70 & $39.5\pm0.8$ &$39.88\pm0.69$&...&Y&...\\
			
			CepA        & 22:56:18.13 & 62:01:46.3 & 0.73&$...$        & 4.92 & $57.0\pm3.0$ & $...$        & Bi&Y&UCHII\\
			
			NGC7538     & 23:13:44.85 & 61:26:50.6 & 2.8 &$920\pm190$  &$...$ & $32.2\pm1.9$ &$47.15\pm0.74$& Bi&...&UCHII\\
			
			W44         & 18:53:18.5  & 61:26:50.6 & 3.7 &$1400\pm600$ &5.48  & $...$        &$41.25\pm0.56$&...&Y&CHII\\
			
			\noalign{\smallskip}\hline
		\end{tabular}\\ 		
		
		Reference. $^a$ \citet{2003ApJS..149..375S}; $^b$ \citet{2004A&A...426..503W}; $^c$ \citet{2010ApJS..188..313W}.
		
	\end{minipage}
\end{sidewaystable}

\clearpage

\clearpage


\begin{table}[htb]
	\caption{Observing information for different HC$_3$N lines.} \label{tabel 2}
	\begin{center} 			
		\begin{tabular}{c c c c c c c }
			\hline
			\hline
			{Transitions}&{$\nu$}&{Date}&{Telescope}&{Beam size}&{RMS}&{$\delta V$}\\ 
			(HC$_3$N)&(GHz)&(UT) & &($''$)&(K)&(km\,s$^{-1}$)  \\
			\hline
			
			24-23 & 218.324 & 2013 May &CSO &  $\sim$\ 35.2  & 0.02-0.08   & 0.34 \\
			17-16 & 154.657 & 2008 Jan &ARO &  $\sim$\ 41.0  & 0.11-0.22   & 0.38 \\
			11-10 & 100.076 & 2008 Jan &ARO &  $\sim$\ 63.3  & 0.06-0.11   & 0.30 \\
			8-7   & 72.784  & 2008 Jan &ARO &  $\sim$\ 87.1  & 0.13-0.26   & 0.40 \\
			
			\noalign{\smallskip}\hline
		\end{tabular}
	\end{center}
\end{table}

\section{Results}
\label{ 3 sect :results }
Multiple transition lines (24-23, 17-16, 11-10, 8-7) of HC$_3$N were detected for most sources. However, only two transitions of $J$\,=\,17-16 and $J$\,=\,11-10 were detected for W44. The observational results of these lines are listed in Table \ref{tabel 3}, including peak intensity, velocity-integrated intensity, central velocity, line width, and line wing information. For most sources, the spectral lines have high S/N\,\textgreater\,5\,$\sigma$. Sources with line wings are plotted in Figure \ref{figure 6}, and the Figure \ref{Figure A1} shows the HC$_3$N (24-23) lines without line wings. It is necessary to note that the $V_{\rm LSR}$ of HC$_3$N (24-23) observed by the CSO telescope seems to have a systematic bias of about $\pm$1 km\,s$^{-1}$ compared with the other three transitions of HC$_3$N (17-16, 11-10, 8-7) observed by the ARO telescope. This is probably due to a doppler tracking problem for the DSB receiver of CSO.


\begin{table}
\caption{Observational results of HC$_3$N (24-23, 17-16, 11-10, 8-7) lines. The columns are: (1) Source name. (2) and (3) Coordinates. (4) Transition. (5) Intensity. (6) Velocity-integrated intensity. (7) Line center. (8) FWHM line width. (9) Line wing (bw, blue wing; rw, red wing; ws, wing on both sides).}
\label{tabel 3} 	
\begin{tabular}{c c c c c c c c c c}
\hline
\hline
{Source}&{RA}&{DEC}  &{Line}& $T_{\rm mb}$ &{$I_{\rm HC_3N}$}&{$V_{\rm LSR}$}   &{FWHM} & {Line wings}\\
&(J2000)&(J2000)&  (HC$_3$N)&    (K)   & (K\,km\,s$^{-1}$)&(km\,s$^{-1}$)&(km\,s$^{-1}$) \\
(1)&(2)&(3)  &(4)& (5) & (6)&(7)&(8)&(9)\\
\hline

W3(OH)       &02:27:04.69&61:52:25.5&24-23&0.2&$0.8\pm0.1$&$-48.1\pm0.2$&$4.4\pm0.4$&...\\
&&&17-16&0.8 &$3.0 \pm0.1 $&$-47.2 \pm0.1 $&$3.3 \pm0.1 $&...\\
&&&11-10&0.9 &$3.1 \pm0.0 $&$-47.0 \pm0.0 $&$3.5 \pm0.1 $&...\\
&&&8-7&0.7 &$2.7 \pm0.1 $&$-47.1 \pm0.1 $&$3.8 \pm0.1 $&...\\
RCW142     &17:50:15.13&$-$28:54:31.5&24-23&0.2 &$1.3 \pm0.1 $&$16.0 \pm0.1 $&$6.1 \pm0.3 $&bw\\
&&&17-16&0.9 &$4.4 \pm0.1 $&$17.2 \pm0.0 $&$4.7 \pm0.1 $&ws\\
&&&11-10&1.4 &$6.4 \pm0.1 $&$17.1 \pm0.0 $&$4.3 \pm0.1 $&ws\\
&&&8-7&1.7 &$6.6 \pm0.2 $&$17.2 \pm0.0 $&$3.7 \pm0.1 $&rw\\
W28A2      &18:00:30.42& $-$24:03:58.5&24-23&1.5 &$8.0 \pm0.1 $&$8.0 \pm0.0 $&$5.0 \pm0.1 $&rw\\
&&&17-16&3.5 &$16.5 \pm0.1 $&$9.0 \pm0.0 $&$4.5 \pm0.0 $&rw\\
&&&11-10&3.0 &$11.4 \pm0.1 $&$9.1 \pm0.0 $&$3.5 \pm0.1 $&rw\\
&&&8-7&3.0 &$11.4 \pm0.1 $&$9.1 \pm0.0 $&$3.5 \pm0.1 $&rw\\
M8E        &18:04:53.25&$-$24:26:42.4&24-23&0.1 &$0.3 \pm0.0 $&$9.8 \pm0.1 $&$2.5 \pm0.3 $&...\\
&&&17-16&0.8 &$2.0 \pm0.0 $&$10.8 \pm0.0 $&$2.2 \pm0.1 $&...\\
&&&11-10&1.4 &$2.9 \pm0.0 $&$10.8 \pm0.0 $&$2.0 \pm0.0 $&...\\
&&&8-7&1.2 &$2.5 \pm0.1 $&$10.6 \pm0.0 $&$2.0 \pm0.1 $&...\\
G8.67-0.36 &18:06:18.87&$-$21:37:37.8&24-23&0.3 &$1.0 \pm0.1 $&$34.0 \pm0.1 $&$3.7 \pm0.2 $&rw\\
&&&17-16&0.6 &$2.5 \pm0.1 $&$34.7 \pm0.0 $&$3.8 \pm0.1 $&rw\\
&&&11-10&1.2 &$5.7 \pm0.0 $&$35.0 \pm0.0 $&$4.3 \pm0.0 $&rw\\
&&&8-7&1.1 &$5.3 \pm0.2 $&$35.0 \pm0.1 $&$4.4 \pm0.2 $&rw\\
W31        &18:08:38.32&$-$19:51:49.7&24-23&0.3 &$3.0 \pm0.1 $&$65.9 \pm0.1 $&$8.9 \pm0.3 $&...\\
&&&17-16&0.5 &$5.3 \pm0.1 $&$66.5 \pm0.1 $&$9.2 \pm0.2 $&...\\
&&&11-10&0.6 &$5.4 \pm0.1 $&$66.8 \pm0.0 $&$7.8 \pm0.1 $&...\\
&&&8-7&0.7 &$5.6 \pm0.2 $&$67.4 \pm0.1 $&$7.5 \pm0.3 $&...\\
G10.6-0.4  &18:10:28.70&$-$19:55:48.6&24-23&0.5 &$3.5 \pm0.1 $&$-4.1 \pm0.1 $&$6.8 \pm0.1 $&...\\
&&&17-16&1.1 &$7.5 \pm0.1 $&$-3.2 \pm0.0 $&$6.7 \pm0.1 $&...\\
&&&11-10&1.4 &$8.9 \pm0.1 $&$-3.0 \pm0.0 $&$6.0 \pm0.0 $&bw\\
&&&8-7&1.2 &$6.9 \pm0.2 $&$-3.0 \pm0.1 $&$5.4 \pm0.2 $&bw\\

\noalign{\smallskip}\hline
\end{tabular}
\end{table}

\addtocounter{table}{-1}
\begin{table}
\caption{$-$ Continued.}
\label{tabel 3} 	
\begin{tabular}{c c c c c c c c c c}
\hline
\hline
{Source}&{RA}&{DEC}  &{Line}& $T_{\rm mb}$ &{$I_{\rm HC_3N}$}&{$V_{\rm LSR}$}   &{FWHM} & {Line wings}\\
&(J2000)&(J2000)&  (HC$_3$N)&    (K)   & (K\,km\,s$^{-1}$)&(km\,s$^{-1}$)&(km\,s$^{-1}$) \\
(1)&(2)&(3)  &(4)& (5) & (6)&(7)&(8)&(9)\\
\hline
W33cont    &18:14:13.67&$-$17:55:25.2&24-23&0.6 &$2.6 \pm0.0 $&$33.8 \pm0.0 $&$4.0 \pm0.1 $&ws\\
&&&17-16&2.1 &$9.5 \pm0.1 $&$34.9 \pm0.0 $&$4.3 \pm0.0 $&rw\\
&&&11-10&2.7 &$12.1 \pm0.0 $&$35.2 \pm0.0 $&$4.2 \pm0.0 $&rw\\
&&&8-7&2.3 &$10.3 \pm0.1 $&$35.2 \pm0.0 $&$4.2 \pm0.1 $&ws\\		
G14.33-0.64  &18:18:54.71&$-$16:47:49.7&24-23&0.2 &$0.5 \pm0.1 $&$21.0 \pm0.2 $&$3.0 \pm0.4 $&...\\
&&&17-16&1.0 &$3.1 \pm0.0 $&$22.0 \pm0.0 $&$2.9 \pm0.1 $&rw\\
&&&11-10&1.7 &$4.8 \pm0.0 $&$22.1 \pm0.0 $&$2.6 \pm0.0 $&rw\\
&&&8-7&1.6 &$3.9 \pm0.1 $&$22.2 \pm0.0 $&$2.2 \pm0.1 $&...\\
W42          &18:36:12.46&$-$07:12:10.1&24-23&0.2 &$1.2 \pm0.1 $&$110.0 \pm0.2 $&$6.6 \pm0.4 $&...\\
&&&17-16&0.5 &$3.3 \pm0.1 $&$110.5 \pm0.1 $&$5.7 \pm0.2 $&rw\\
&&&11-10&0.7 &$2.9 \pm0.0 $&$110.1 \pm0.0 $&$3.8 \pm0.1 $&rw\\
&&&8-7&1.0 &$3.7 \pm0.1 $&$110.0 \pm0.1 $&$3.6 \pm0.2 $&...\\
S76E         &18:56:10.43&07:53:14.1&24-23&0.2 &$0.5 \pm0.0 $&$31.8 \pm0.1 $&$3.1 \pm0.3 $&...\\
&&&17-16&0.9 &$2.7 \pm0.1 $&$32.8 \pm0.0 $&$2.7 \pm0.1 $&...\\
&&&11-10&1.8 &$5.4 \pm0.0 $&$32.7 \pm0.0 $&$2.8 \pm0.0 $&...\\
&&&8-7&1.6 &$4.9 \pm0.1 $&$32.8 \pm0.0 $&$2.8 \pm0.1 $&...\\
G35.20-0.74  & 18:58:12.73 & 01:40:36.5 &24-23 & 0.1 & $0.4\pm0.0$ & $32.5\pm0.2$ & $4.0\pm0.5$ &...\\
&&&17-16&0.6 &$2.9 \pm0.1 $&$34.2 \pm0.1 $&$4.5 \pm0.1 $&...\\
&&&11-10&1.2 &$5.5 \pm0.0 $&$34.0 \pm0.0 $&$4.1 \pm0.0 $&...\\
&&&8-7&1.3 &$5.3 \pm0.1 $&$34.0 \pm0.0 $&$3.9 \pm0.1 $&...\\
W51M         &19:23:43.86&14:30:29.4&24-23&0.7 &$6.8 \pm0.1 $&$56.2 \pm0.1 $&$8.7 \pm0.1 $&...\\
&&&17-16&1.5 &$13.9 \pm0.1 $&$57.2 \pm0.0 $&$9.0 \pm0.1 $&ws\\
&&&11-10&1.3 &$13.5 \pm0.1 $&$57.2 \pm0.0 $&$9.9 \pm0.1 $&ws\\
&&&8-7&1.0 &$11.4 \pm0.2 $&$56.9 \pm0.1 $&$10.3 \pm0.2 $&...\\
W75N         & 20:38:36.93&42:37:37.4&24-23&0.4 &$2.0 \pm0.1 $&$8.5 \pm0.1 $&$5.2 \pm0.2 $&...\\
&&&17-16&0.9 &$4.4 \pm0.1 $&$9.3 \pm0.0 $&$4.4 \pm0.1 $&rw\\
&&&11-10&1.4 &$4.8 \pm0.1 $&$9.4 \pm0.0 $&$3.3 \pm0.0 $&rw\\
&&&8-7&1.1 &$3.1 \pm0.1 $&$9.5 \pm0.0 $&$2.6 \pm0.1 $&bw\\
		
\noalign{\smallskip}\hline
\end{tabular}	
\end{table}

\addtocounter{table}{-1}
\begin{table}
\caption{$-$ Continued.}
\label{tabel 3} 	
\begin{tabular}{c c c c c c c c c c}
\hline
\hline
{Source}&{RA}&{DEC}  &{Line}& $T_{\rm mb}$ &{$I_{\rm HC_3N}$}&{$V_{\rm LSR}$}   &{FWHM} & {Line wings}\\
&(J2000)&(J2000)&  (HC$_3$N)&    (K)   & (K\,km\,s$^{-1}$)&(km\,s$^{-1}$)&(km\,s$^{-1}$) \\
(1)&(2)&(3)  &(4)& (5) & (6)&(7)&(8)&(9)\\
\hline
DR21S        &20:39:00.80&42:19:29.7&24-23&0.5 &$1.8 \pm0.0 $&$-3.0 \pm0.0 $&$3.4 \pm0.1 $&bw\\
&&&17-16&2.0 &$5.9 \pm0.1 $&$-2.1 \pm0.0 $&$2.8 \pm0.0 $&bw\\
&&&11-10&3.0 &$7.8 \pm0.1 $&$-2.1 \pm0.0 $&$2.5 \pm0.0 $&bw\\
&&&8-7&2.1 &$5.0 \pm0.1 $&$-2.0 \pm0.0 $&$2.2 \pm0.1 $&bw\\		
W75(OH)      &20:39:01.00&42:22:49.8&24-23&0.3 &$1.9 \pm0.1 $&$-3.8 \pm0.1 $&$5.5 \pm0.2 $&...\\
&&&17-16&0.9 &$4.8 \pm0.1 $&$-3.0 \pm0.0 $&$5.0 \pm0.1 $&...\\
&&&11-10&2.2 &$9.9 \pm0.1 $&$-3.1 \pm0.0 $&$4.2 \pm0.0 $&rw\\
&&&8-7&1.7 &$6.7 \pm0.2 $&$-3.2 \pm0.0 $&$3.8 \pm0.1 $&rw\\			
Cep	A   &22:56:18.13&62:01:46.3&24-23&0.3 &$1.6 \pm0.1 $&$-11.4 \pm0.1 $&$4.4 \pm0.3 $&...\\
&&&17-16&0.8 &$2.5 \pm0.1 $&$-10.0 \pm0.0 $&$2.8 \pm0.1 $&bw\\
&&&11-10&1.3 &$4.1 \pm0.0 $&$-10.4 \pm0.0 $&$3.0 \pm0.0 $&bw\\
&&&8-7&1.2 &$3.6 \pm0.1 $&$-10.5 \pm0.0 $&$2.9 \pm0.1 $&...\\
NGC7538 & 23:13:44.85&61:26:50.6&24-23&0.1 &$0.6 \pm0.1 $&$-56.6 \pm0.3 $&$5.1 \pm0.7 $&...\\
&&&17-16&0.5 &$2.4 \pm0.1 $&$-55.6 \pm0.1 $&$4.4 \pm0.2 $&...\\
&&&11-10&1.1 &$5.1 \pm0.0 $&$-56.0 \pm0.0 $&$4.4 \pm0.0 $&...\\
&&&8-7&1.1 &$4.8 \pm0.1 $&$-56.2 \pm0.0 $&$4.1 \pm0.1 $&...\\
W44     &18:53:18.5 & 61:26:50.6 & 24-23 & $...$ & $<$ 0.20$^{*}$ & $...$ &$...$&...\\
&&& 17-16 & 1.1 & $7.6\pm0.1$ & $58.6\pm0.0$&$6.5\pm0.1$&...\\ 
&&& 11-10 & 1.7 & $9.4\pm0.1$ & $58.3\pm0.0$&$5.2\pm0.0$&...\\
&&& 8-7 & $...$ & $<$ 0.71$^{*}$ & $...$ & $...$ &...\\
				
\noalign{\smallskip}\hline
\end{tabular}

Notes: The asterisk $^{*}$ shows 3\,$\sigma$ upper limit, $\sigma$ = RMS $\sqrt{\delta V\cdot\Delta V}$. We assume $\Delta V$(24-23) = 6.5 and $\Delta V$(8-7) = 5.2 for W44 using line widths of HC$_3$N 17-16 and 11-10, respectively.  
	
\end{table}

\subsection{Line wings}
Line wings  can be seen in  RCW142, W28A2, G8.67-0.36, G10.6-0.4, W33cont, G14.33-0.64, W42, W51M, W75N, DR21S, W75(OH) and CepA. They are marked in Figure \ref{figure 6}. Detailed information on them is listed in Table \ref{tabel 3}.

Among these 12 clumps, G10.6-0.4 and W51M contain CHII regions. The remaining clumps contain UCHII regions, except for W75(OH) \citep{2003ApJS..149..375S}. The evolutionary status of the HII region of W75(OH) is unclear. Considering the dynamical signatures, we also marked the type of outflows based on the outflow catalog of \citet {2004A&A...426..503W} and present the infall signatures from \citet{2010ApJS..188..313W} in Table \ref{tabel 1}. \citet {2004A&A...426..503W} detected bipolar CO molecular outflows in W28A2, W75N, and CepA, and a blue monopolar CO molecular outflow in DR21S. The remaining clumps were not included in their catalog. Infall signatures were reported in RCW142, G8.67-0.36, G10.6-0.4, W51M, DR21S, W75(OH), and CepA \citep{2010ApJS..188..313W}.

\subsection{Line widths}
\begin{figure}
	\vskip5pt 
	
	\begin{minipage}[t]{0.495\linewidth}
		\centering
		\includegraphics[width=90mm]{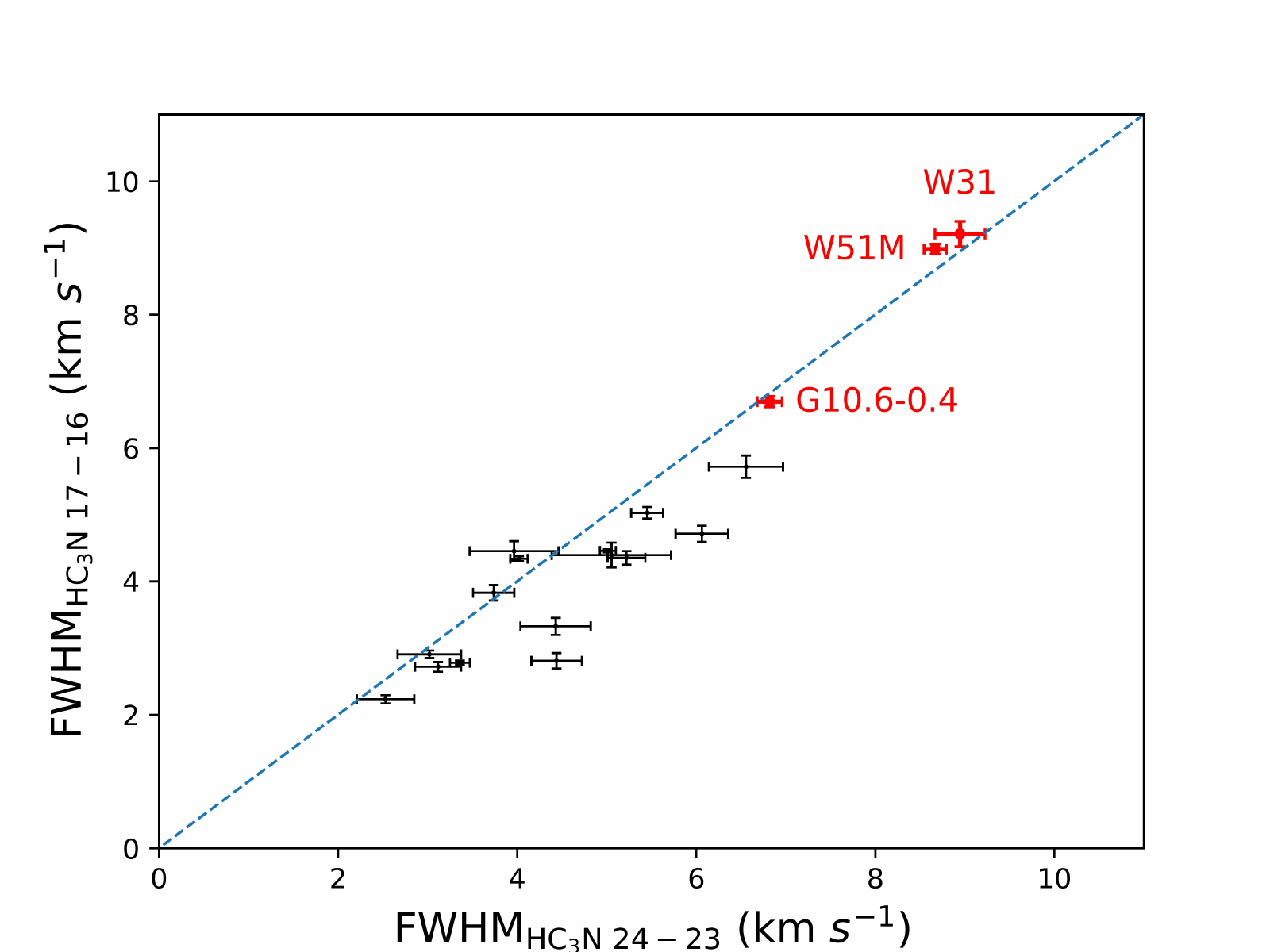}
	\end{minipage}%
	\begin{minipage}[t]{0.495\textwidth}
		\centering
		\includegraphics[width=90mm]{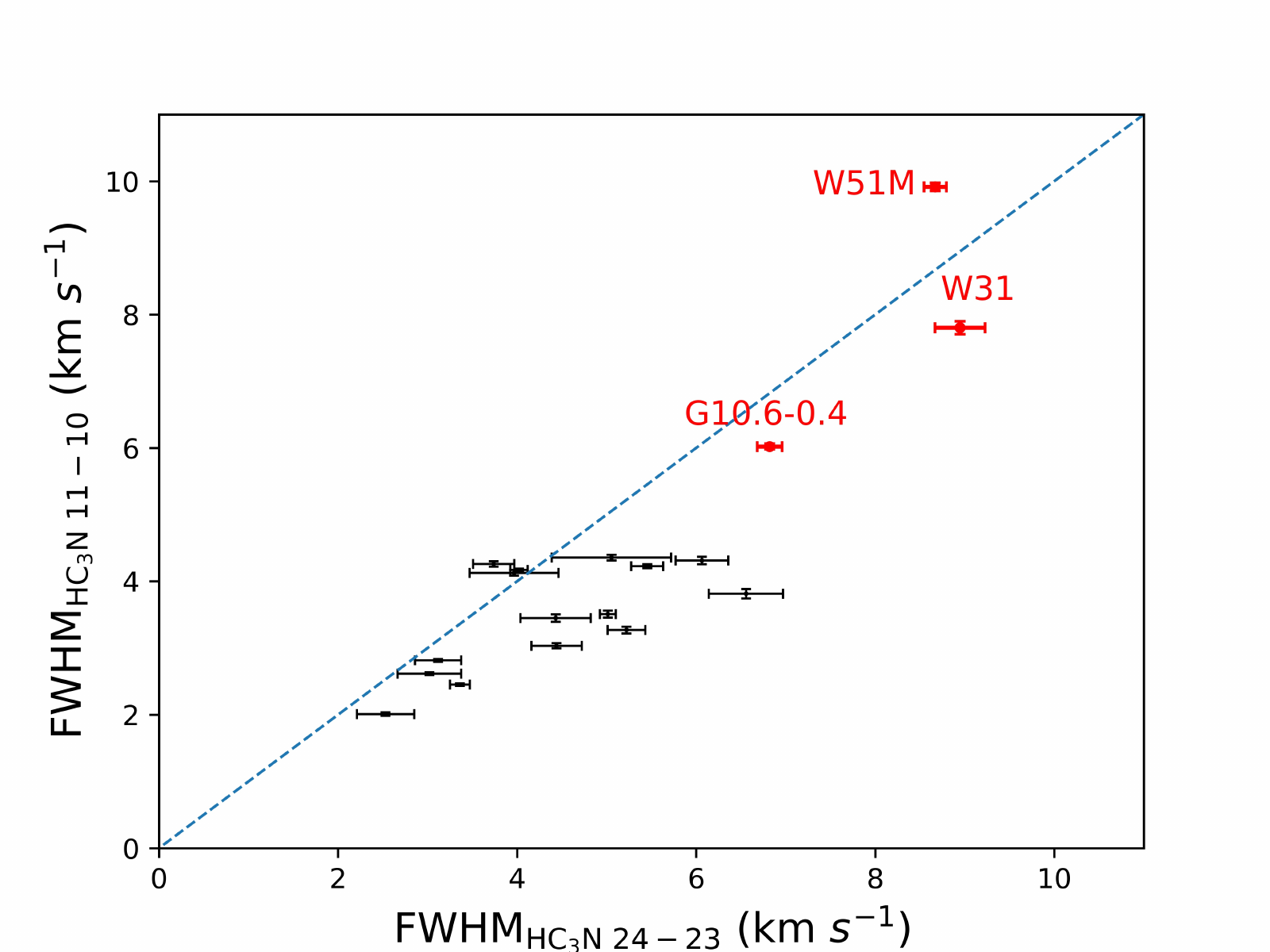}
	\end{minipage}%
	
	\begin{minipage}[t]{0.495\linewidth}
		\centering
		\includegraphics[width=90mm]{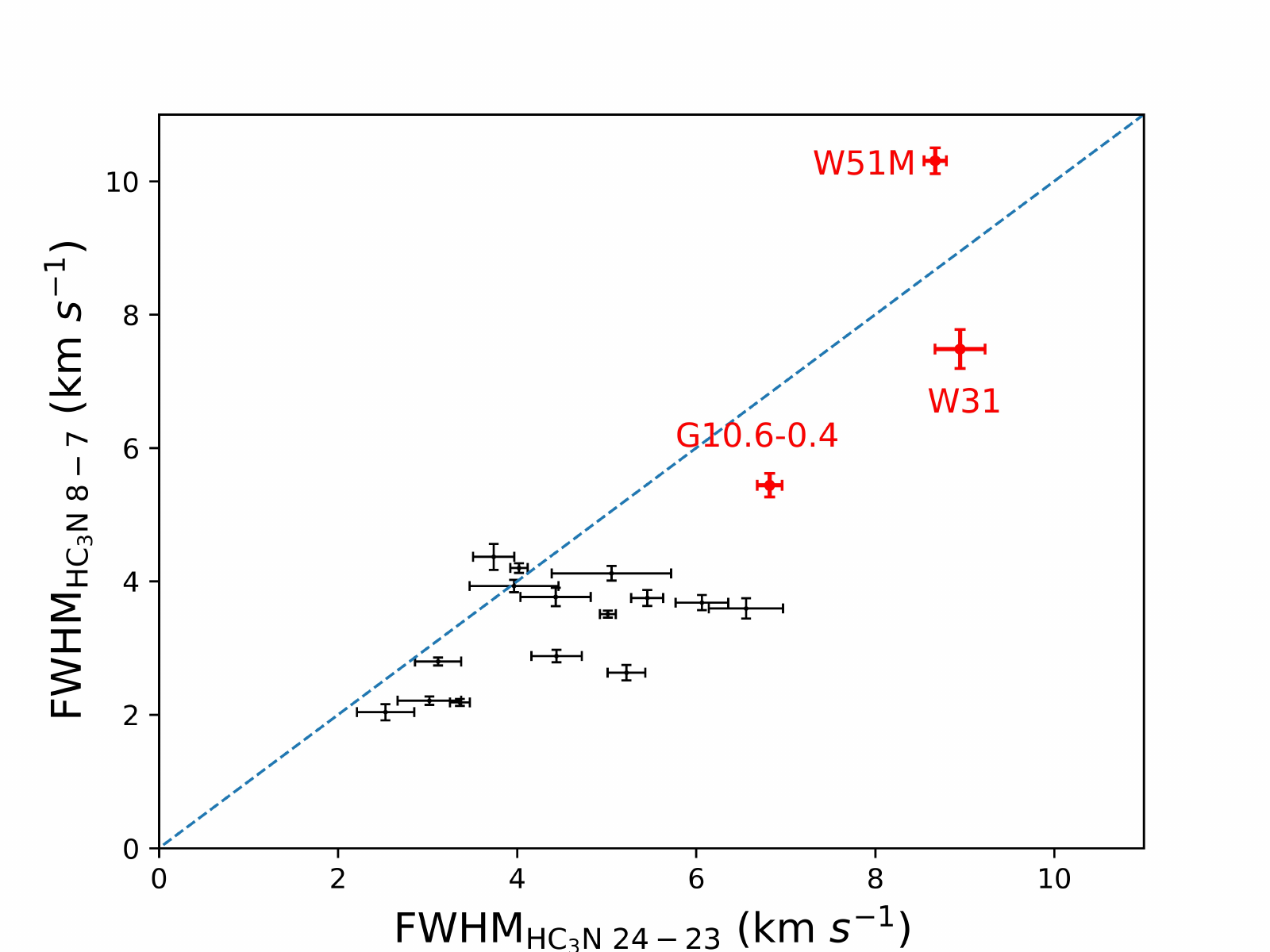}
	\end{minipage}%
	\begin{minipage}[t]{0.495\textwidth}
		\centering
		\includegraphics[width=90mm]{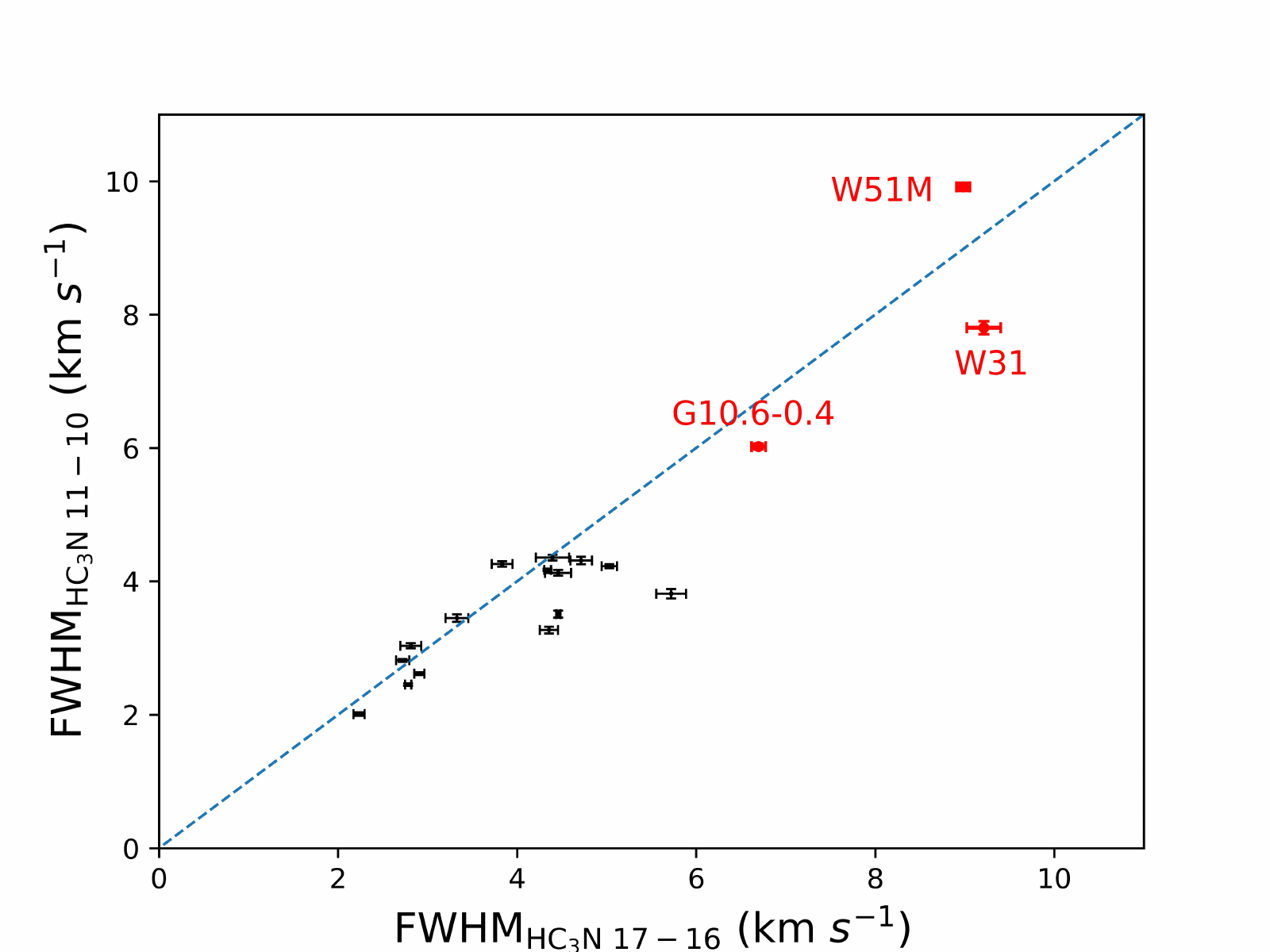}
	\end{minipage}%
	
	\begin{minipage}[t]{0.495\linewidth}
		\centering
		\includegraphics[width=90mm]{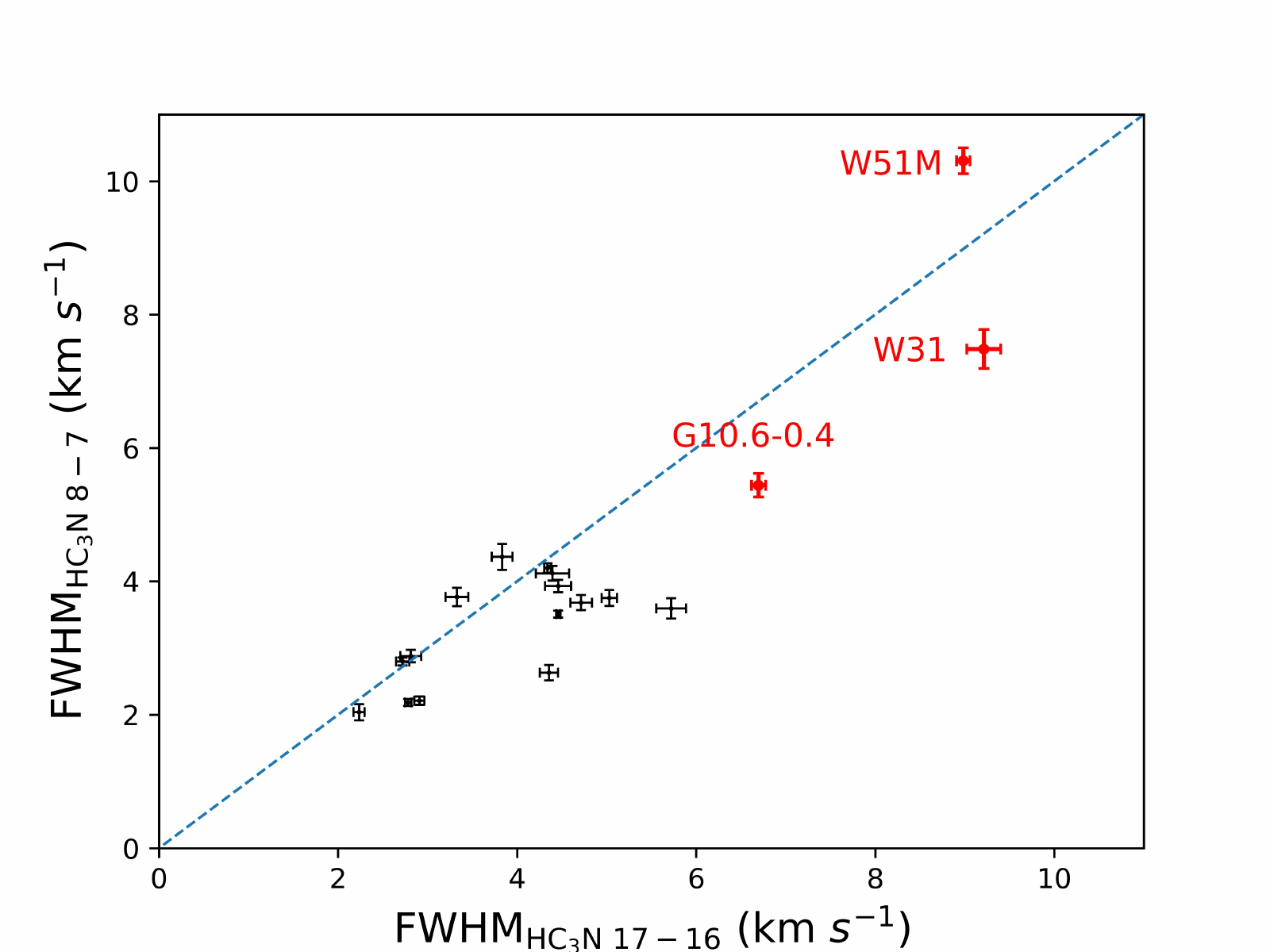}
	\end{minipage}%
	\begin{minipage}[t]{0.495\textwidth}
		\centering
		\includegraphics[width=90mm]{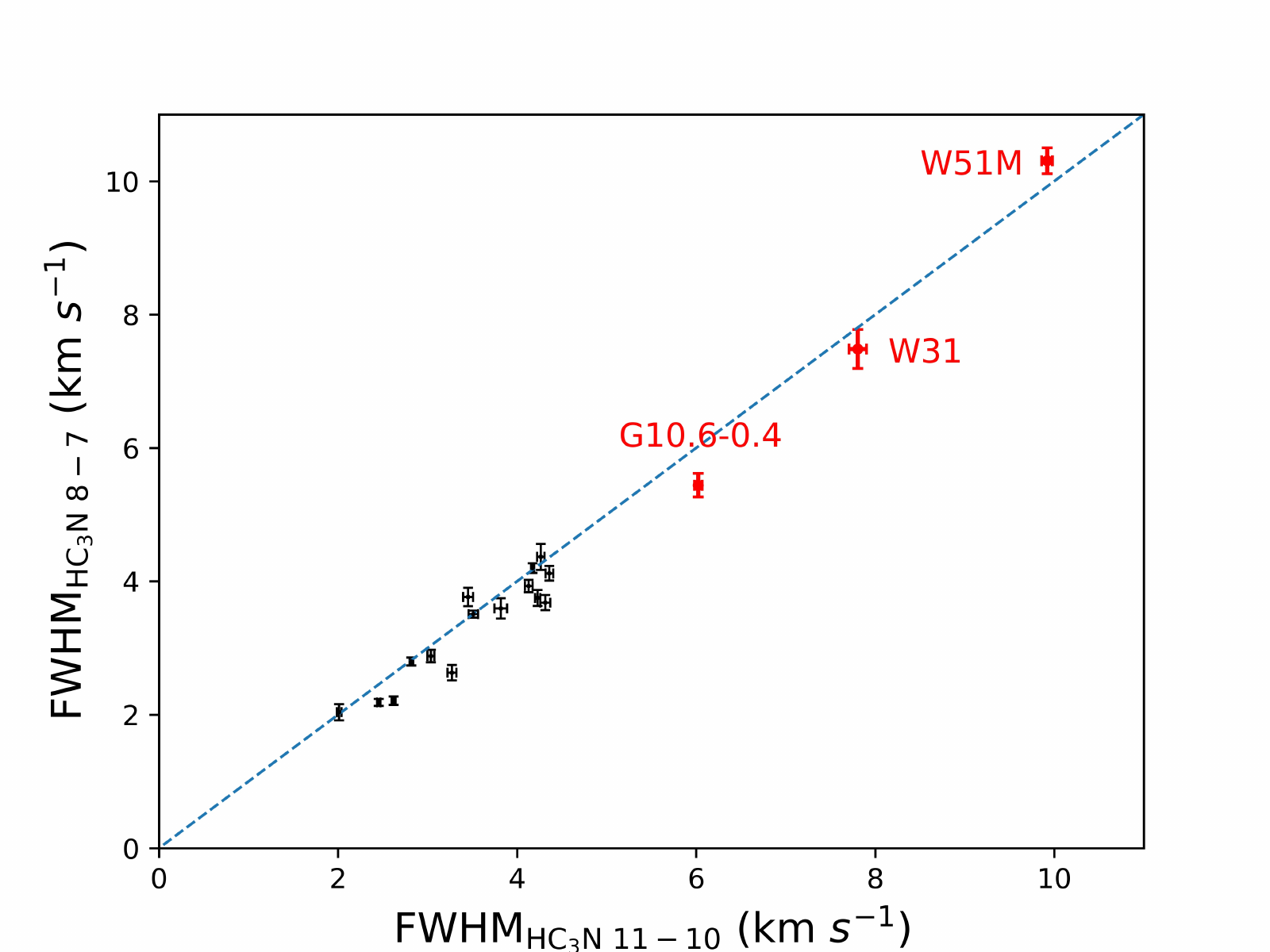}
	\end{minipage}%
	\caption{Comparisons of line widths between each of four transitions for the 19 sources except W44. The slope of the blue dashed line is 1.}
	
	\label{figure 1}	
\end{figure}

For all sources except W44, comparisons of line widths between each pair of four transitions are presented in Figure \ref{figure 1}. For most of these sources, the line widths of high-$J$ transitions are larger than the low-$J$ ones, which can also be seen in the statistics of mean and median line widths of each transition (See Table \ref{tabel 4}). The mean and median line widths are 5.02$\pm$0.27, 4.71$\pm$0.1, 4.28$\pm$0.04, 4.04$\pm$0.12 and 4.73, 4.4, 4.13, 3.64 for HC$_3$N 24-23, 17-16, 11-10, and 8-7, respectively. In Figure \ref{figure 1}, W31, W51M, and G10.6-0.4 are highlighted because their line widths are noticeably wider than the others. In addition, these three sources also present infall signatures referred from \citet{2010ApJS..188..313W} (see Table \ref{tabel 1}).

\begin{figure*}
	
	\vskip0pt 
	\begin{center}
		\includegraphics[width=10cm]{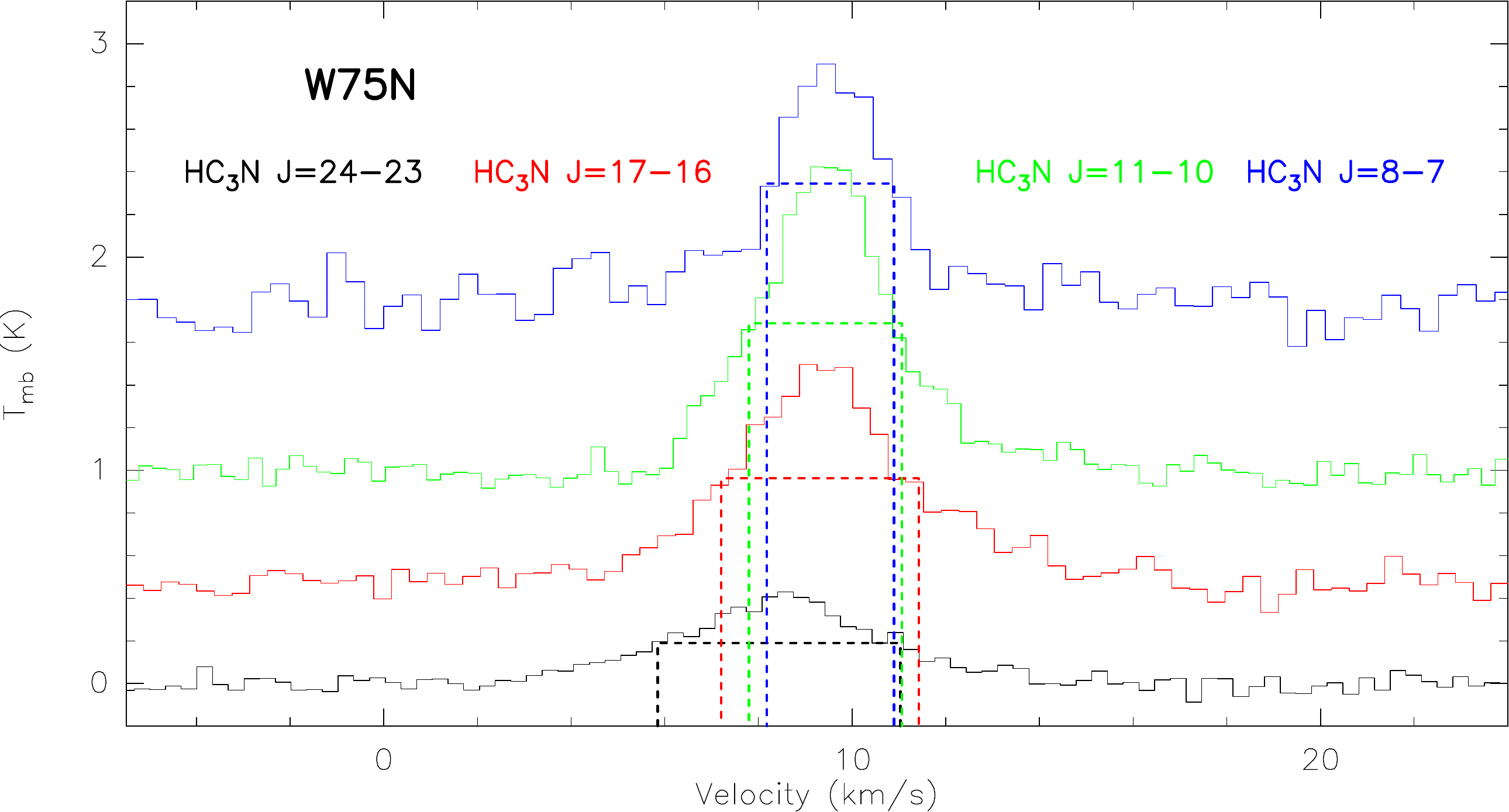}
	\end{center}
	
	\vskip0pt 
	\begin{center}
		\includegraphics[width=10cm]{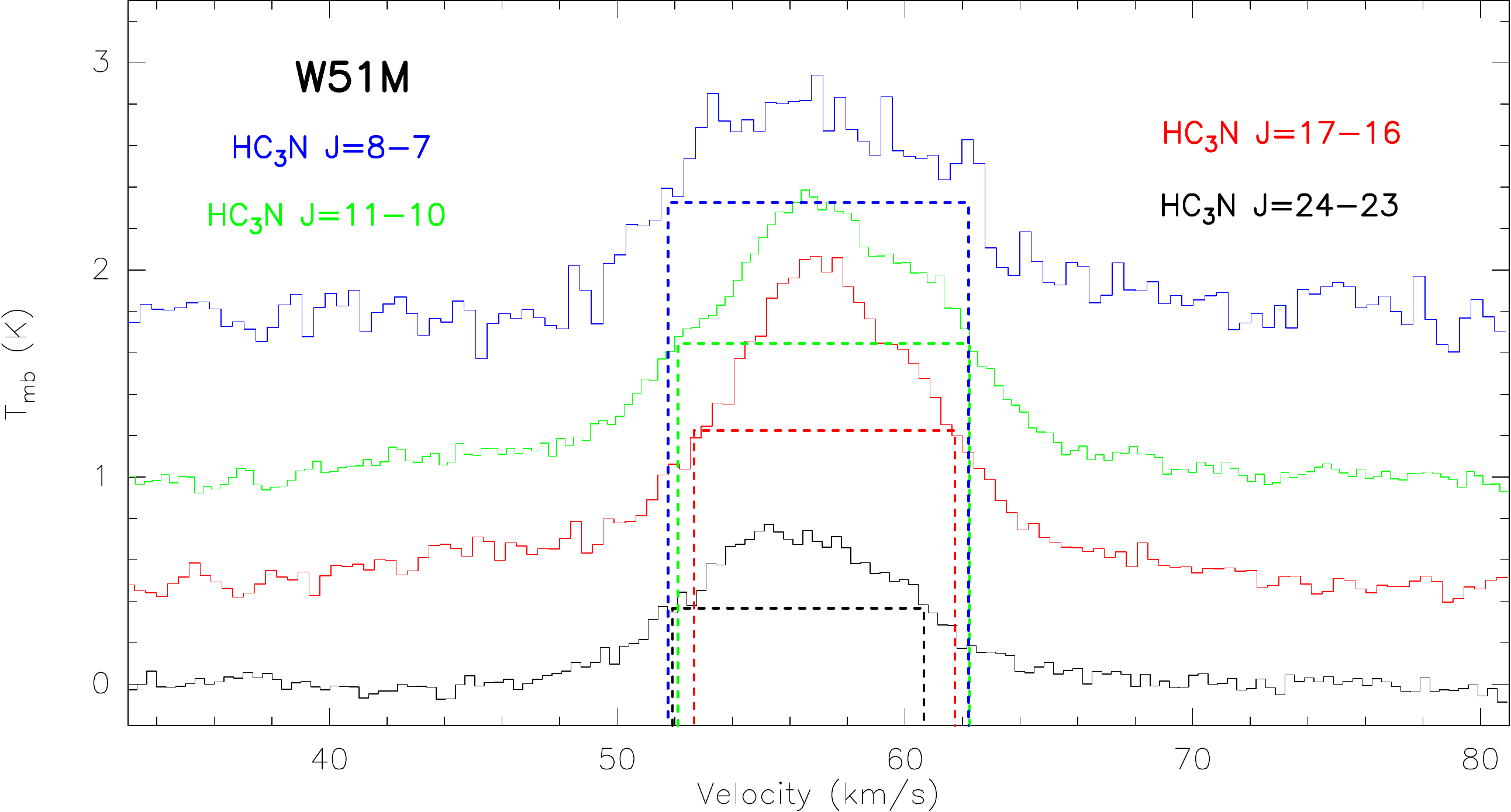}
	\end{center}
	
	\caption{Comparisons of FWHM line widths of four transitions of HC$_3$N for W75N and W51M. The horizontal dashed lines represent the FWHM line width. The vertical dashed lines are used to compare the FWHM line widths.}
	
	\label{figure 2}	
\end{figure*}

Some sources in our sample show regularity between their $J$-levels and HC$_3$N line widths. The line widths of HC$_3$N for W75N (figure \ref{figure 2} upper panel), RCW142, W75(OH), and DR21S increase with the $J$-levels. The line widths of HC$_3$N for W51M increase as the $J$-levels decrease (figure \ref{figure 2} lower panel). For S76E and W33cont, their HC$_3$N line widths keep roughly constant for different $J$-levels (table \ref{tabel 1}). For other sources, their HC$_3$N line widths show no obvious trend with change of HC$_3$N $J$-levels.

\begin{table}[h]
	\caption{Statistics of line widths for HC$_3$N (24-23, 17-16, 11-10, 8-7) transitions.}  
	\label{tabel 4}
	\begin{center}
		\begin{tabular}{c c c c c }
			\hline
			\hline
			{HC$_3$N}&{$\Delta V$(24-23)} &{$\Delta V$(17-16)}&{$\Delta V$(11-10)}&{$\Delta V$(8-7)}\\
			&(km\,s$^{-1}$)       &(km\,s$^{-1}$)      &(km\,s$^{-1}$)      &(km\,s$^{-1}$)   \\
			\hline
			Mean  & $5.02\pm0.27$ & $4.71\pm0.10$   &  $4.28\pm0.04$  & $4.04\pm0.12$   \\ 
			Median& 4.73 & 4.40 &  4.13   & 3.64  \\
			\noalign{\smallskip}\hline
			
		\end{tabular}
	\end{center}
\end{table}


\subsection{Excitation temperature}

We calculated the excitation temperature ($T_{\rm ex}$) of 18 sources using the velocity-integrated intensity ratio between two different transitions of the same molecule through \citep{Beijing: Publisher of Chinese Science and Technology}

\begin{equation}
$$R_{\rm nj} = (\frac{n}{j})\ {\rm exp}(-\frac{\Delta E_{\rm nj}}{k T_{\rm ex}})$$ 
\label{eq:equation 1}
\end{equation}
 
where $R_{\rm nj}$ is the ratio of the velocity-integrated intensity in the brightness temperature of HC$_3$N$\,J\,=\,n\rightarrow n-1$ to that of HC$_3$N$\,J\,=\,j\rightarrow j-1$ (n \textgreater j). $\Delta E_{\rm nj}$ is the energy difference between $J$ = $n$ and $J$ = $j$. $k$ and the $T_{\rm ex}$ are the Boltzmann constant and the excitation temperature, respectively. We used the velocity-integrated intensity ratio of  HC$_3$N (24-23) to (17-16) to obtain the excitation temperature of HC$_3$N for these clumps, which are listed in Table \ref{tabel 1}. The excitation temperatures range from 22.9 to 57 K, with a median value of 35.9 K.

Because HC$_3$N (24-23) and (17-16) have similar beam size (See Table \ref{tabel 2}), $R_{\rm 24,17}$ can partly avoid the effect of the filling factor. In addition, the angular extent of CS (5-4) mapping observations (See Table \ref{tabel 1}) is more extended than the beam sizes of HC$_3$N (24-23) and (17-16) \citep{2010ApJS..188..313W}. If we assume HC$_3$N (24-23) and (17-16) emissions extend to regions that are similar to those of CS (5-4), it is reasonable to expect similar filling factors of HC$_3$N (24-23) and (17-16) lines.

\section{Discussions}
\label{ 4 sect :discus }

\subsection{The origin of line wings of optically thin dense gas tracers}

The line wings should be from non-Gaussian motions of the molecular gas, including outflows, infall, etc. Since the HC$_3$N lines are normally optically thin over the entire  velocity range, the relative fraction of high-speed non-Gaussian motions to the total gas is proportional to the flux ratio. Thus, gas with non-Gaussian motion is significant if line wings can be identified with such optically thin dense gas tracers. For most of our sources with line wings, the wings appear in high-$J$ levels of $J$\,=\,24-23 or $J$\,=\,17-16, which implies that non-Gaussian motions may be from denser and warmer regions for these sources. Among the 12 sources with line wings, CO molecular outflows were reported in 4 \citep{2004A&A...426..503W}, while infall signatures were reported in 7 \citep{2010ApJS..188..313W}. Both molecular outflows and infalls may produce line wings, while molecular outflow is one type of feedback of massive star formation to the surrounding molecular cloud.

We also checked possible line blending for line wings of HC$_3$N 8-7, 11-10, 17-16, and 24-23 from  other molecules with Splatalogue\footnote{https://splatalogue.online}. After ruling out lines with upper-level energy ($E_{up}$) higher than 300\,K, isotopic lines, or complex molecules with extremely low expected abundance,  we found only one line, CCS\,$10_{11}-10_{10}$ at a rest frequency of 72.783\,GHz, which may cause line blending for HC$_3$N (8-7). No such lines have effects on other HC$_3$N transitions. It is possible that the red wings of G8.67-0.36, W33cont, and W75(OH) were affected by line blending from CCS\,$10_{11}-10_{10}$, as shown in Figure \ref{figure 6}.

\subsection{Line broadening in regions with different densities }
The line information about the state of turbulence and the main gas-emitting region can be inferred from the FWHM line width of Gaussian fits to the observed profiles \citep{2002ApJ...566..945B}. The thermal broadening contribution to the line width of massive clumps is negligible. Assuming $T_{\rm k}$\,=\,50\,K and a mean molecular weight of $\mu$\,=\,51, thermal broadening accounts for only 0.16\,km\,${\rm s}^{-1}$, whereas the smallest line width in our sample is 2\,km\,${\rm s}^{-1}$ (M8E). Thermal broadening only accounts for a small fraction of this line width. The turbulent line width of HC$_3$N (24-23, 17-16, 11-10, 8-7) for the sources in our sample is clearly supersonic.

The HC$_3$N lines trace ``hot dense" gas. Crudely speaking, the excitation of higher-$J$ transitions requires higher temperature and higher critical density, which traces denser, warmer and inner parts of the core, than the excitation of lower-$J$ transitions. Based  on the line width comparisons in Figure \ref{figure 1} and the statistics shown in Table \ref{tabel 4}, we can see that the line width increases with increasing $J$-level in most of the sources, which indicates that the inner dense, warm regions have more violent turbulence, or other motions like rotation, than outer regions.  Such a line broadening trend can be clearly seen in RCW142, W75N, DR21S, and W75(OH) (upper panel of Figure \ref{figure 2}). This trend had been found with multiple CS transitions, and has been explained by star formation feedback \citep{2010ApJS..188..313W}. However, the line width of W51M increases with decreasing $J$-level (bottom panel of Figure \ref{figure 2}), while different HC$_3$N lines have almost the same line widths in S76E and W33cont. Line broadening in different density regions in massive star-forming regions cannot be explained with one simple model. Sources with larger line widths in the inner denser, warmer regions might need more time to dissipate turbulence and/or angular momentum and initiate dynamical collapse than others.

W31, W51M and G10.6-0.4, which are highlighted in Figure \ref{figure 1}, showed noticeably broader line widths than other sources. Such broad line widths indicate that there is more violent turbulence in these sources than in others. Infall signatures in these sources have been reported previously \citep{2010ApJS..188..313W}.

\begin{figure*}
	\vskip5pt 
	
	\begin{minipage}[t]{0.495\linewidth}
		\centering
		\includegraphics[width=90mm]{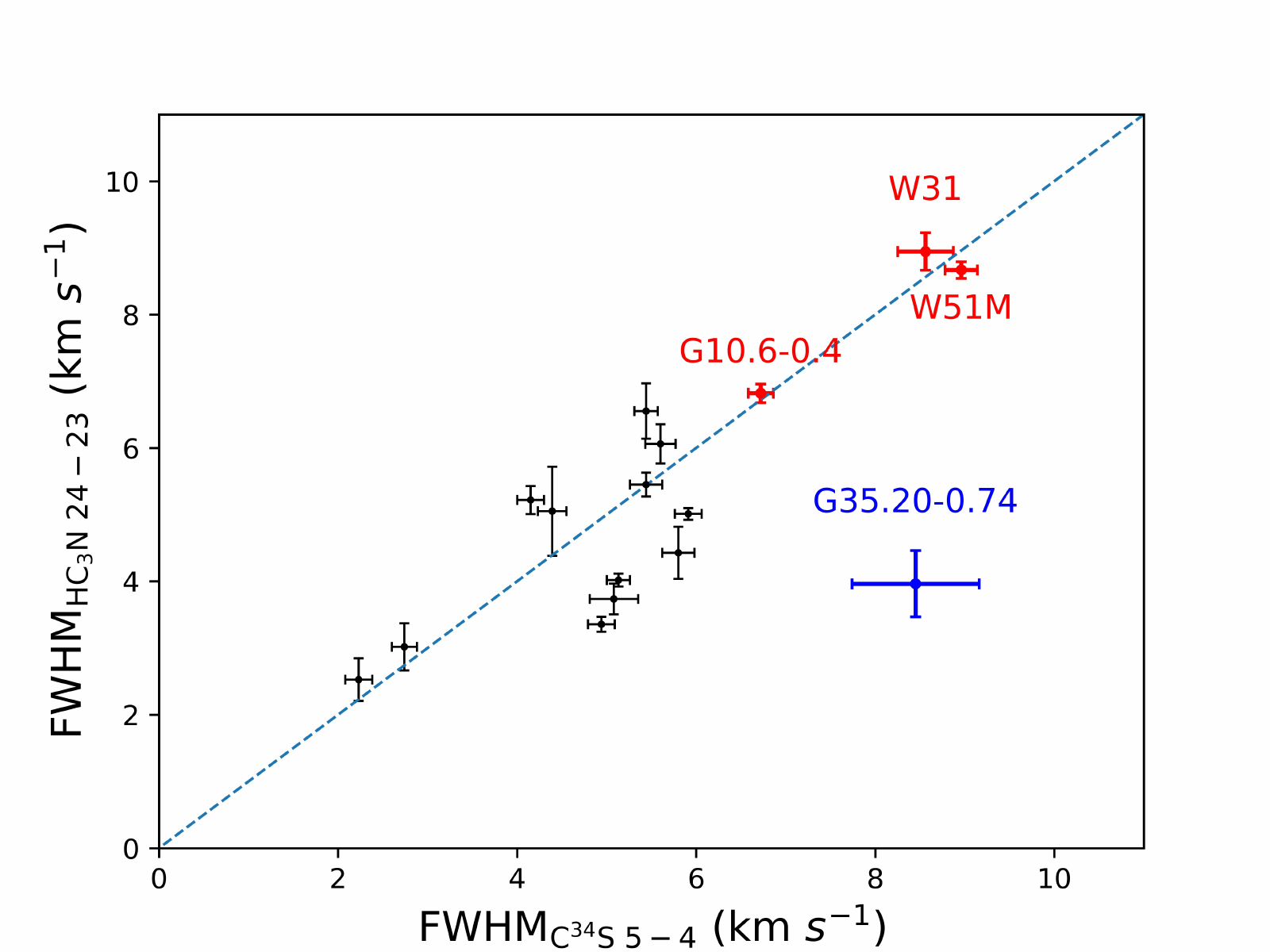}
	\end{minipage}%
	\begin{minipage}[t]{0.495\textwidth}
		\centering
		\includegraphics[width=90mm]{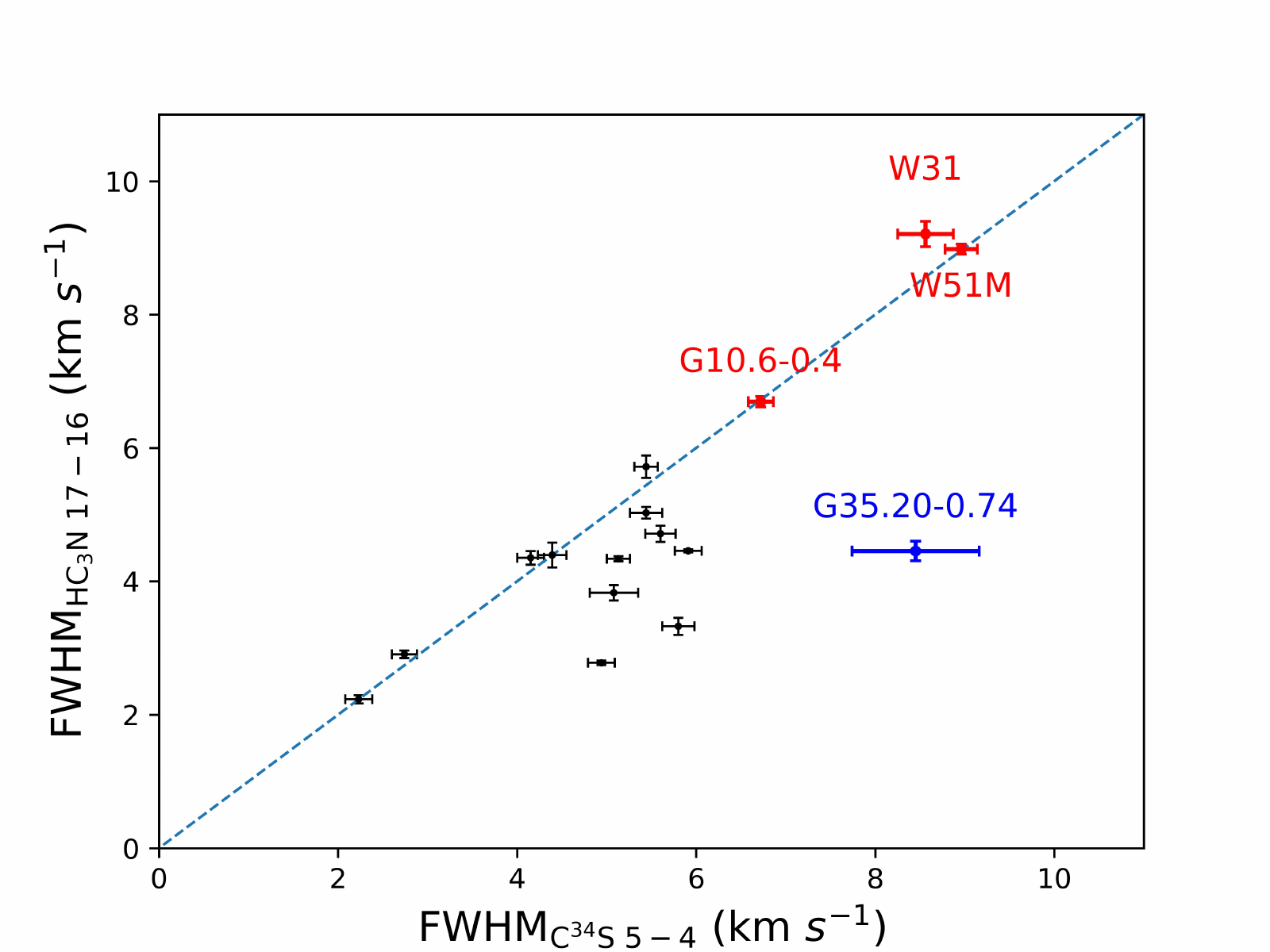}
	\end{minipage}%
	
	\begin{minipage}[t]{0.495\linewidth}
		\centering
		\includegraphics[width=90mm]{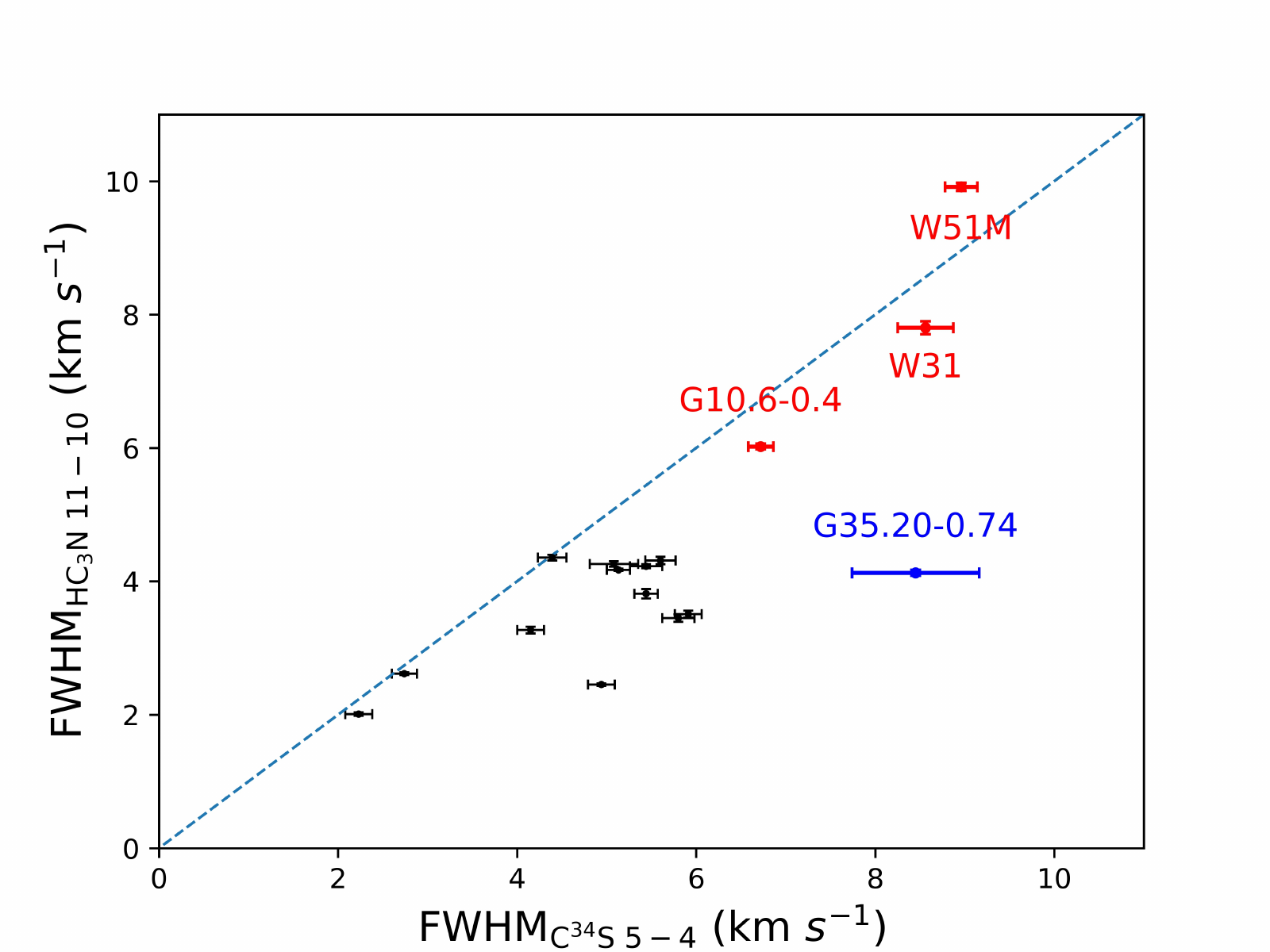}
	\end{minipage}%
	\begin{minipage}[t]{0.495\textwidth}
		\centering
		\includegraphics[width=90mm]{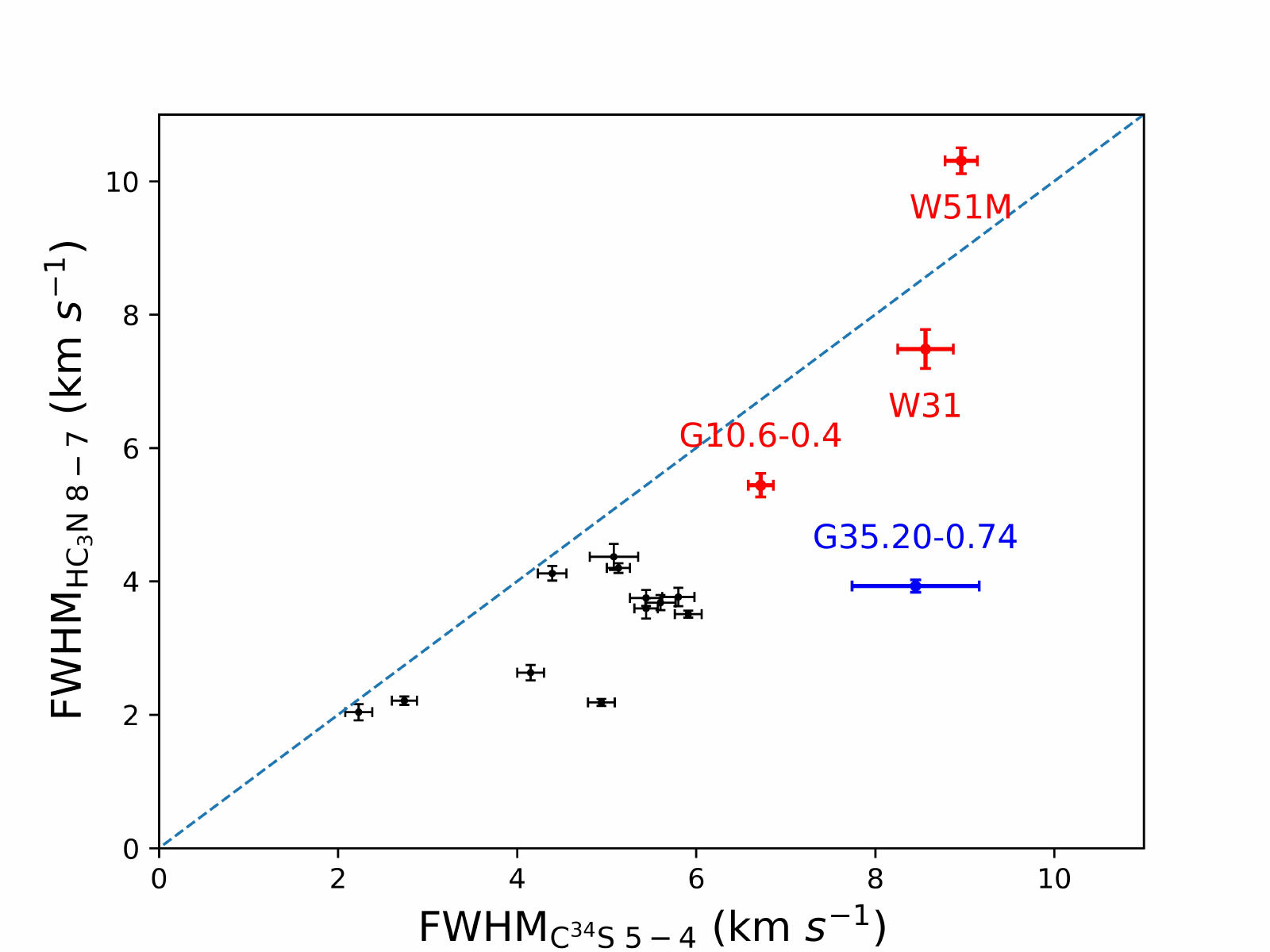}
	\end{minipage}%

	\caption{Comparisons of FWHM line width between HC$_3$N (24-23, 17-16, 11-10, 8-7) and C$^{34}$S (5-4) for the 19 sources except W44, CepA and S76E. The slope of the blue dashed line is 1.}
	
	\label{figure 3}	
\end{figure*}

C$^{34}$S (5-4) was used for deriving the virial mass of clumps in \citet{2003ApJS..149..375S}. The FWHM line width is one of the parameters for calculating virial mass ($\Delta V \propto M_{\rm vir}$). In Figure \ref{figure 3}, we compare the line widths (FWHM) of HC$_3$N (24-23, 17-16, 11-10, 8-7) with C$^{34}$S (5-4). The Line widths of the four transitions show approximately linear correlation with the line widths of C$^{34}$S (5-4). However, G35.20-0.74 shows clear deviation from the nearly linear correlations. The virial mass in G35.20-0.74 may need to be recalculated with the line widths of the HC$_3$N lines instead of C$^{34}$S (5-4).

\subsection{Possible cutoff between log$L_{\rm IR}$ and the line width relation}
\begin{figure*}[hp]
	\vskip0pt 
	
	\begin{minipage}[t]{0.495\linewidth}
		\centering
		\includegraphics[width=80mm]{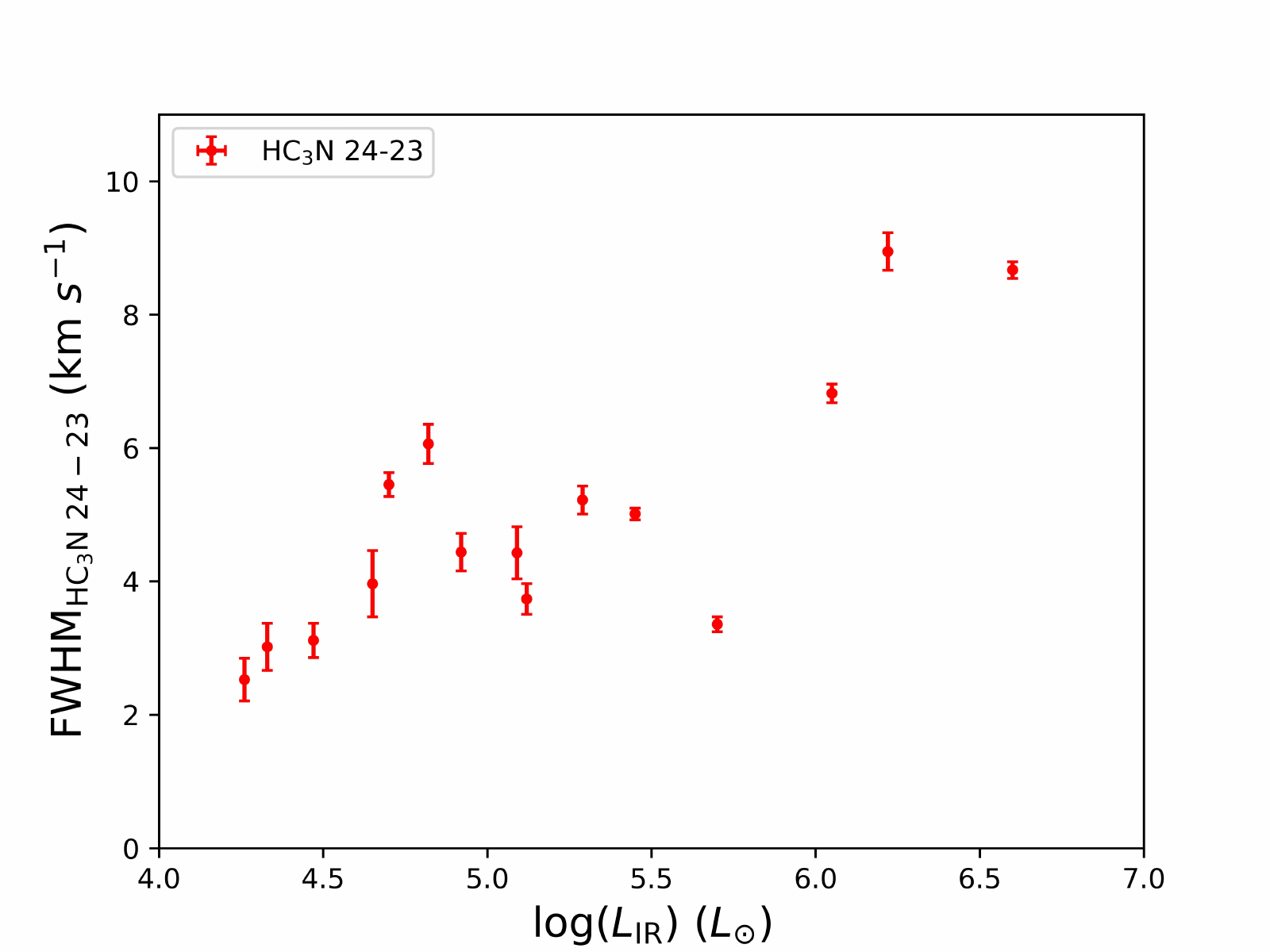}
	\end{minipage}%
	\begin{minipage}[t]{0.495\textwidth}
		\centering
		\includegraphics[width=80mm]{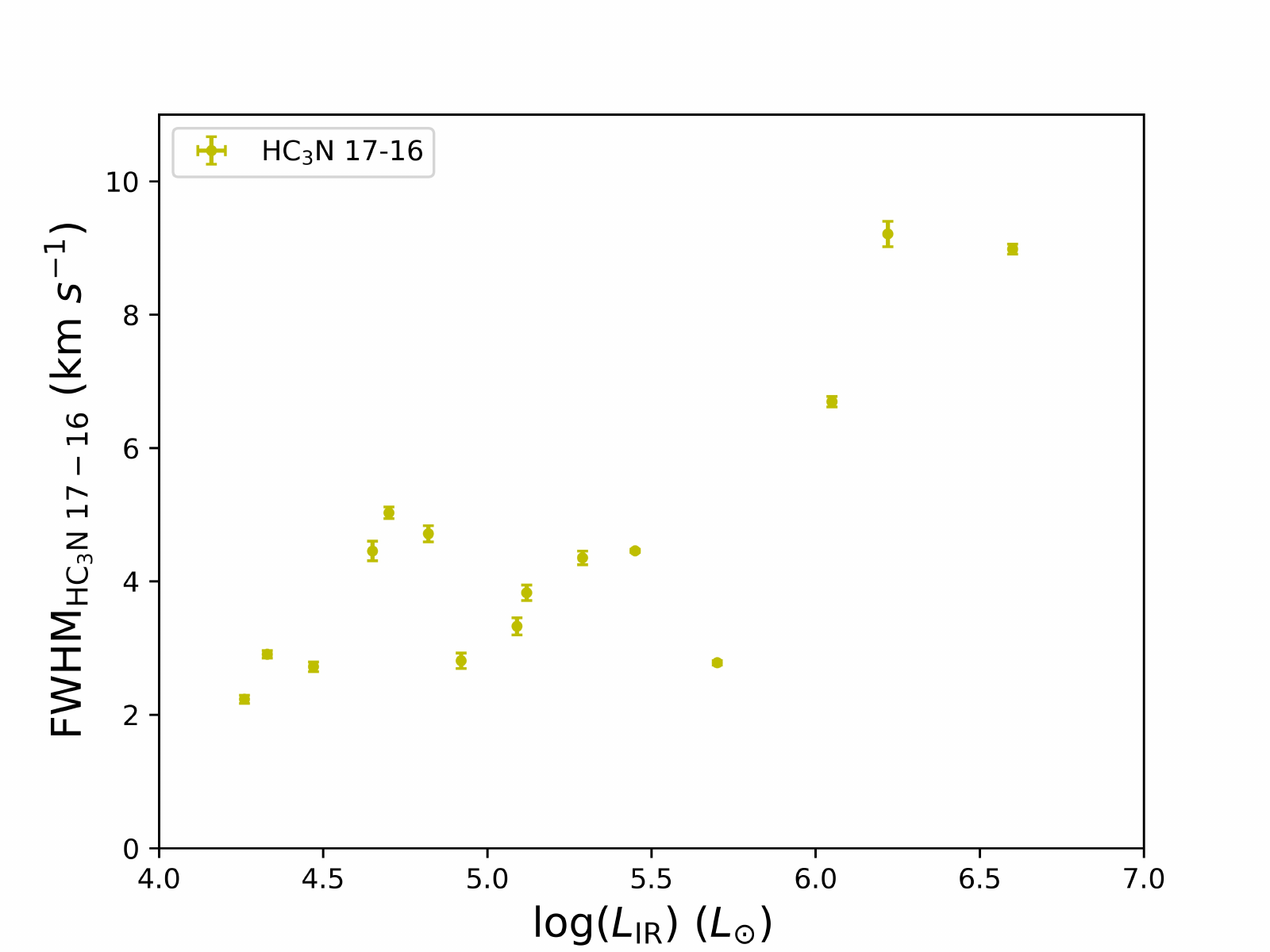}
	\end{minipage}%
	
	\begin{minipage}[t]{0.495\linewidth}
		\centering
		\includegraphics[width=80mm]{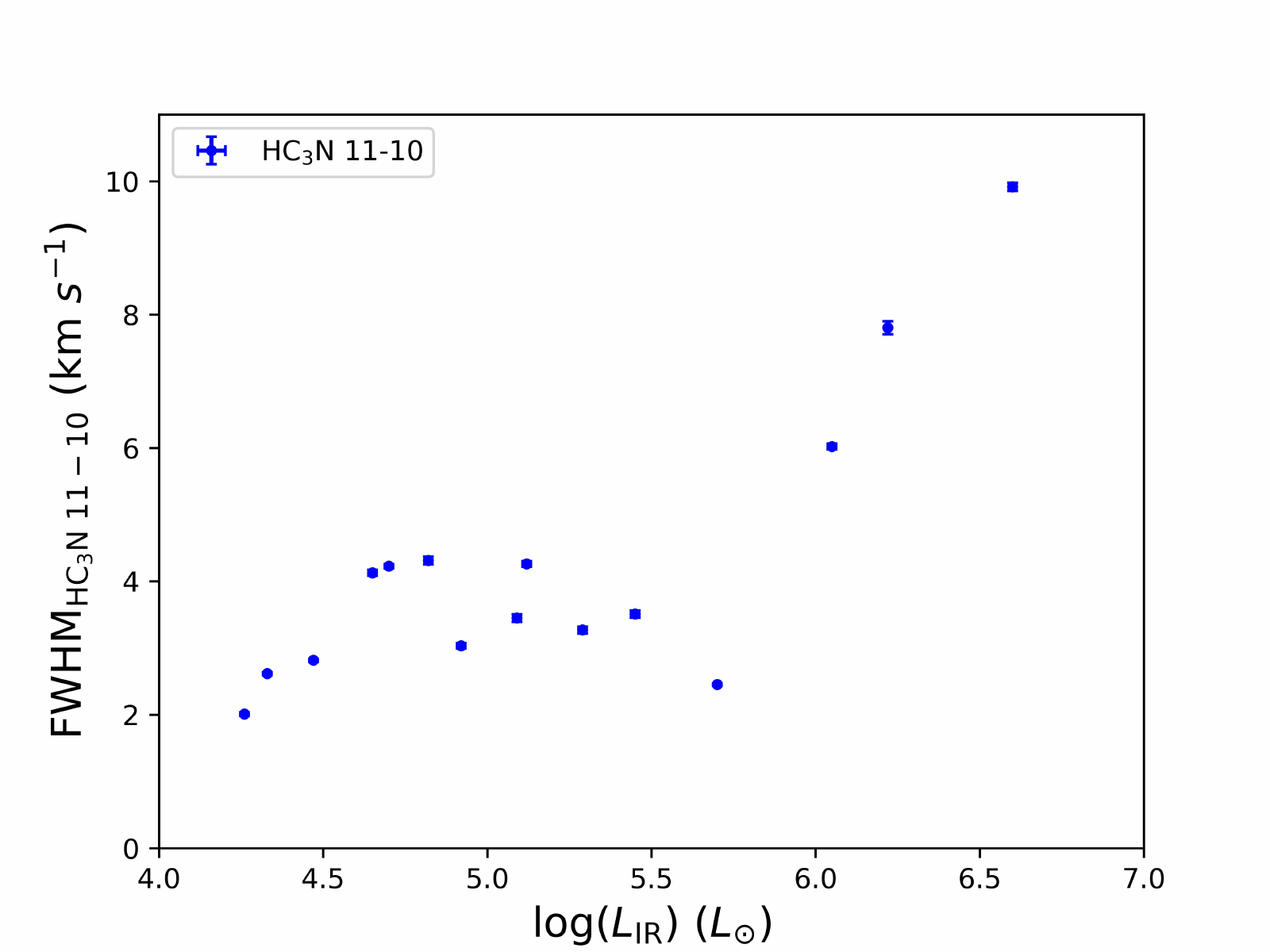}
	\end{minipage}%
	\begin{minipage}[t]{0.495\textwidth}
		\centering
		\includegraphics[width=80mm]{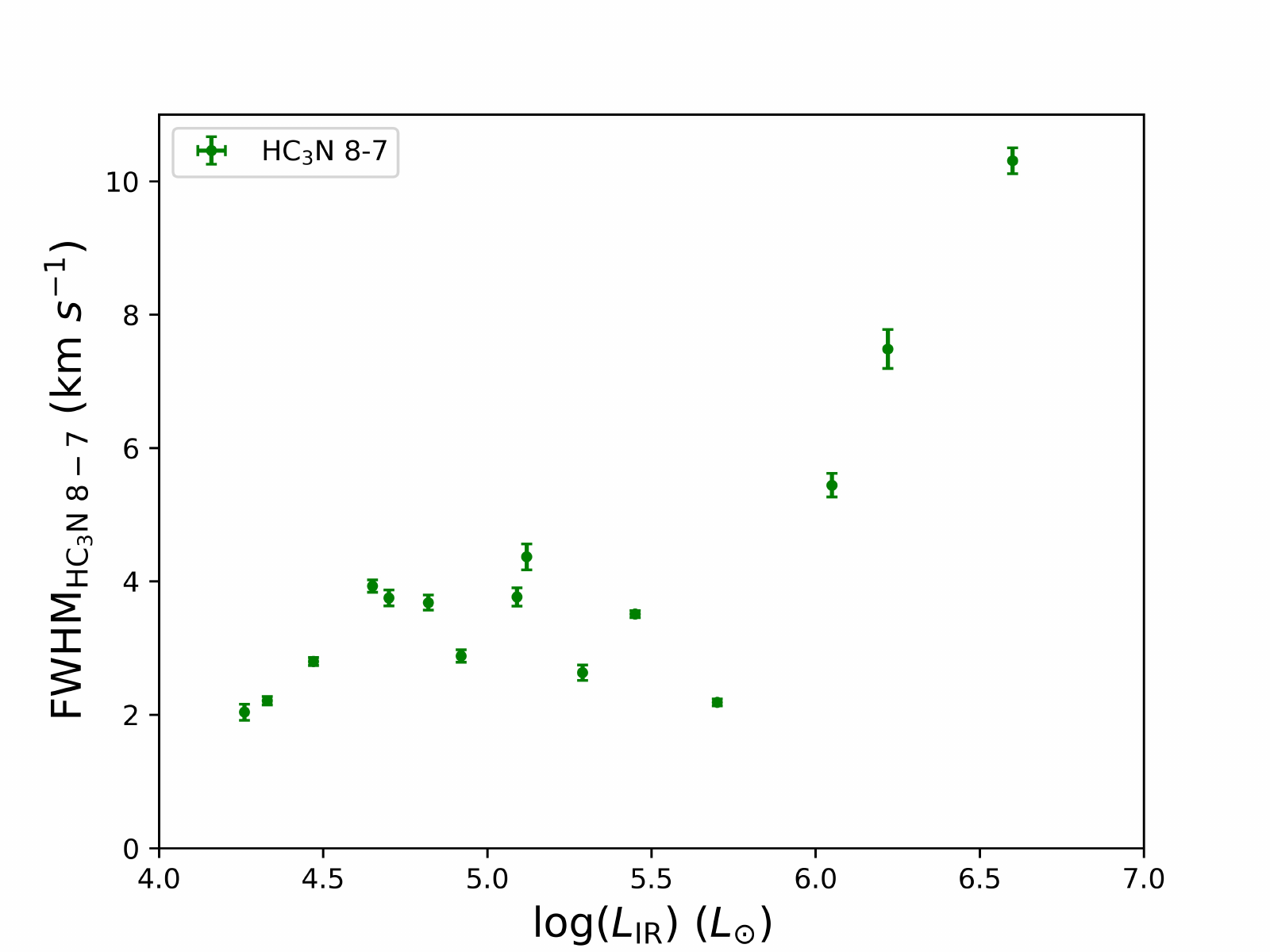}
	\end{minipage}%
	
	\begin{minipage}[t]{1\textwidth}
		\centering
		\includegraphics[width=90mm]{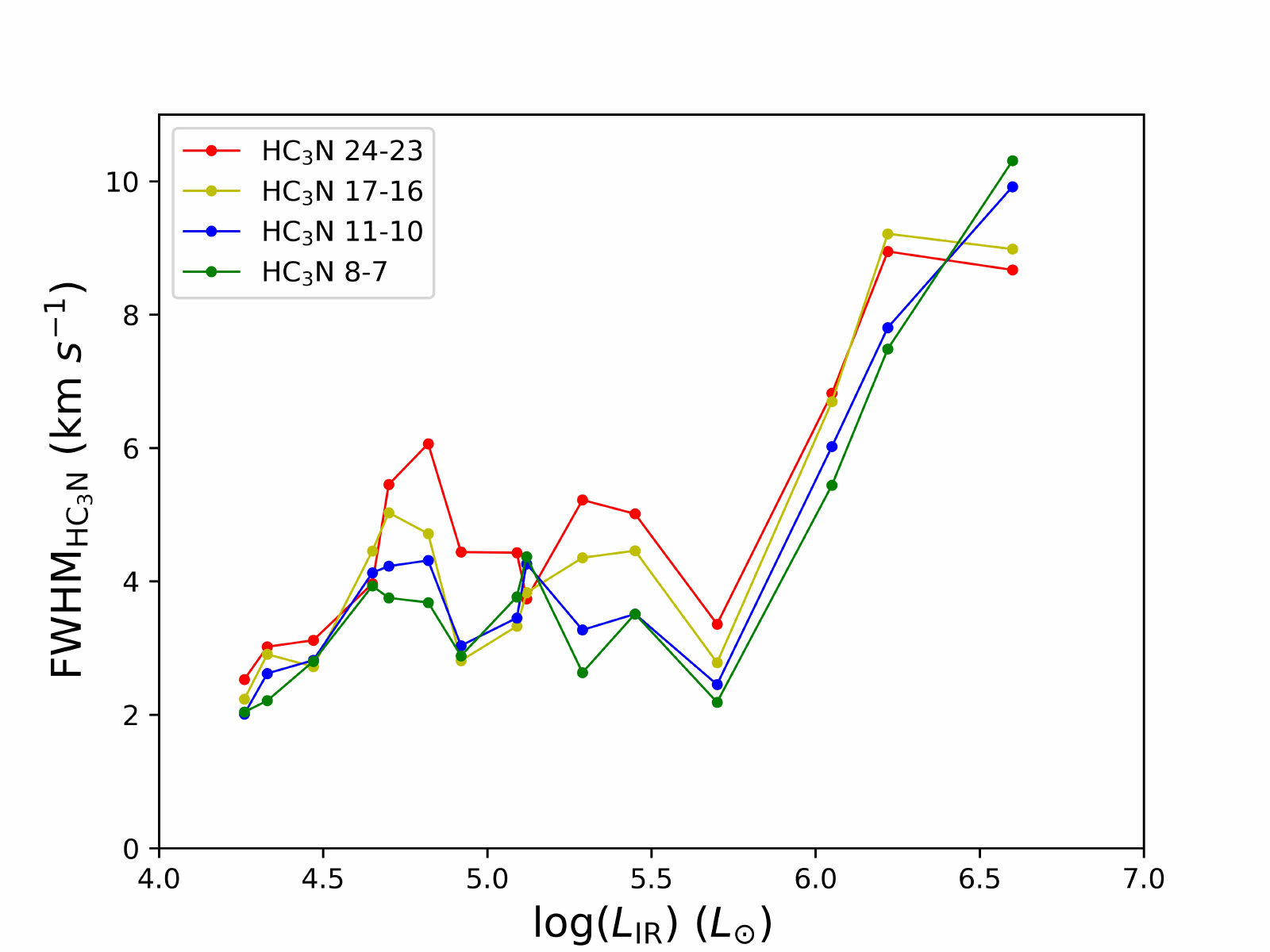}
	\end{minipage}%
	
	\caption{Comparisons between total infrared luminosity and FWHM line widths of HC$_3$N (24-23, 17-16, 11-10, 8-7) for the 19 sources except W44, W42, NGC7538 and W33cont.}
	
	\label{figure 4}	
\end{figure*}

Figure \ref{figure 4} shows log$L_{\rm IR}$ against the line width of 15 sources. Because the line width of HC$_3$N (24-23 and 8-7) for W44 were not obtained and W33cont, W42 and NGC7538 were not included in the infrared luminosity catalog of \citet{2010ApJS..188..313W}, these four sources are excluded in the comparisons. As can be seen in Figure \ref{figure 4}, we found a possible cutoff around $L_{\rm IR}$ $\sim$ 10$^{5.5}$\,$L_\odot$ for all four transitions. When infrared luminosities are  less than 10$^{5.5}$\,$L_\odot$, the FWHM line widths are roughly flat for various $L_{\rm IR}$, while, the FWHM line widths show a deep upward trend with increasing $L_{\rm IR}$ when infrared luminosities ($L_{\rm IR}$) are greater than 10$^{5.5}$\,$L_\odot$. However, such an upward trend is not persuasive. Further extensive samples are necessary for more convincing confirmation of such a cutoff. The possible cutoff trend might imply star formation feedback or the initial conditions of star formation.

\clearpage

\subsection{Excitation temperature vs other properties?}
The excitation temperature ($T_{\rm ex}$) is determined by the velocity-integrated intensity ratio between HC$_3$N (24-23) and HC$_3$N (17-16). $T_{\rm ex}$ for massive star-formation regions was measured in the range of 30 - 50\,K for CO molecule \citep{1996ApJ...472..225S, 2002A&A...383..892B, 2004A&A...426..503W}. The median value of $T_{\rm ex}$ in our sample is 35.9\,K, which is consistent with previous works. On the other hand, $T_{\rm ex}$ for low-mass sources ranges from 10\,K to 15\,K \citep{1980ApJ...239L..17S, 1984ApJ...286..599G}. 

\begin{figure*}[hp]
	\vskip5pt 
	
	\begin{minipage}[t]{0.495\linewidth}
		\centering
		\includegraphics[width=90mm]{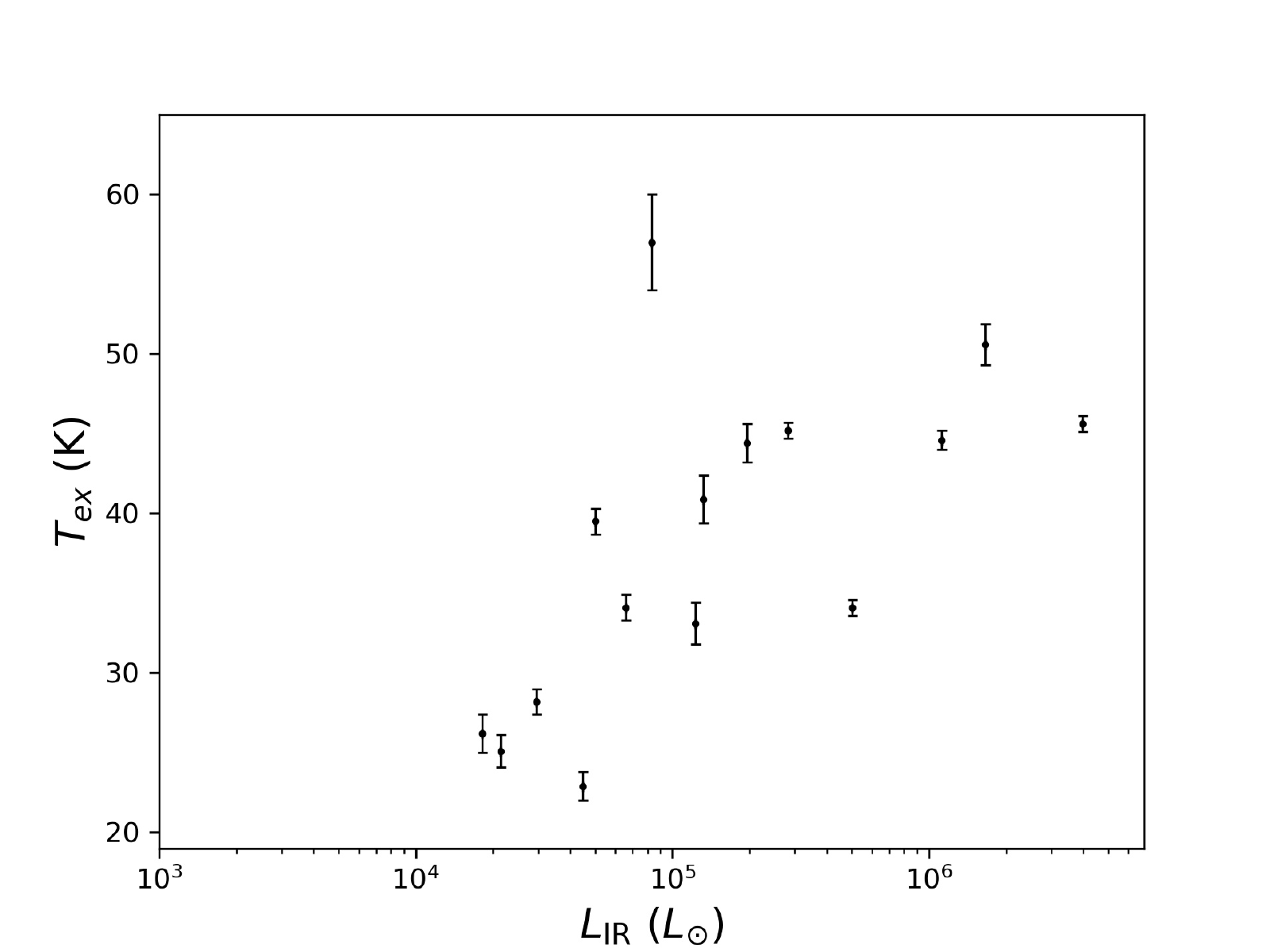}
	\end{minipage}%
	\begin{minipage}[t]{0.495\textwidth}
		\centering
		\includegraphics[width=90mm]{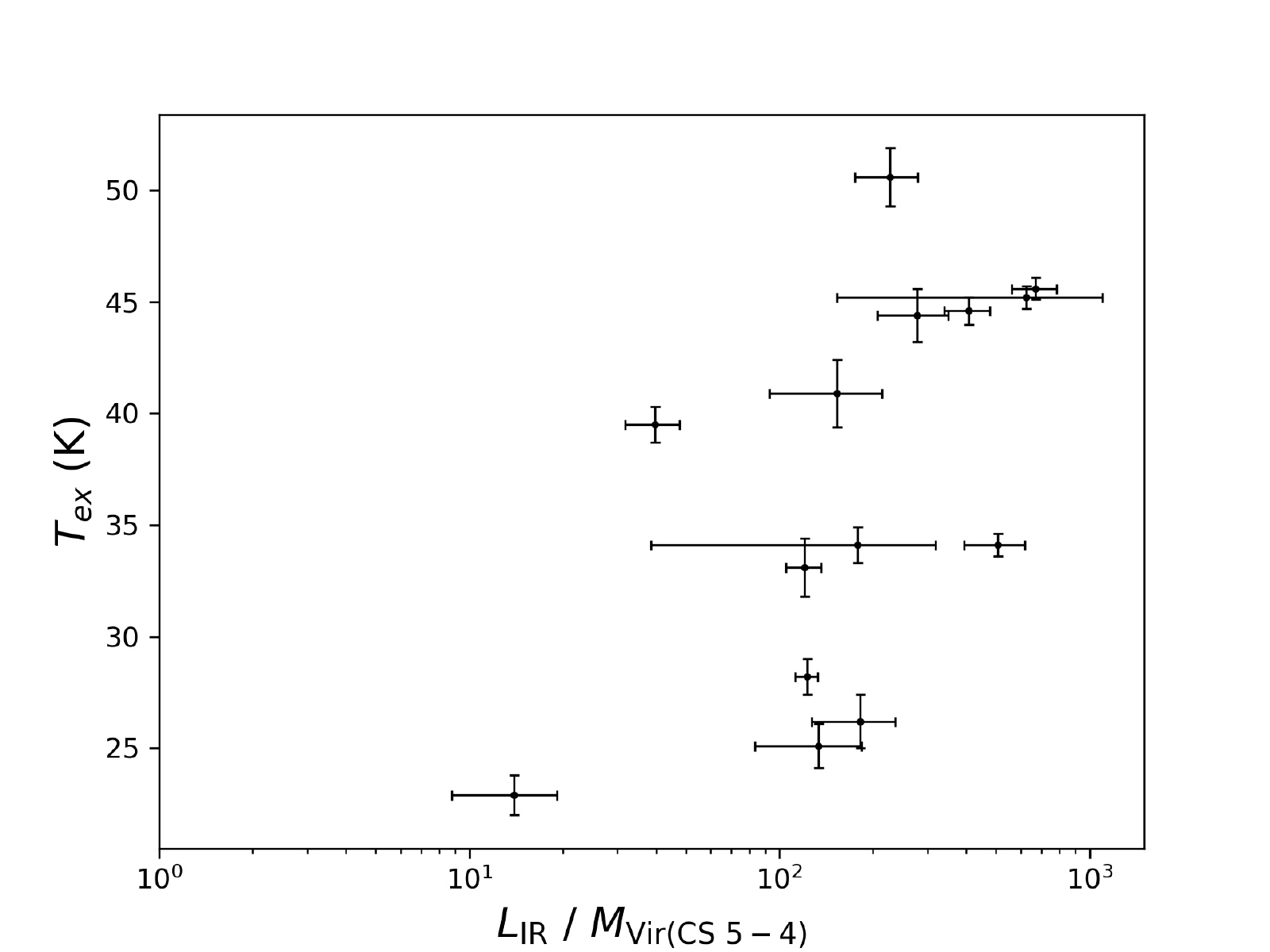}
	\end{minipage}%

	\caption{Comparisons between excitation temperature and infrared luminosity $L_{\rm IR}$ and $L_{\rm IR}$/$M_{\rm vir}$.}

\label{figure 5}	
\end{figure*}

\begin{figure*}
	\vskip25pt 
	
	\begin{minipage}[t]{0.495\linewidth}
		\centering
		\includegraphics[width=77mm]{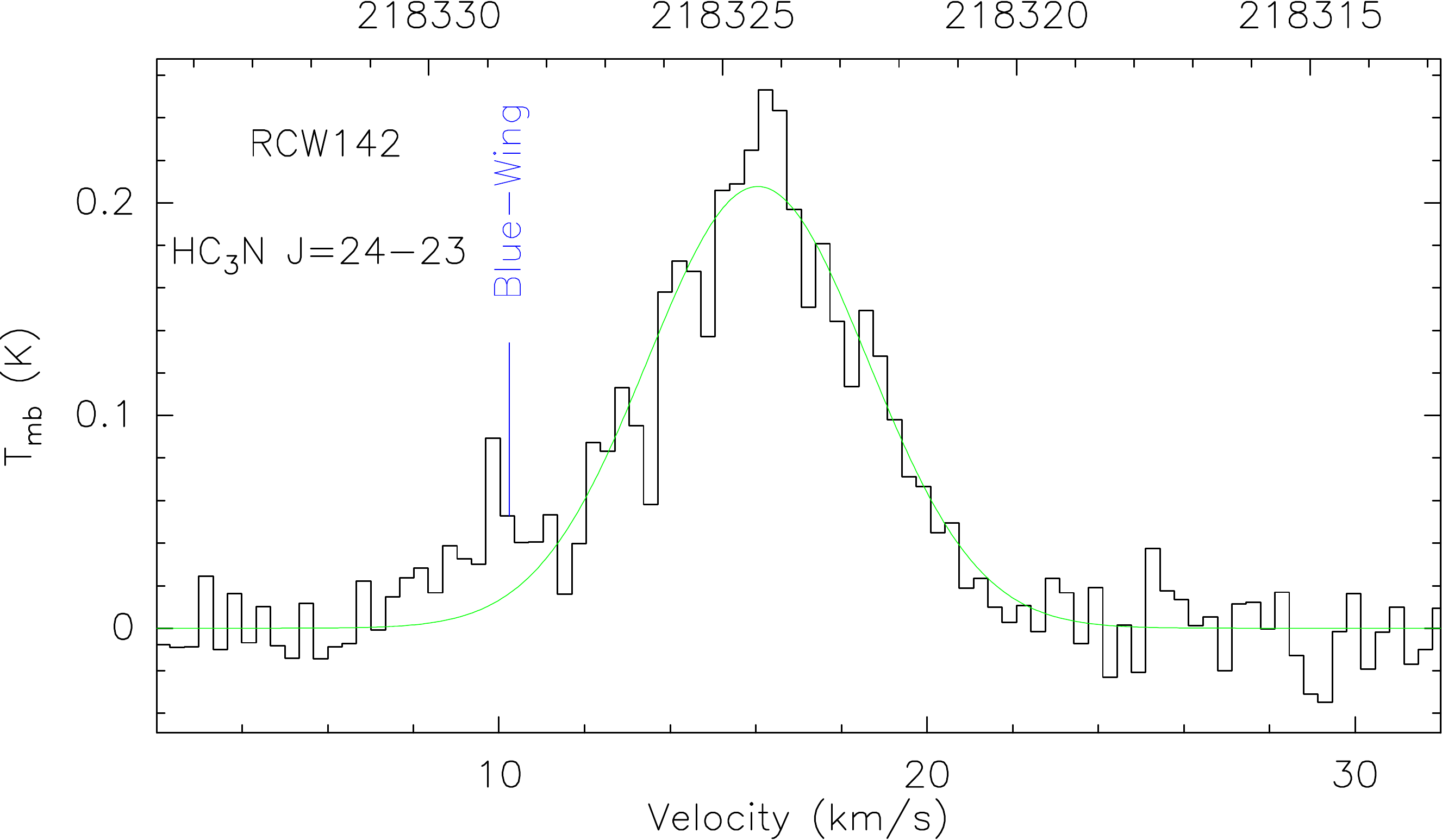}
	\end{minipage}%
	\begin{minipage}[t]{0.495\textwidth}
		\centering
		\includegraphics[width=77mm]{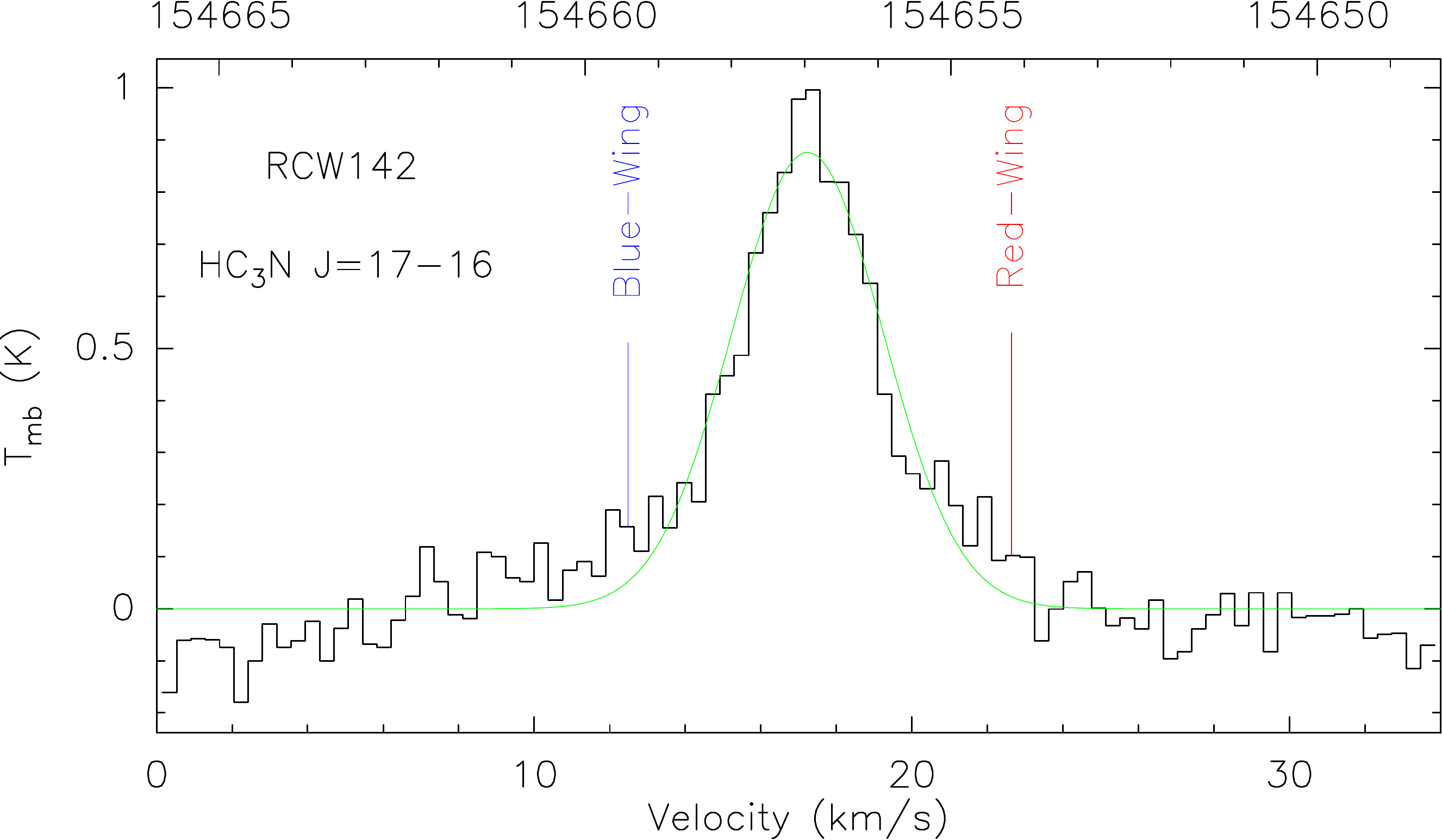}
	\end{minipage}%
	
	\vskip20pt 
	
	\begin{minipage}[t]{0.495\linewidth}
		\centering
		\includegraphics[width=80mm]{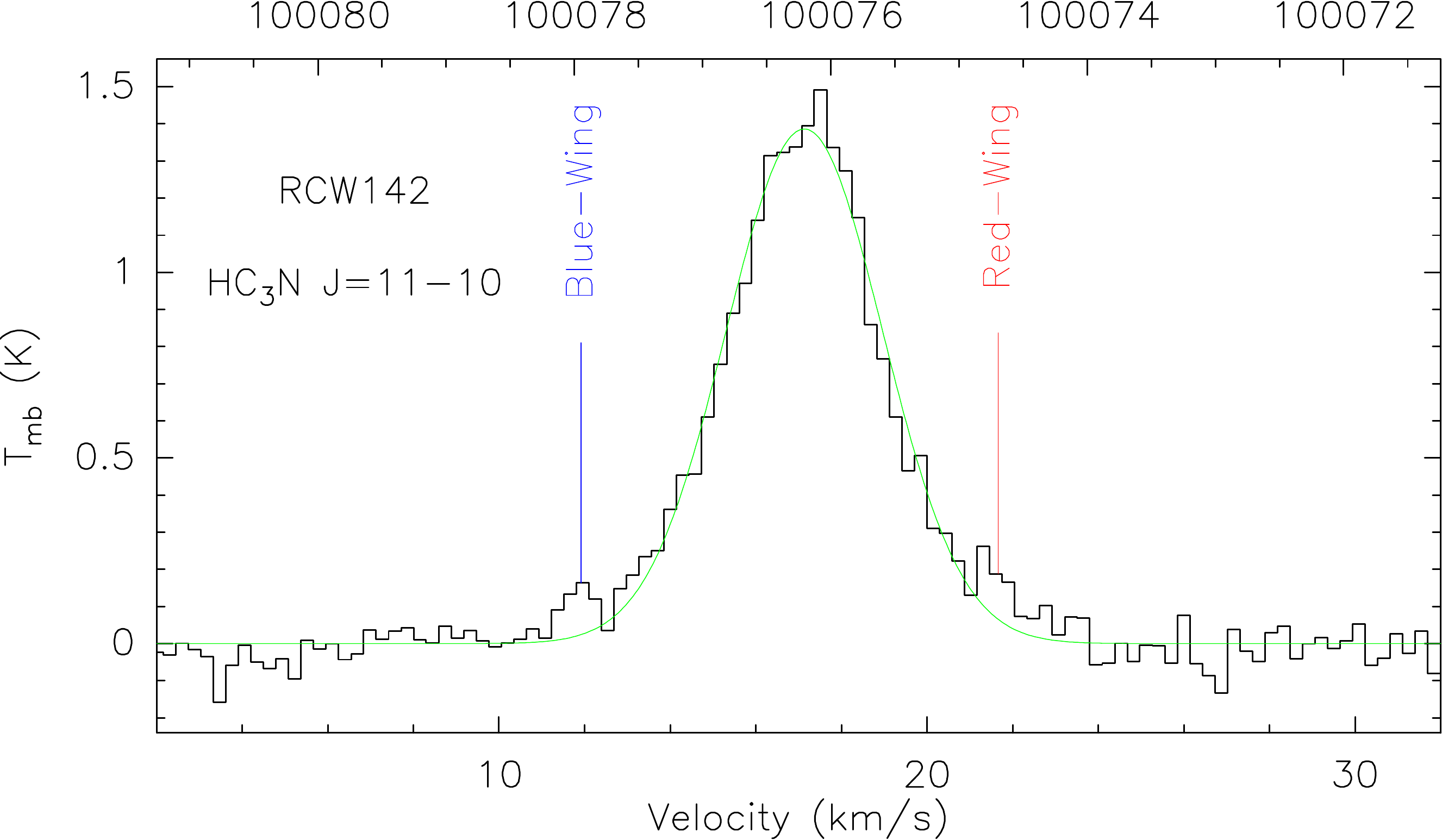}
	\end{minipage}%
	\begin{minipage}[t]{0.495\textwidth}
		\centering
		\includegraphics[width=80mm]{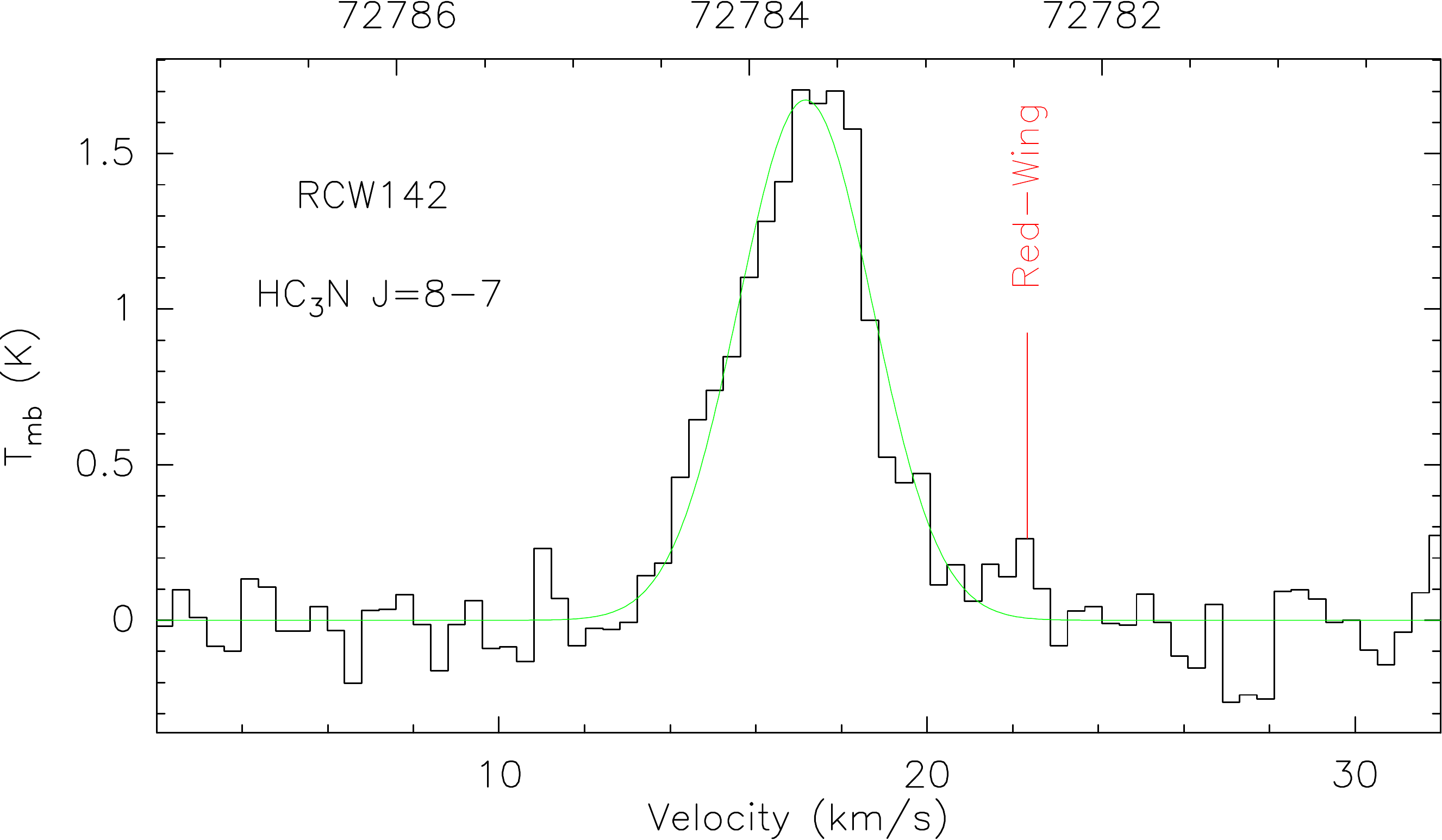}
	\end{minipage}%
	
	\vskip20pt 
	
	\begin{minipage}[t]{0.495\linewidth}
		\centering
		\includegraphics[width=80mm]{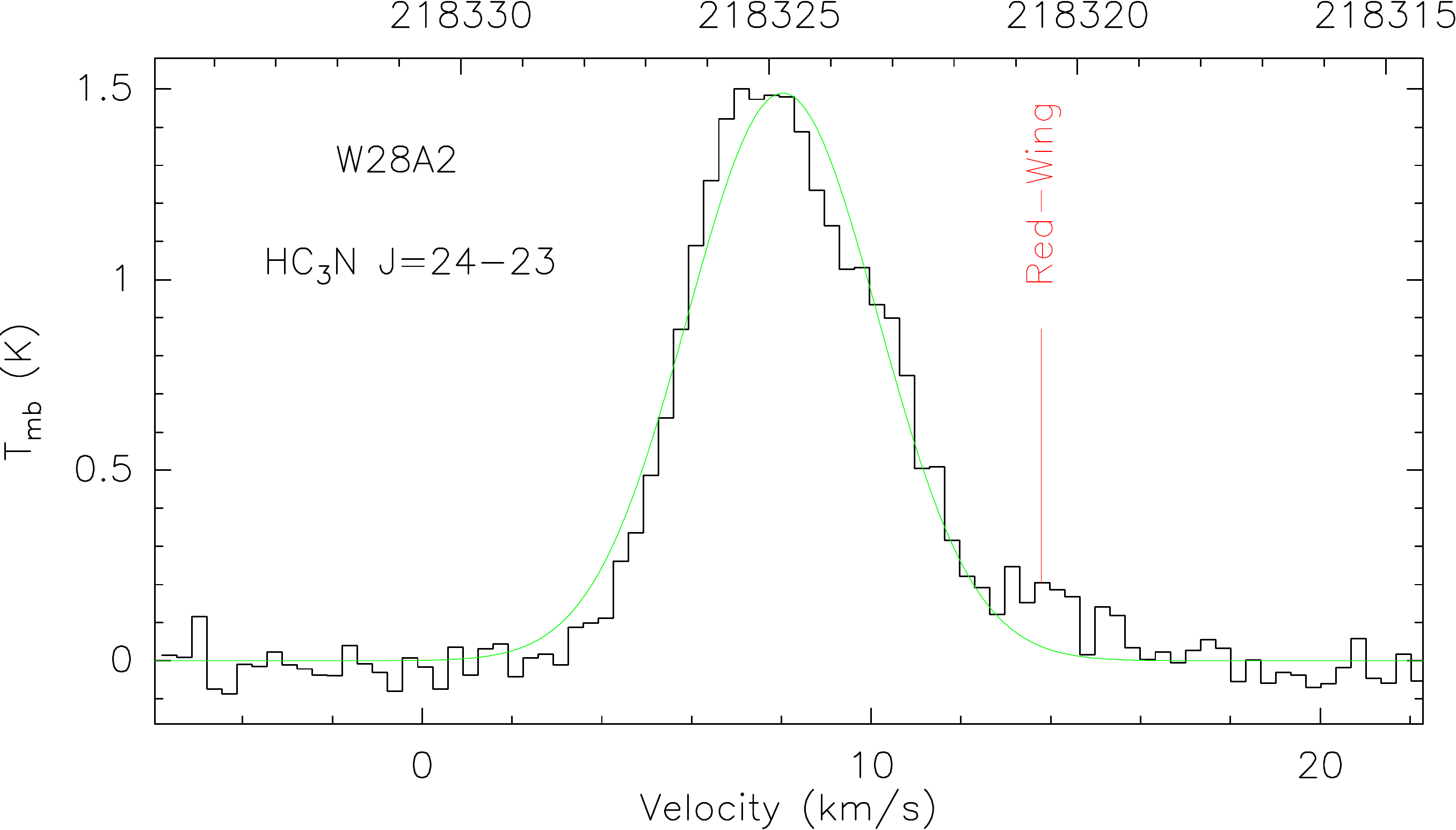}
	\end{minipage}%
	\begin{minipage}[t]{0.495\textwidth}
		\centering
		\includegraphics[width=80mm]{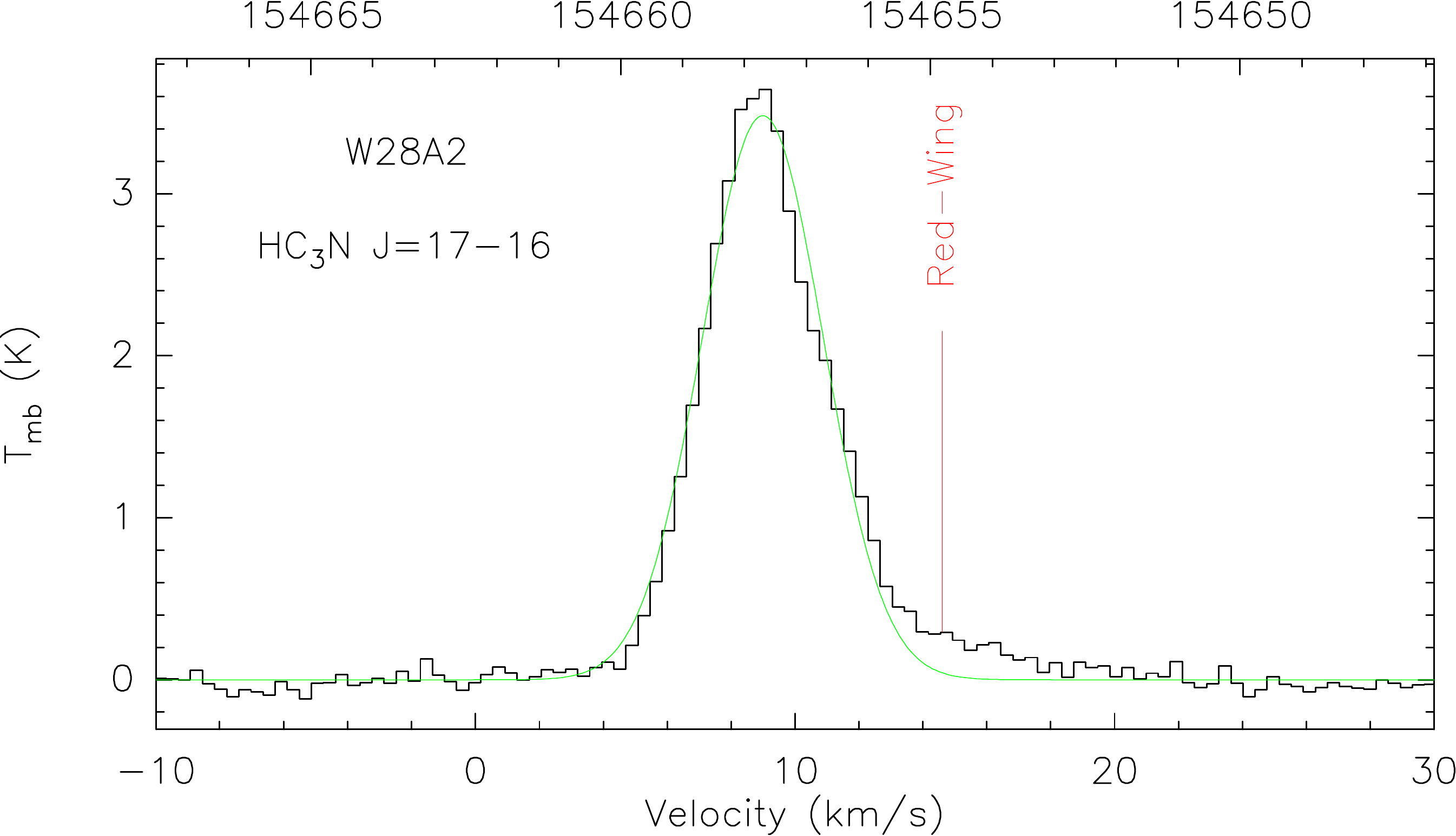}
	\end{minipage}%
	
	\vskip20pt 
	
	\begin{minipage}[t]{0.495\linewidth}
		\centering
		\includegraphics[width=80mm]{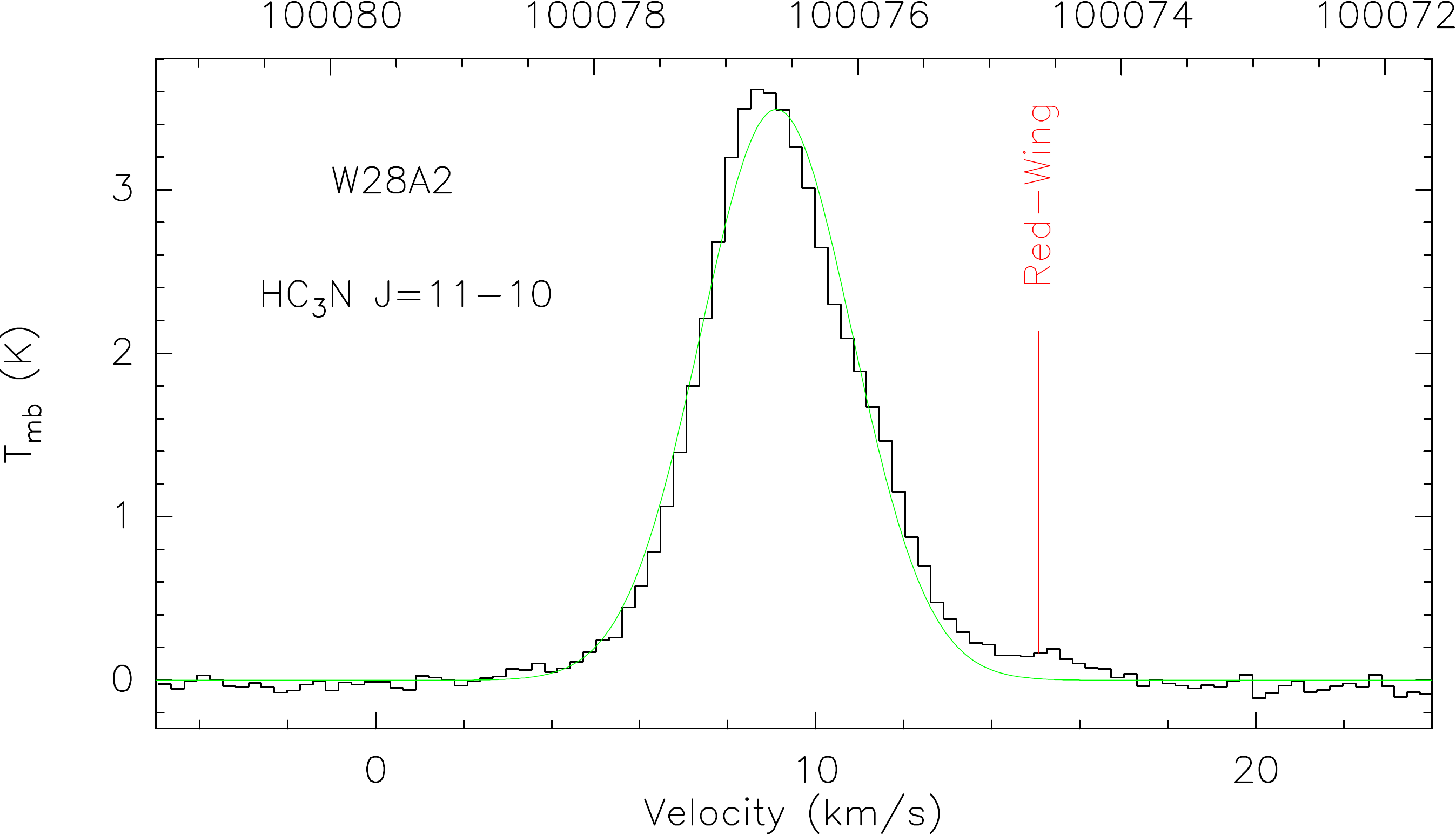}
	\end{minipage}%
	\begin{minipage}[t]{0.495\textwidth}
		\centering
		\includegraphics[width=80mm]{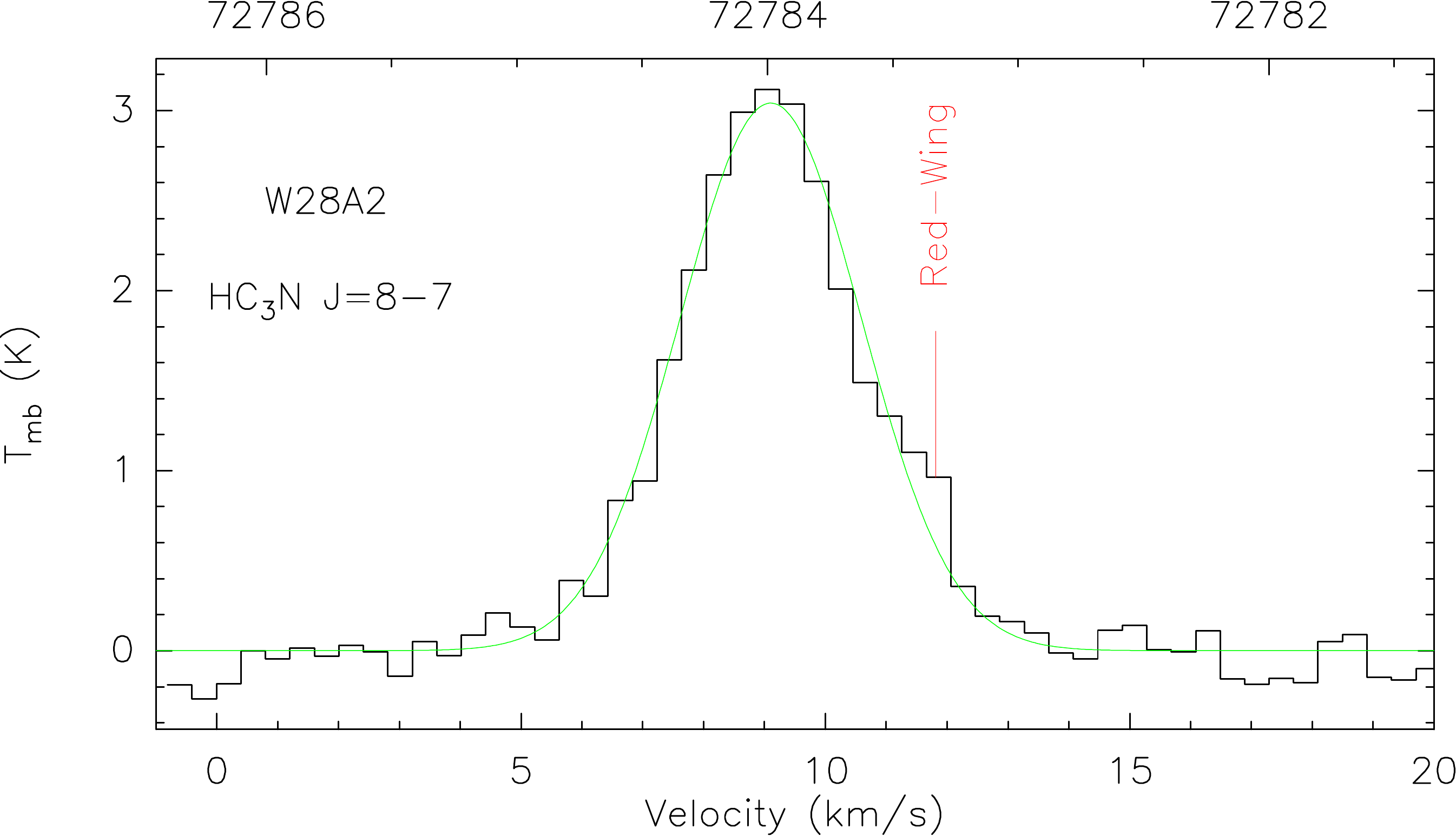}
	\end{minipage}%
	\caption{Spectral lines for sources with line wings. The possible line blending of CCS\,$10_{11}-10_{10}$ at rest frequency of 72.783\,GHz has also been marked. The identifications of transition are labeled on the upper left of figures. Velocity resolution is $\sim$\,0.35\,km s$^{-1}$, overlapping with a Gaussian fit (green line).}
	\label{figure 6}	
\end{figure*}

The left panel of Figure \ref{figure 5} shows a comparison between $T_{\rm ex}$ and $L_{\rm IR}$. There is a roughly positive correlation between $T_{\rm ex}$ and $L_{\rm IR}$. \citet{2010ApJS..188..313W} proposed that higher luminosity may lead to a greater fraction of gas being warm enough to emit higher-$J$ lines. However, there is only a weak correlation between the logarithmic line luminosity ratio of CS (7-6) to CS (5-4) and log\,$L_{\rm IR}$. As shown in Figure \ref{figure 5}, there is a strong correlation between $T_{\rm ex}$ and $L_{\rm IR}$ with a correlation coefficient $r$ of 0.76. According to equation (\ref{eq:equation 1}), log\,$T_{\rm ex}$ is proportional to log\,$R_{\rm 24,17}$. Such a correlation implies that gas in more luminous sources is warmer than in low-luminosity ones. What was proposed above by \citet{2010ApJS..188..313W} may be another explanation for the correlation between $T_{\rm ex}$ and $L_{\rm IR}$.

However, taking the errors in the virial mass into consideration, no obvious trend can be found for the relation between  $T_{\rm ex}$ and star-formation efficiency ($L_{\rm IR}$/$M_{\rm vir}$), as shown in the right panel of Figure \ref{figure 5}. This implies that star-formation efficiency is not proportional to the excitation conditions of dense gas.


\clearpage

\section{Summary}
\label{ 5 sect :sum }

We used the ARO 12-m and CSO 10.5-m radio telescopes to observe HC$_3$N (24-23, 17-16, 11-10, 8-7) lines towards a sample consisting of 19 Galactic massive star-forming regions. Our main results are summarized as follows:

1. In our 19 sources, we obtained the HC$_3$N (17-16, 11-10) lines of W44 and HC$_3$N (24-23, 17-16, 11-10, 8-7) lines of the remaining 18 sources. For most sources, the spectral lines have S/N \textgreater\,5\,$\sigma$. 

2. Twelve of the sources, RCW142, W28A2, G8.67-0.36, G10.6-0.4, W33cont, G14.33-0.64, W42, W51M, W75N, DR21S, W75(OH) and CepA, have line wings in their lines. 

3. The line widths of higher-$J$ HC$_3$N lines are generally larger than lower-$J$ ones , which may imply more violent turbulence or other motions in the inner regions of massive star formation than the outer regions for most sources.

4. Excitation temperatures derived from the line ratio of HC$_3$N (24-23) to HC$_3$N (17-16) for 18 sources range from 23 K to 57 K, with a median value of 35.9 K. 

5. A possible cutoff tendency was found around $L_{\rm IR}$ $\sim$ 10$^{5.5}$\,$L_\odot$ in the relation between $L_{\rm IR}$ and the line widths of four lines.

\section*{Acknowledgements}
\label{ 6 sect :acknow }
This work is supported by the National Key Basic Research and Development Program of China (grant no. 2017YFA0402704), National Natural Science Foundation of China grants 11590783 and U1731237, and China Postdoctoral Science Foundation grant no.\,2020M671267. We thank the staff at the CSO 10.4\,m telescope and the ARO 12\,m telescope for their kind help and support during our observations.

{}

\clearpage

\clearpage

\clearpage

\addtocounter{figure}{-1}
\begin{figure*}
	\vskip25pt 
	
	\begin{minipage}[t]{0.495\linewidth}
		\centering
		\includegraphics[width=80mm]{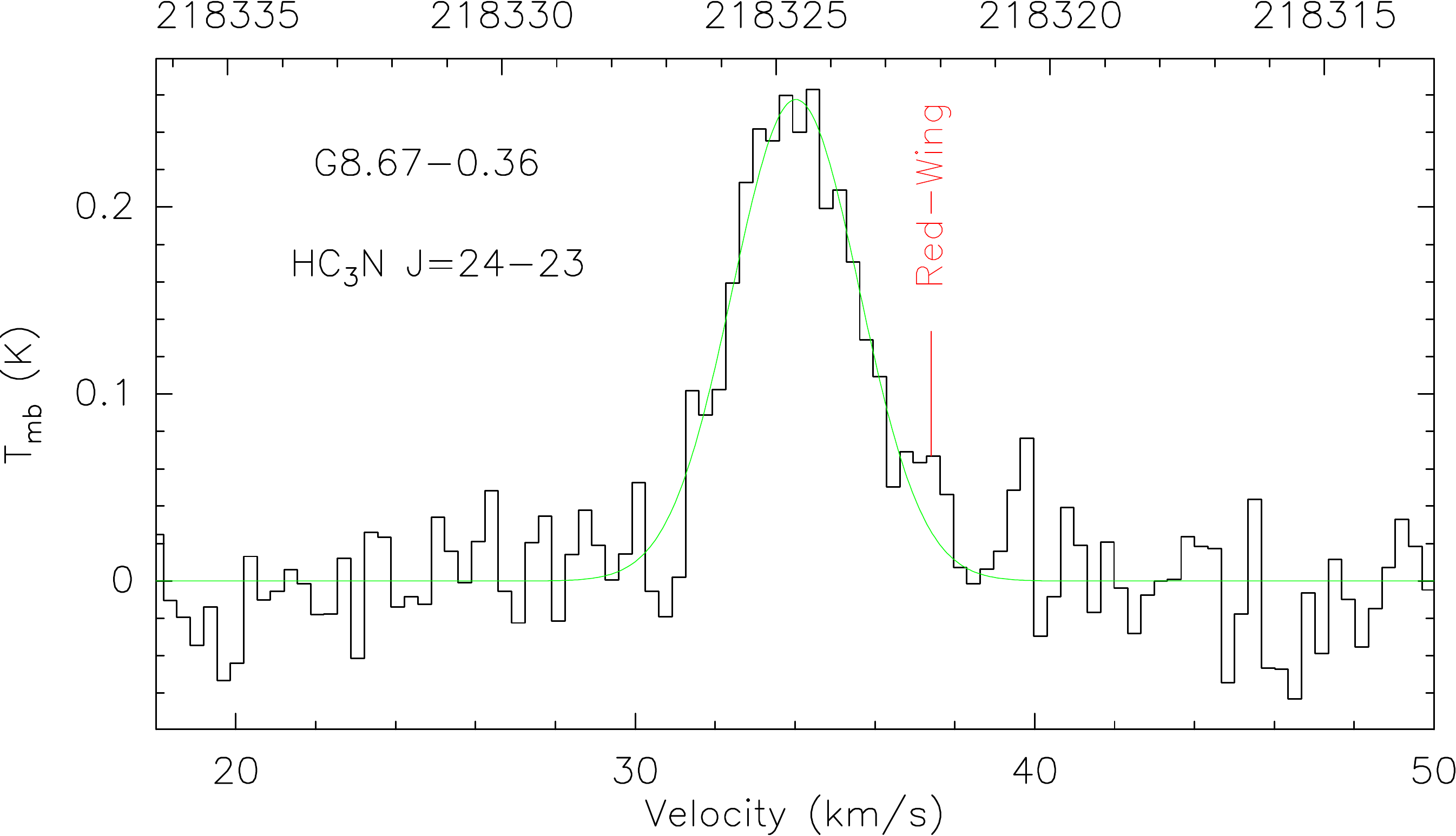}
	\end{minipage}%
	\begin{minipage}[t]{0.495\textwidth}
		\centering
		\includegraphics[width=80mm]{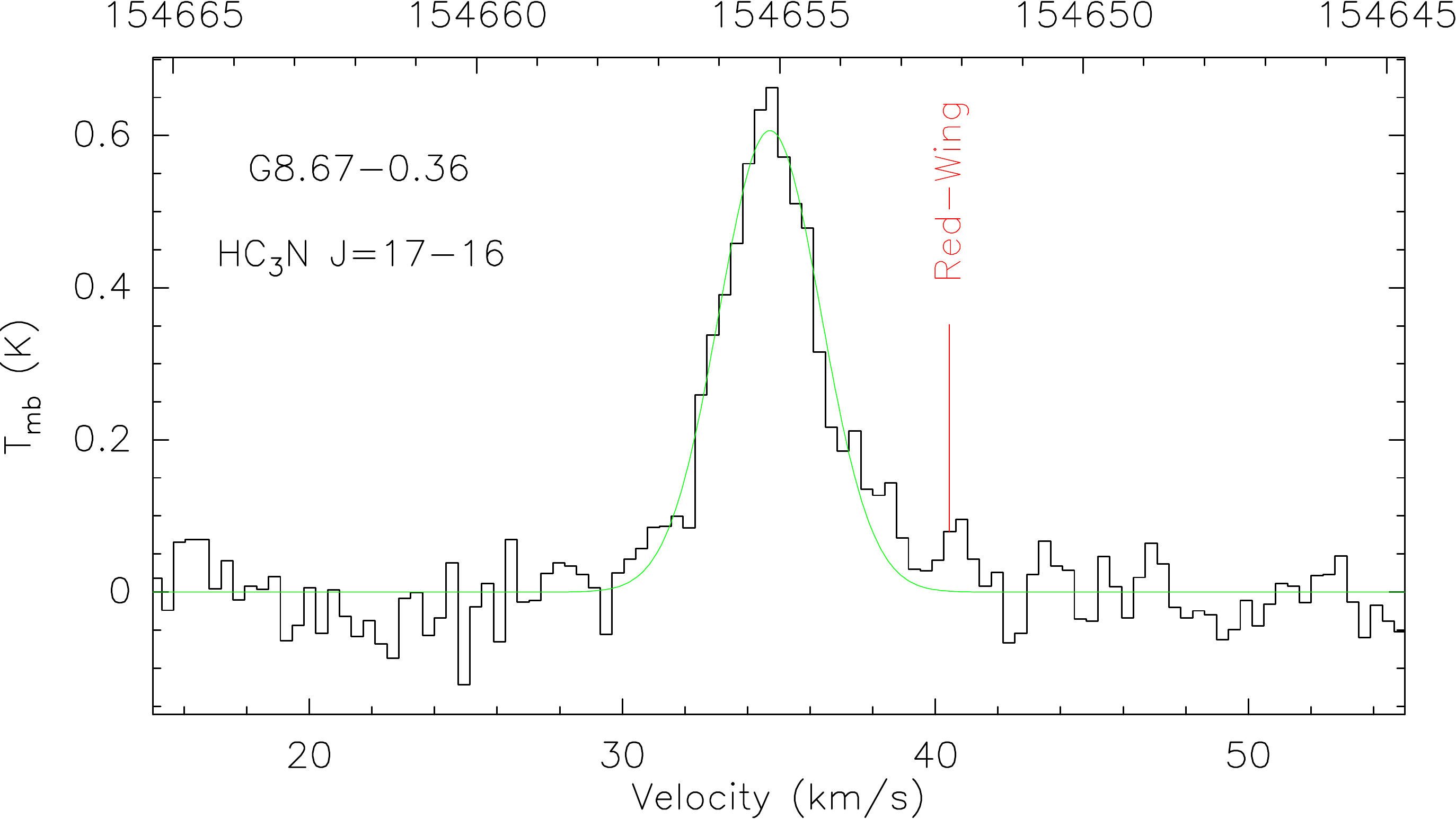}
	\end{minipage}%
	
	\vskip20pt 
	
	\begin{minipage}[t]{0.495\linewidth}
		\centering
		\includegraphics[width=80mm]{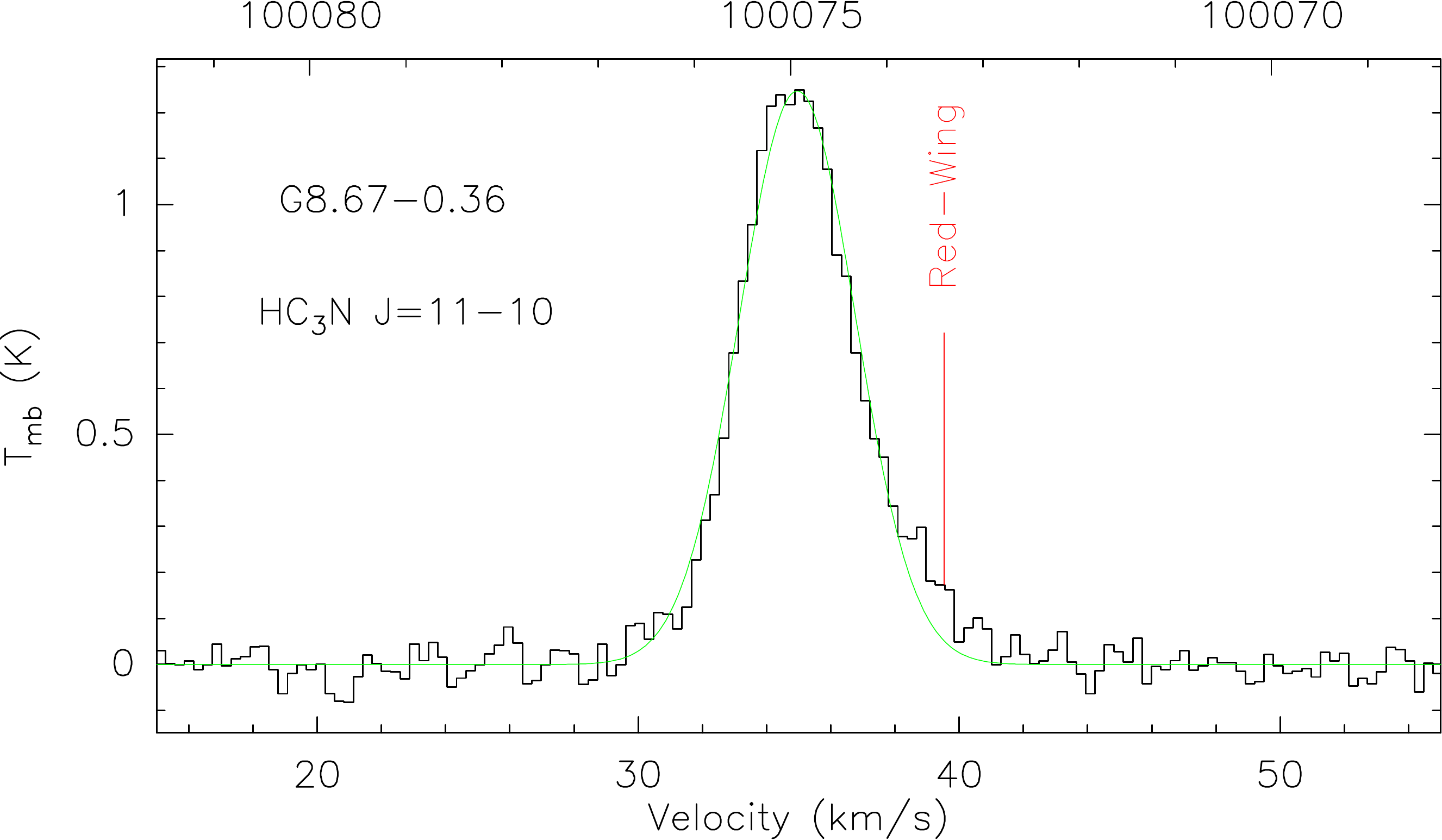}
	\end{minipage}%
	\begin{minipage}[t]{0.495\textwidth}
		\centering
		\includegraphics[width=80mm]{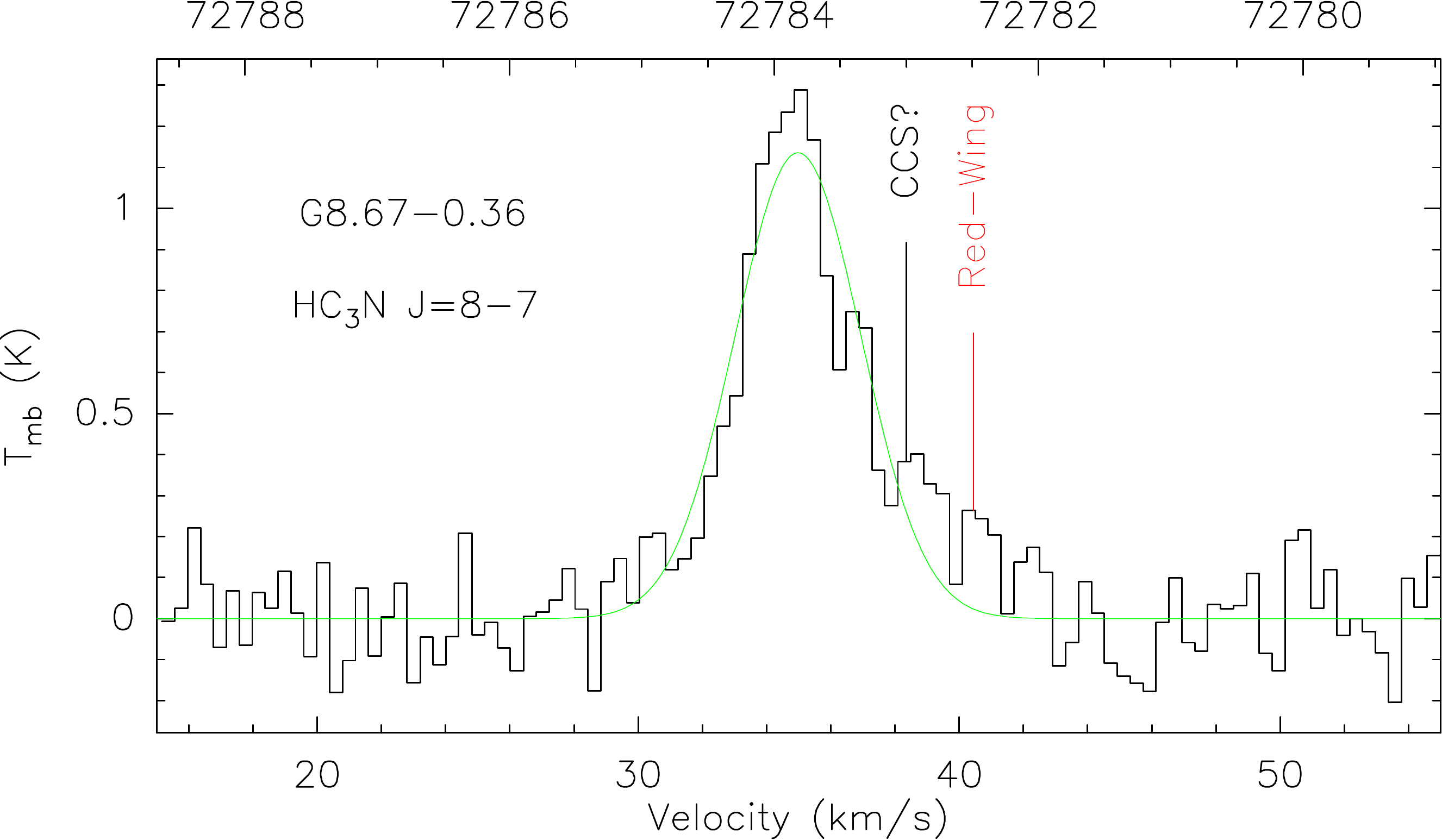}
	\end{minipage}%
	
	\vskip20pt 
	
	\begin{minipage}[t]{0.495\linewidth}
		\centering
		\includegraphics[width=80mm]{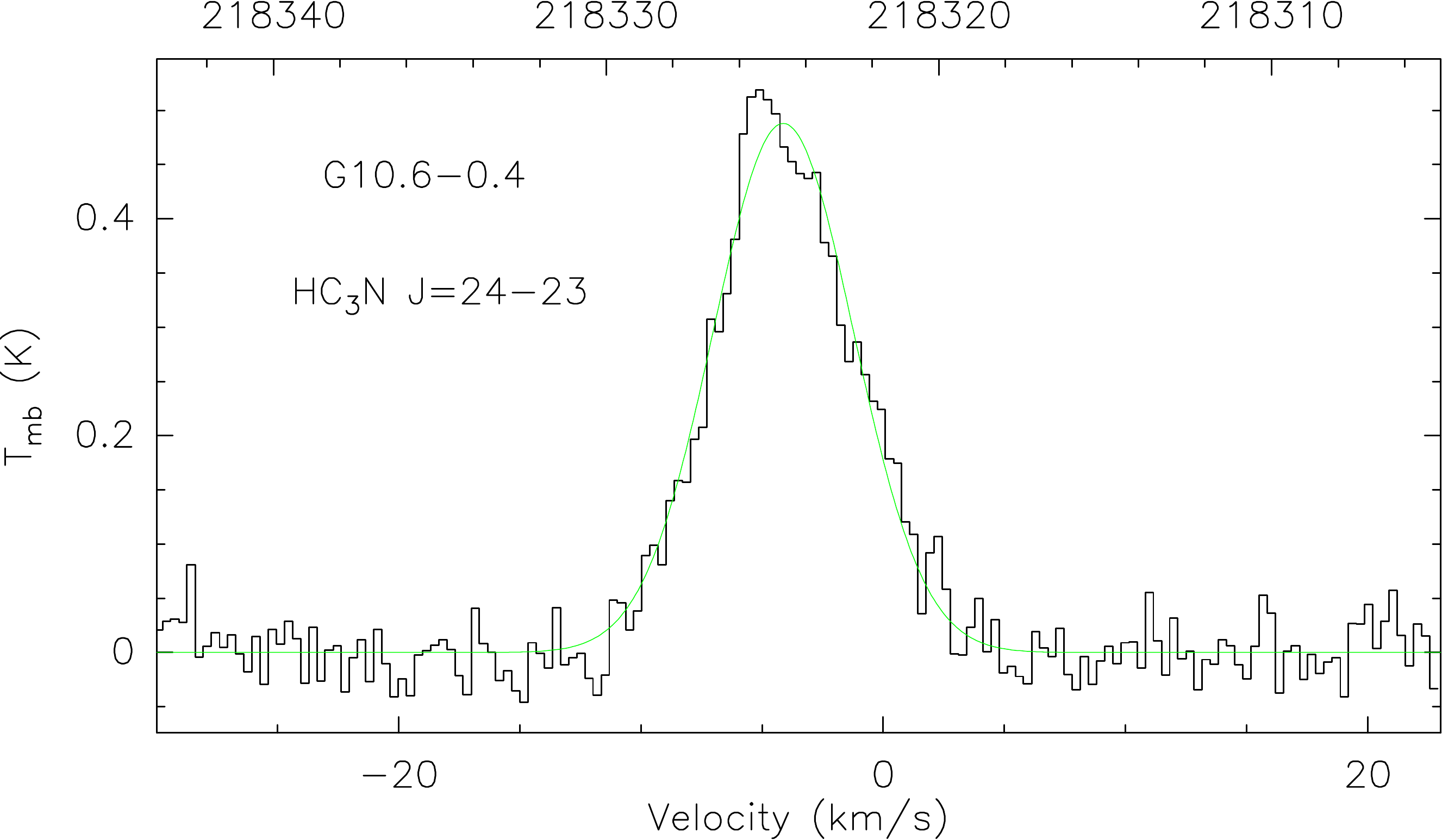}
	\end{minipage}%
	\begin{minipage}[t]{0.495\textwidth}
		\centering
		\includegraphics[width=80mm]{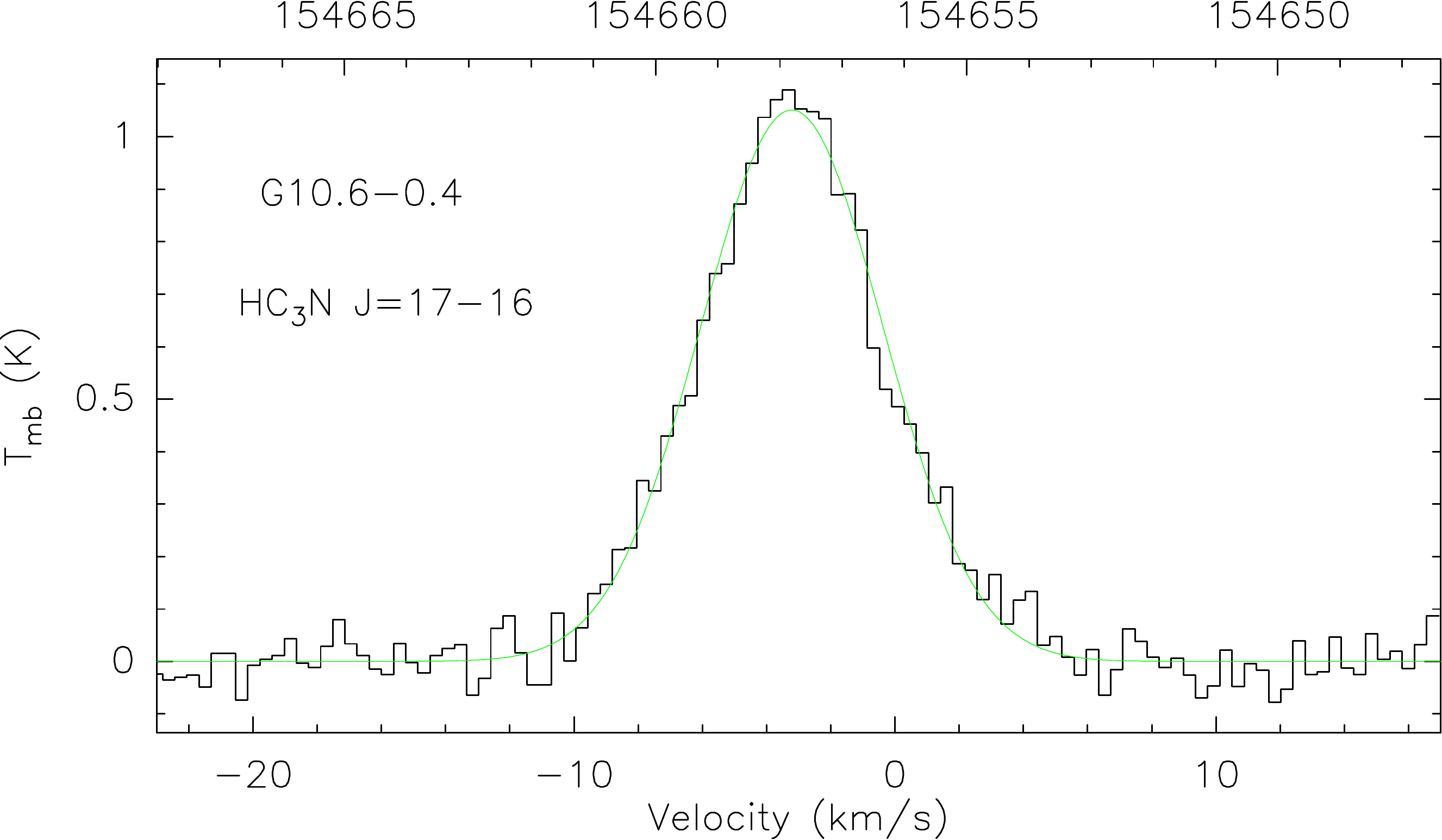}
	\end{minipage}%
	
	\vskip20pt 
	
	\begin{minipage}[t]{0.495\linewidth}
		\centering
		\includegraphics[width=80mm]{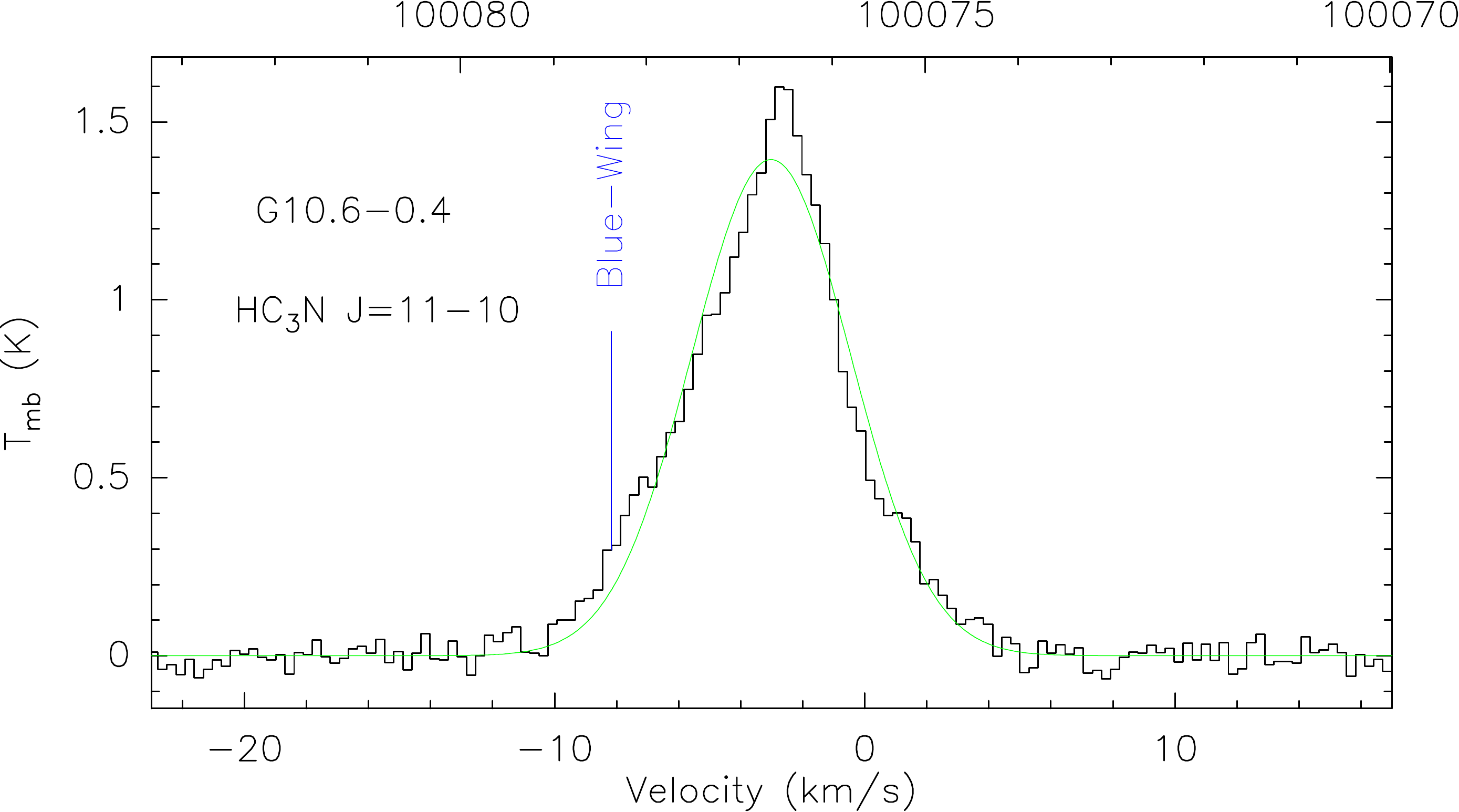}
	\end{minipage}%
	\begin{minipage}[t]{0.495\textwidth}
		\centering
		\includegraphics[width=80mm]{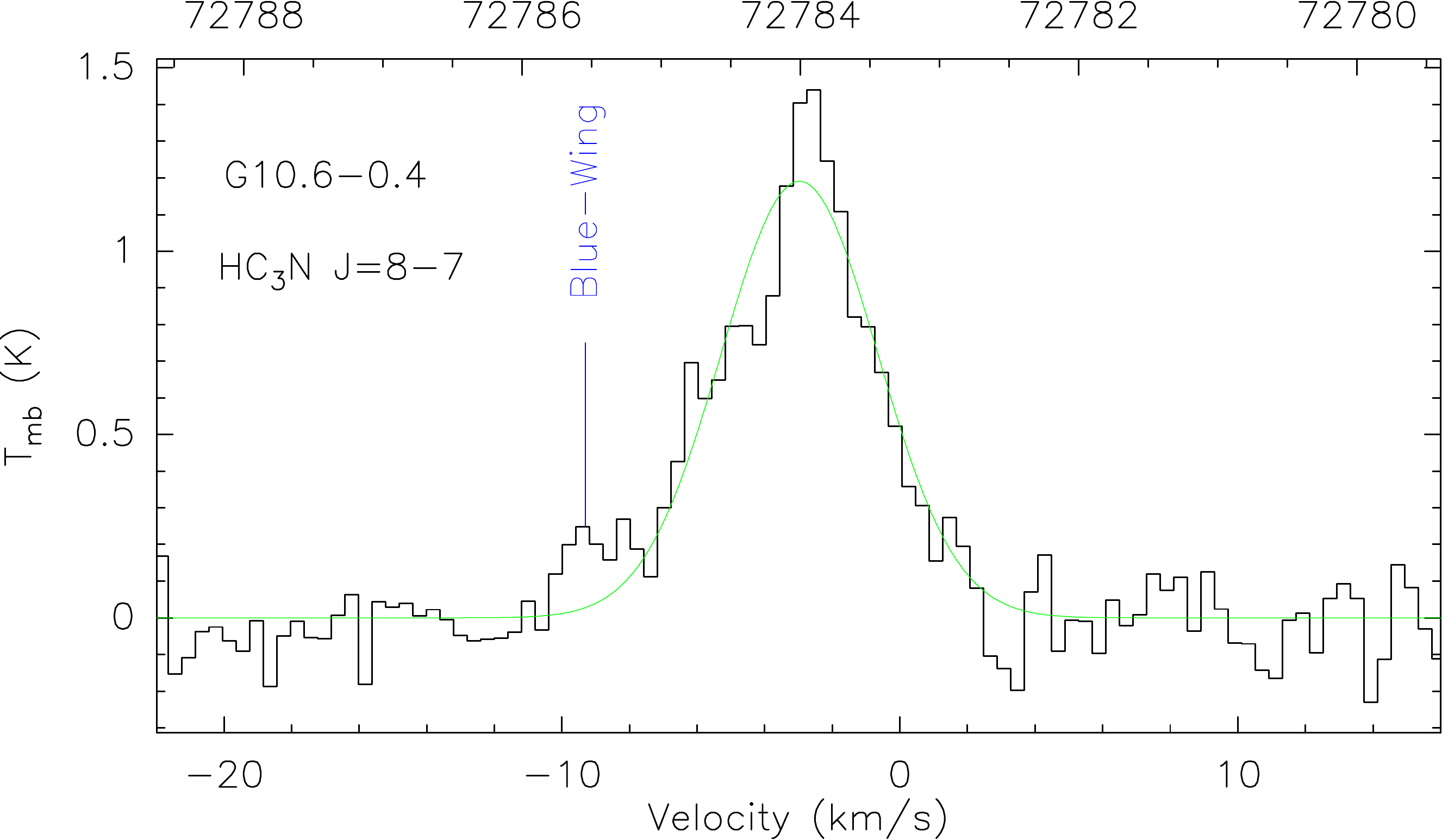}
	\end{minipage}%
	\caption{$-$ Continued.}   
\end{figure*}

\clearpage

\addtocounter{figure}{-1}
\begin{figure*}
	\vskip25pt 
	
	\begin{minipage}[t]{0.495\linewidth}
		\centering
		\includegraphics[width=80mm]{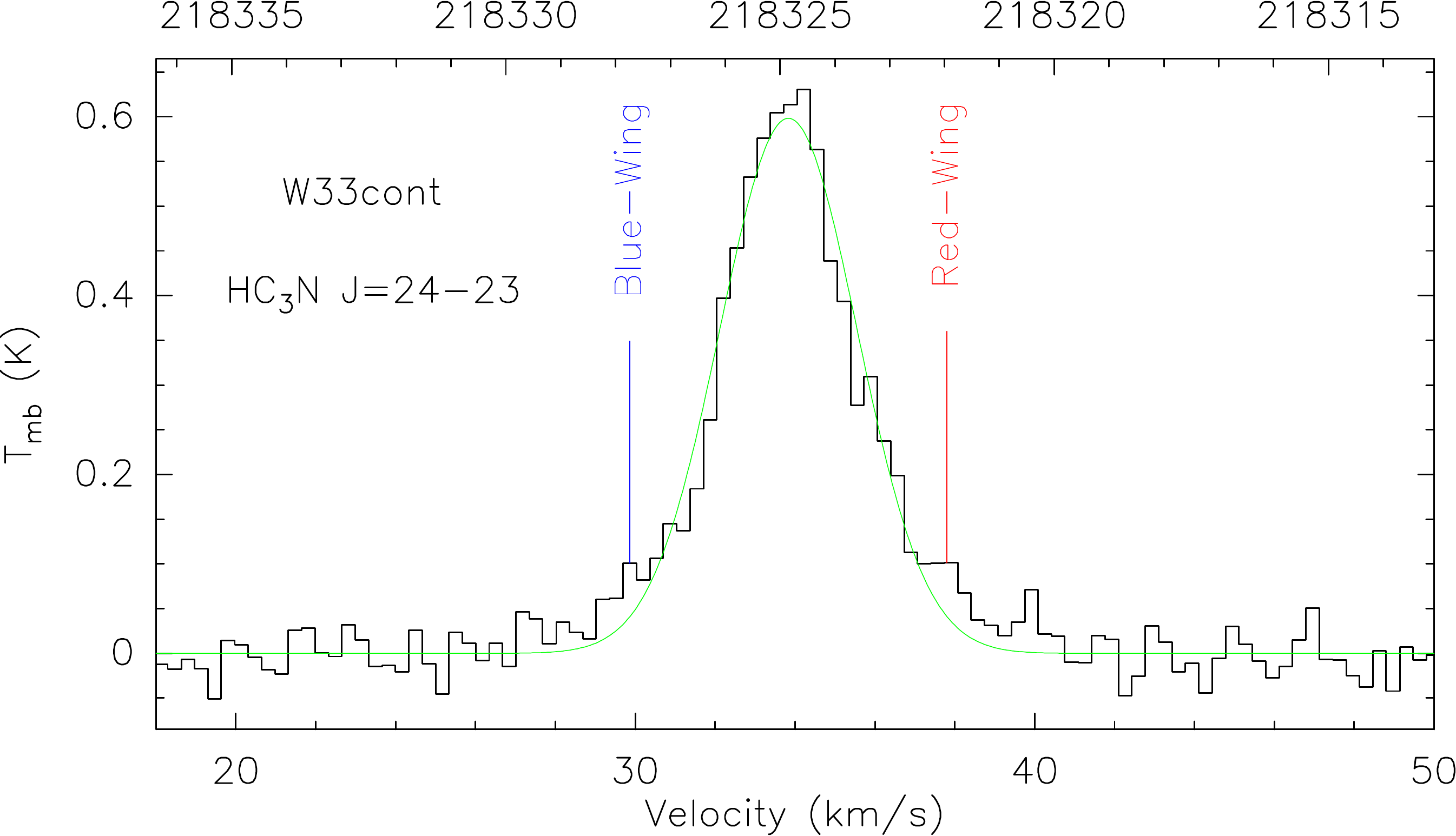}
	\end{minipage}%
	\begin{minipage}[t]{0.495\textwidth}
		\centering
		\includegraphics[width=80mm]{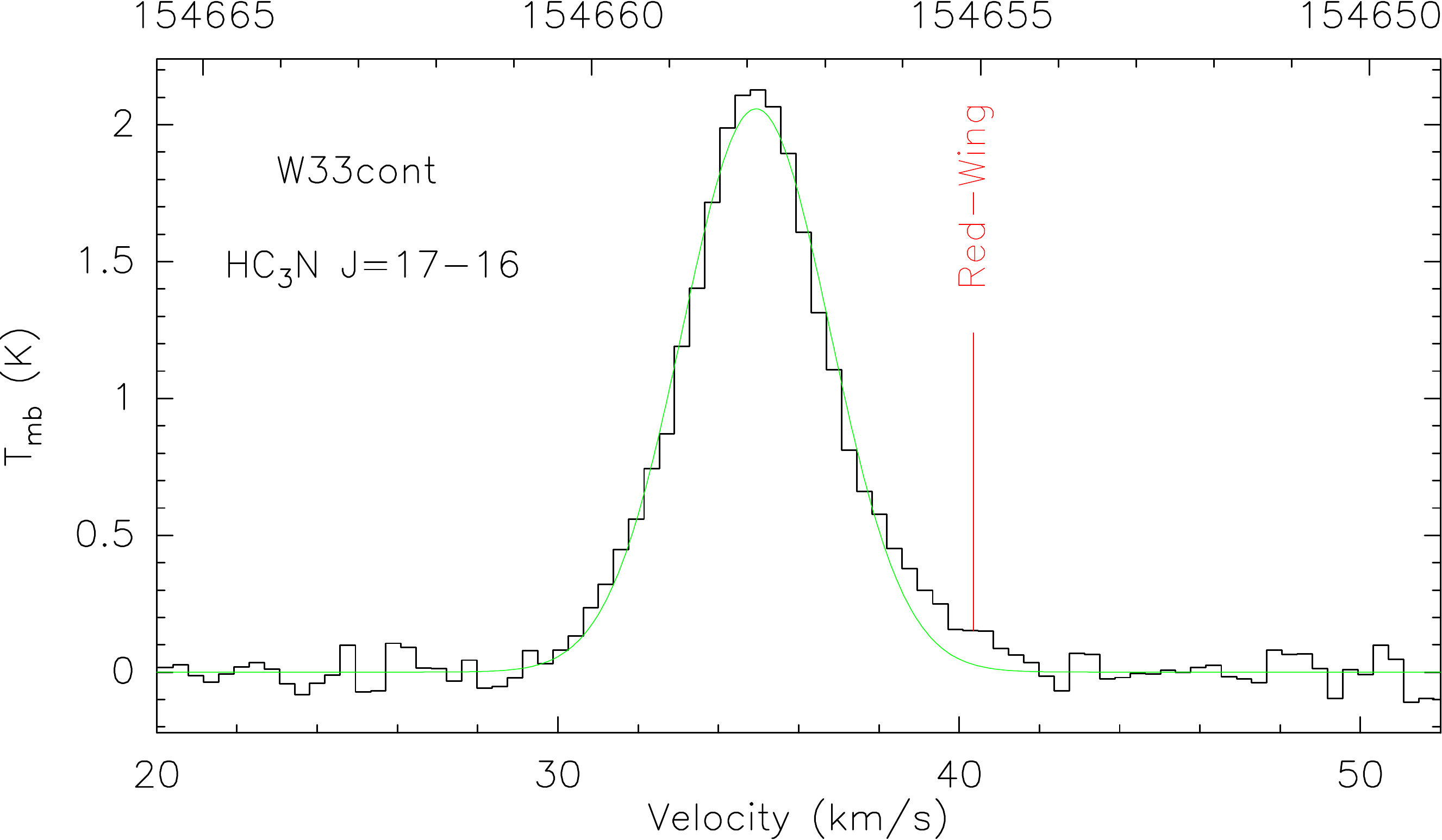}
	\end{minipage}%
	
	\vskip20pt 
	
	\begin{minipage}[t]{0.495\linewidth}
		\centering
		\includegraphics[width=80mm]{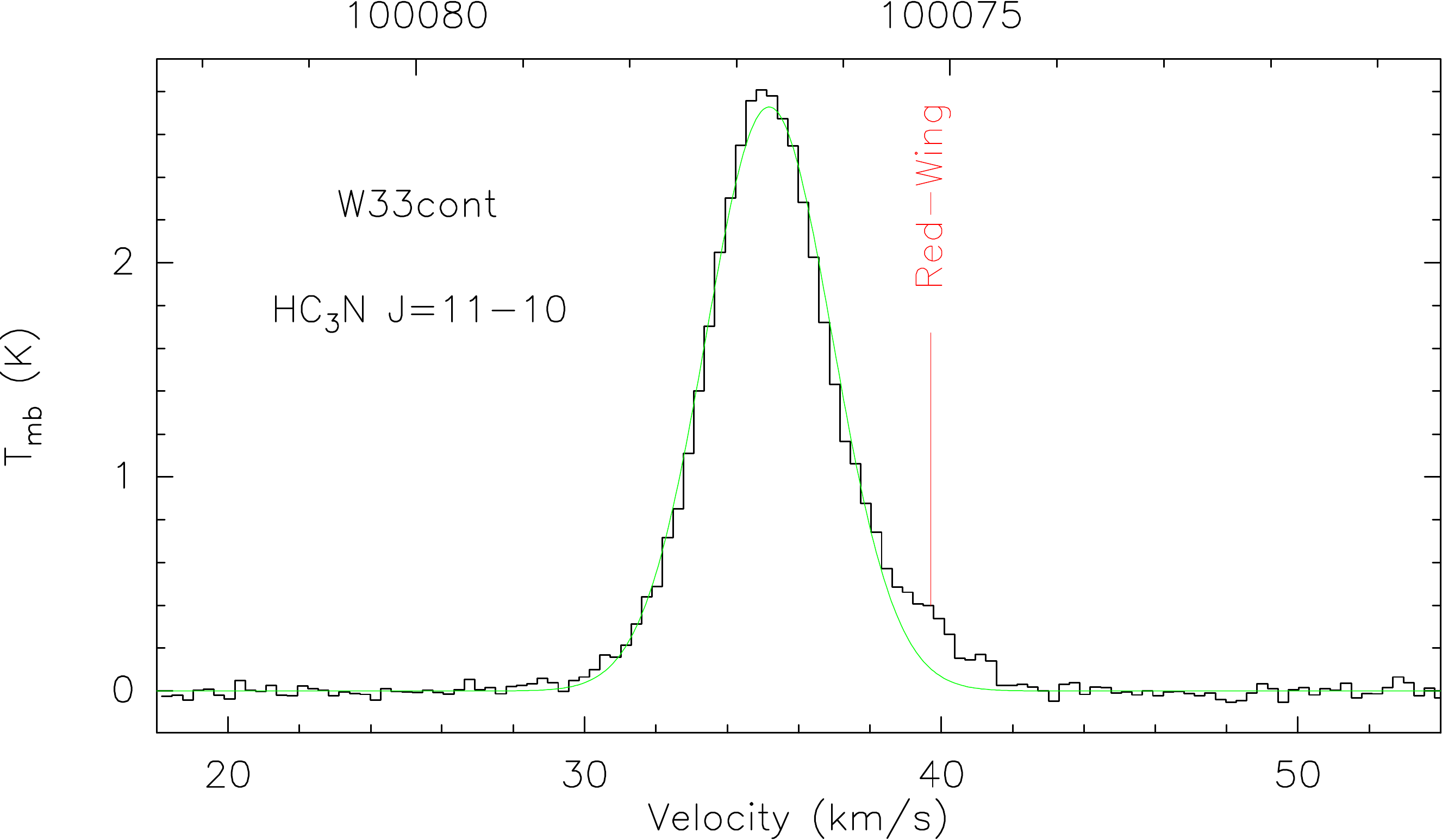}
	\end{minipage}%
	\begin{minipage}[t]{0.495\textwidth}
		\centering
		\includegraphics[width=80mm]{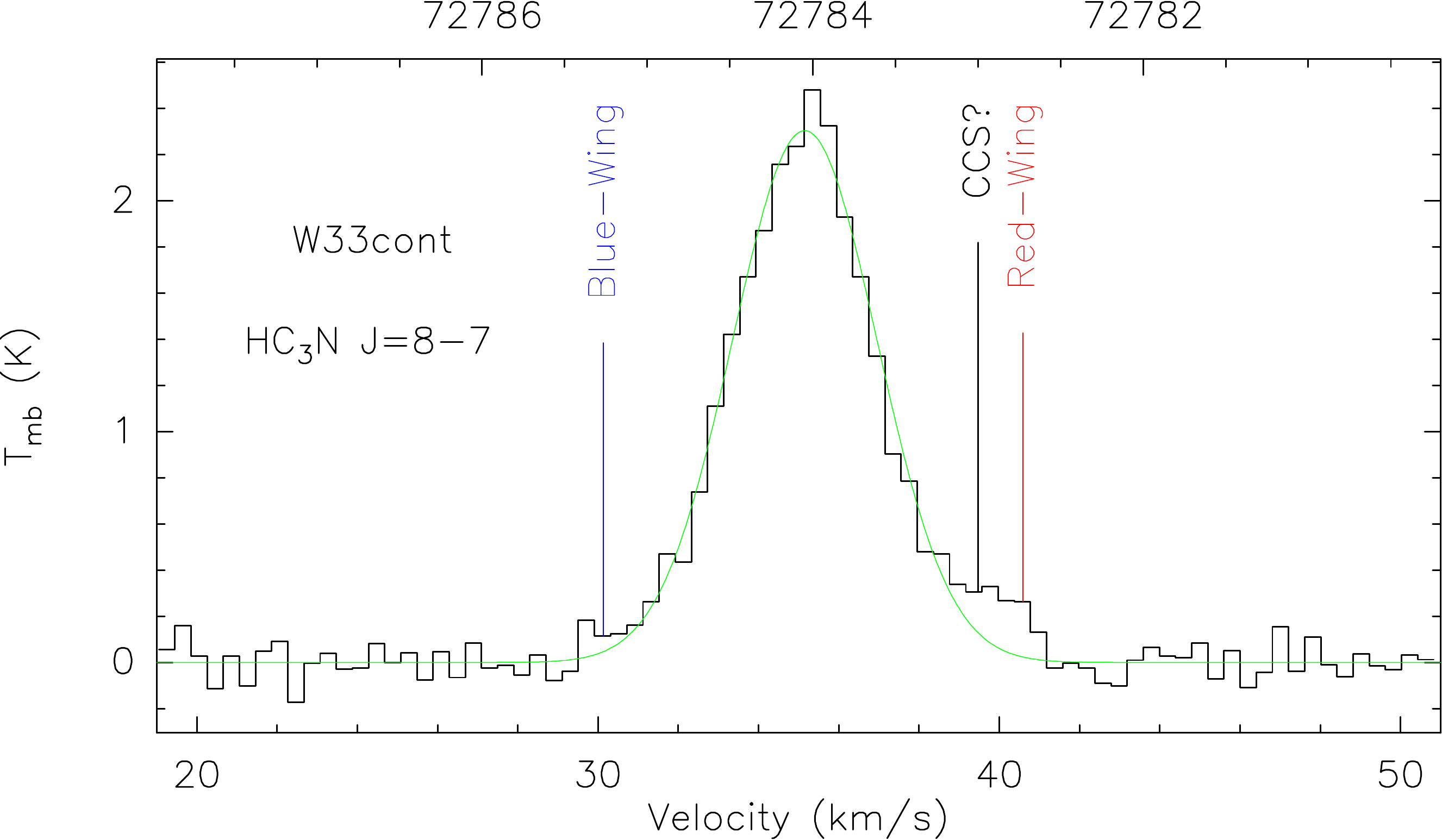}
	\end{minipage}%
	
	\vskip20pt 
	
	\begin{minipage}[t]{0.495\linewidth}
		\centering
		\includegraphics[width=80mm]{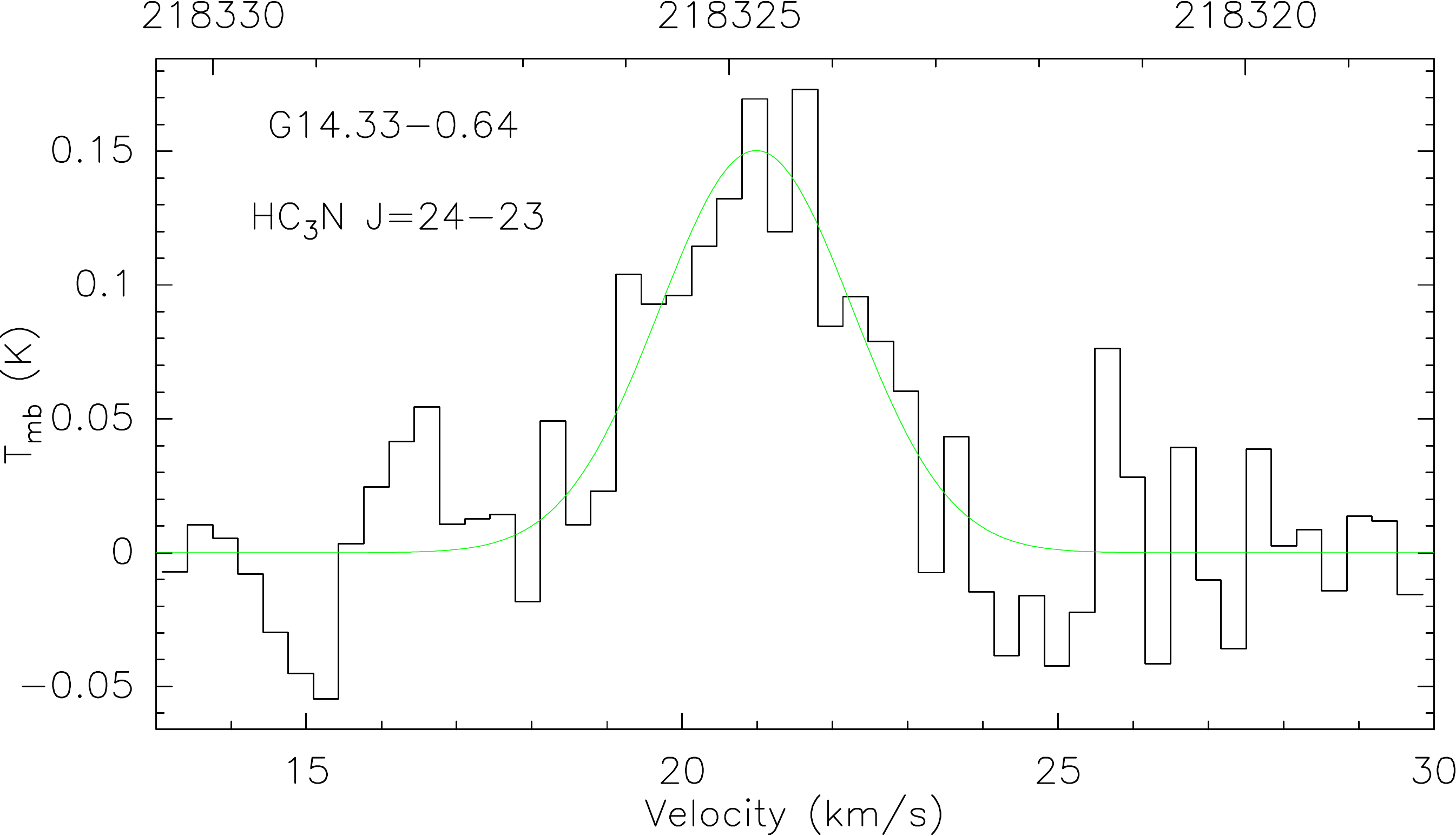}
	\end{minipage}%
	\begin{minipage}[t]{0.495\textwidth}
		\centering
		\includegraphics[width=80mm]{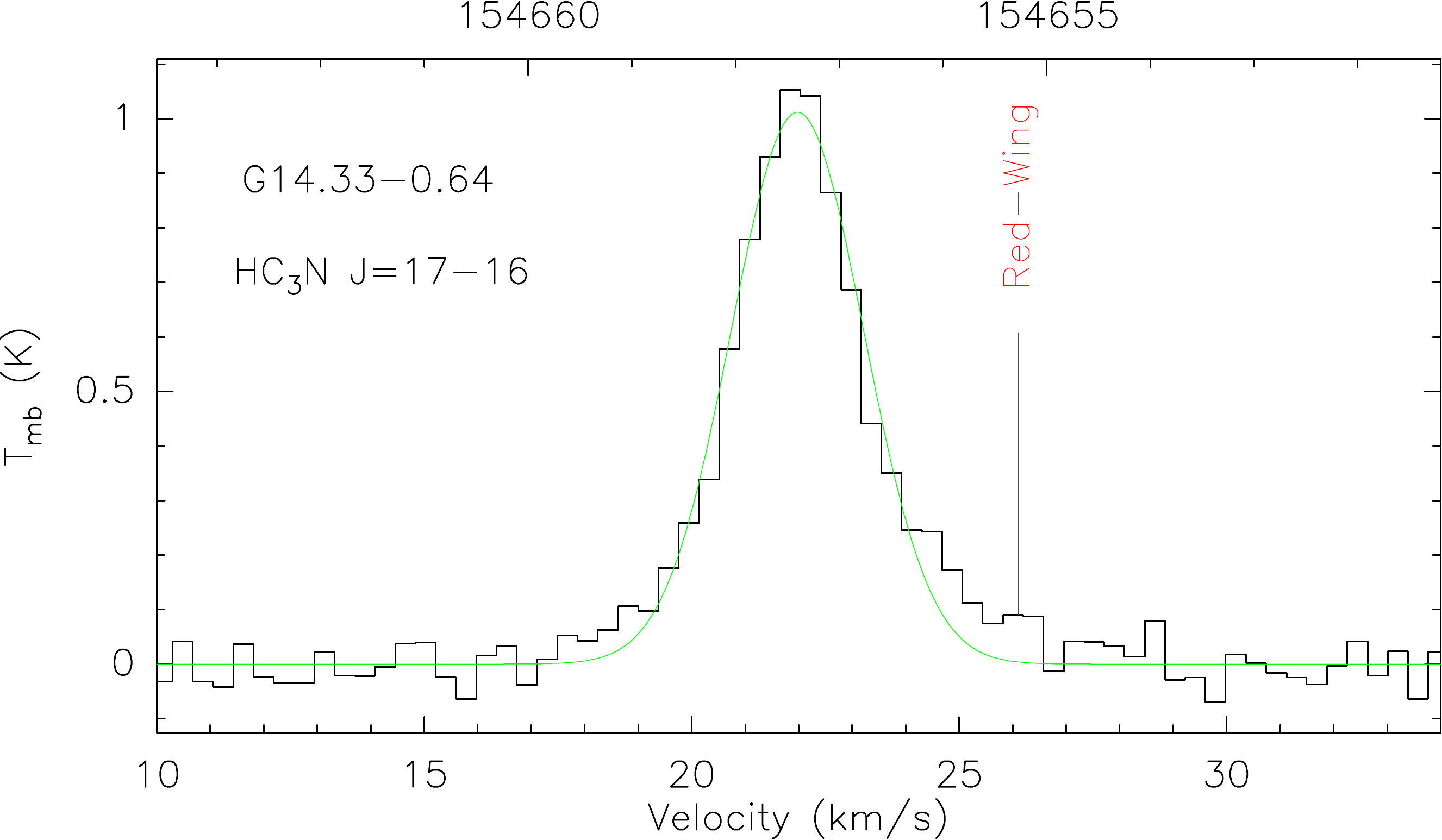}
	\end{minipage}%
	
	\vskip20pt 
	
	\begin{minipage}[t]{0.495\linewidth}
		\centering
		\includegraphics[width=80mm]{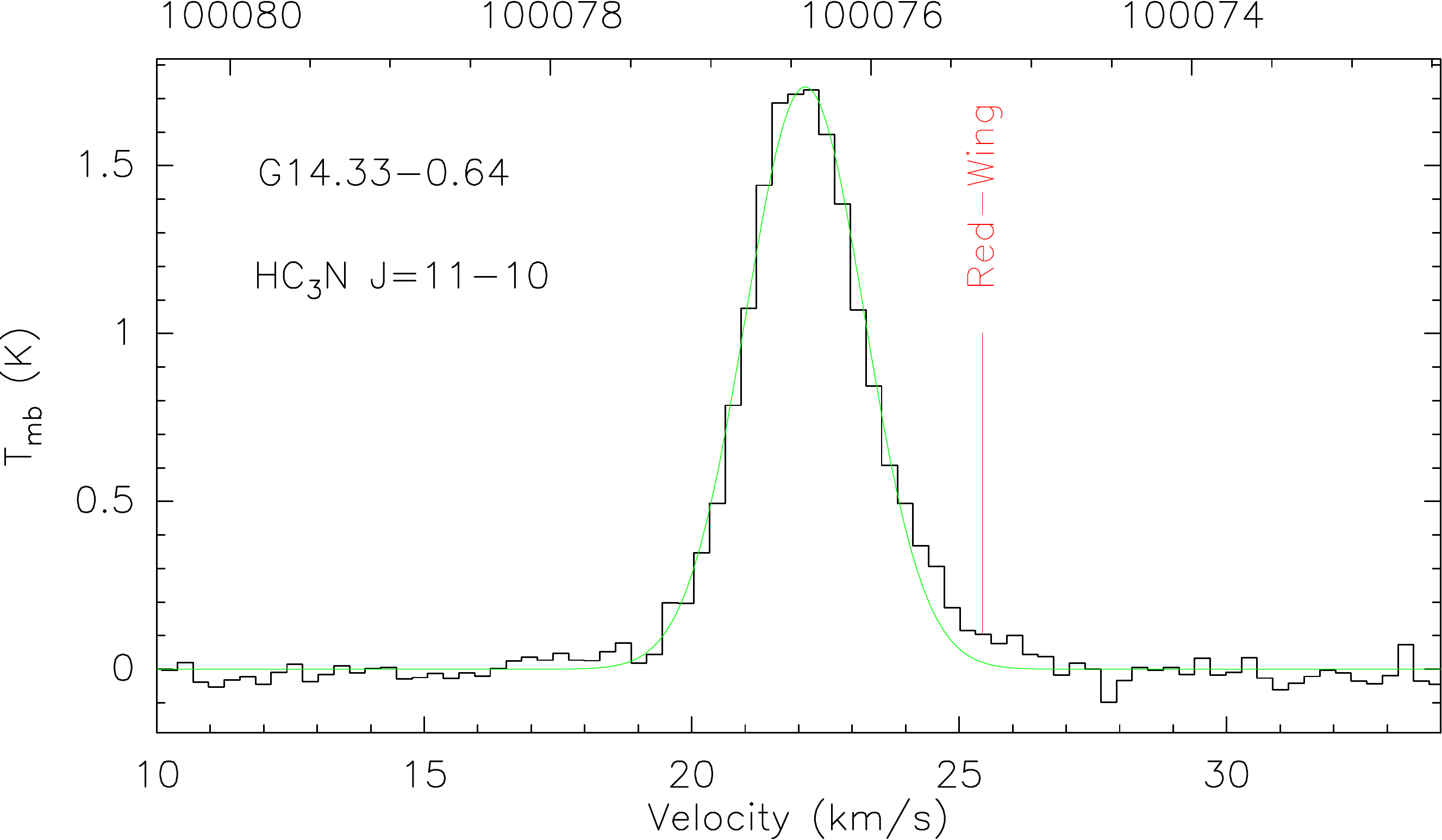}
	\end{minipage}%
	\begin{minipage}[t]{0.495\textwidth}
		\centering
		\includegraphics[width=80mm]{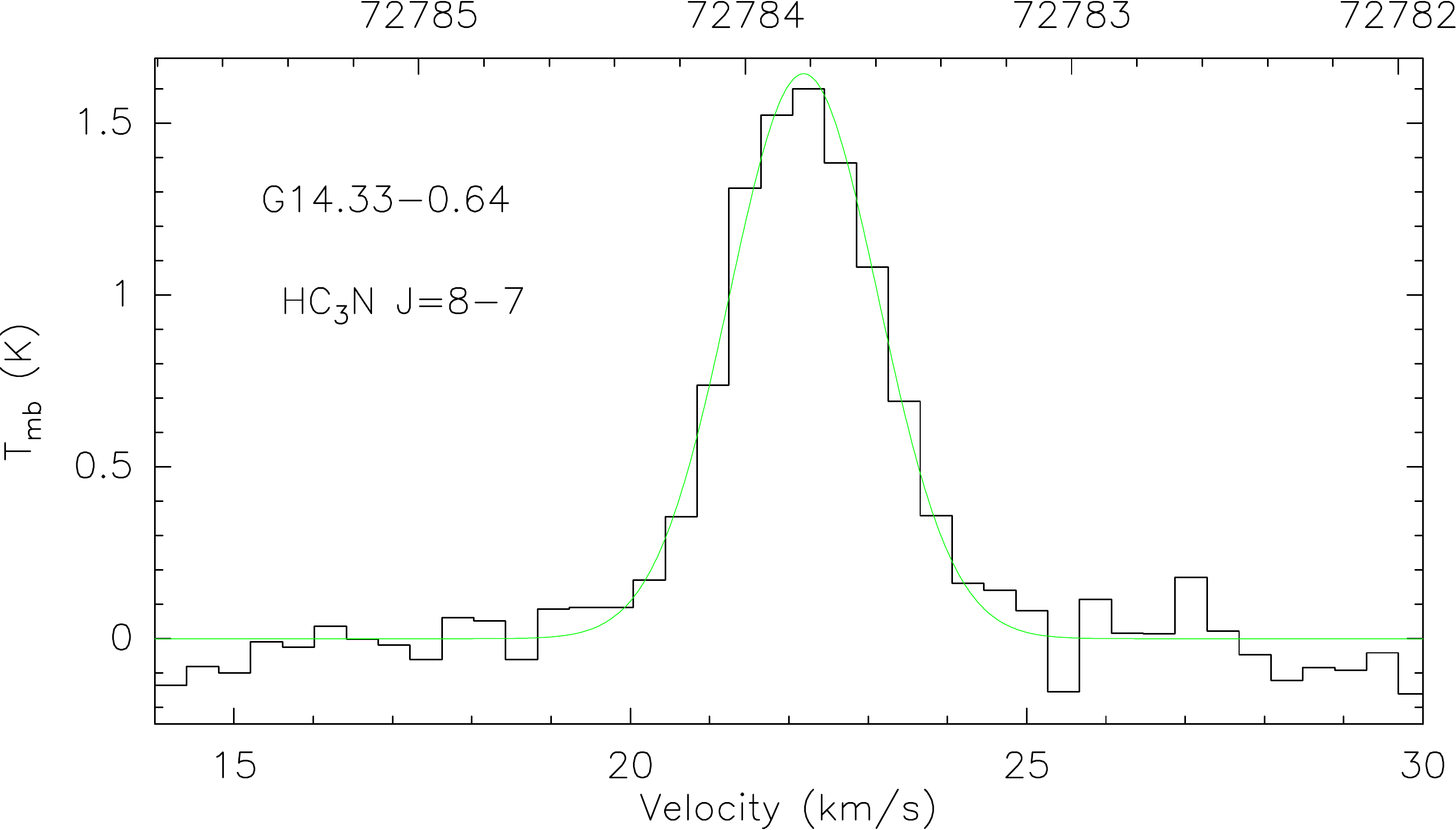}
	\end{minipage}%
	\caption{$-$ Continued.}  
	
\end{figure*}

\clearpage

\addtocounter{figure}{-1}
\begin{figure*}
	\vskip25pt 
	
	\begin{minipage}[t]{0.495\linewidth}
		\centering
		\includegraphics[width=80mm]{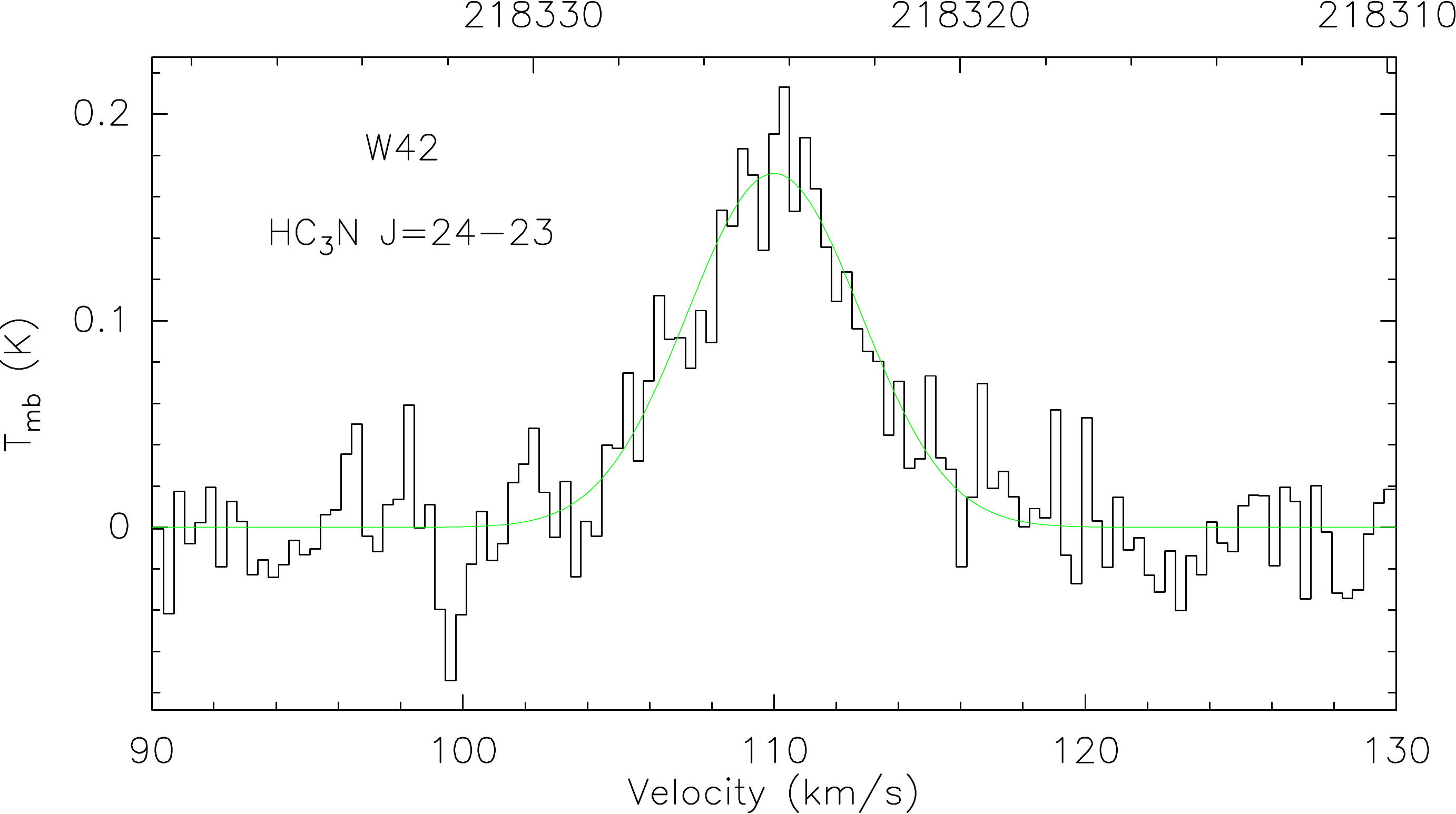}
	\end{minipage}%
	\begin{minipage}[t]{0.495\textwidth}
		\centering
		\includegraphics[width=80mm]{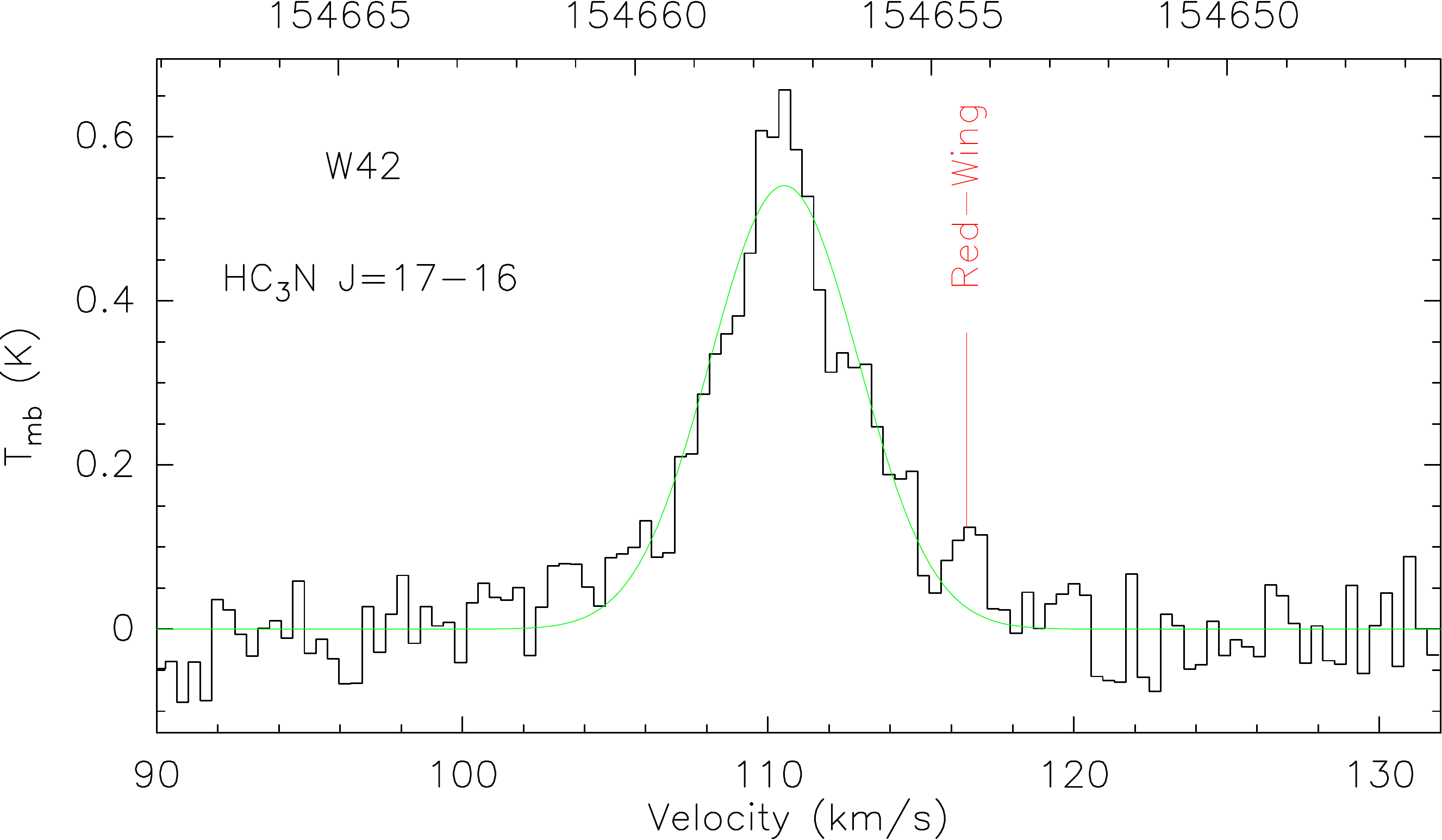}
	\end{minipage}%
	
	\vskip20pt 
	
	\begin{minipage}[t]{0.495\linewidth}
		\centering
		\includegraphics[width=80mm]{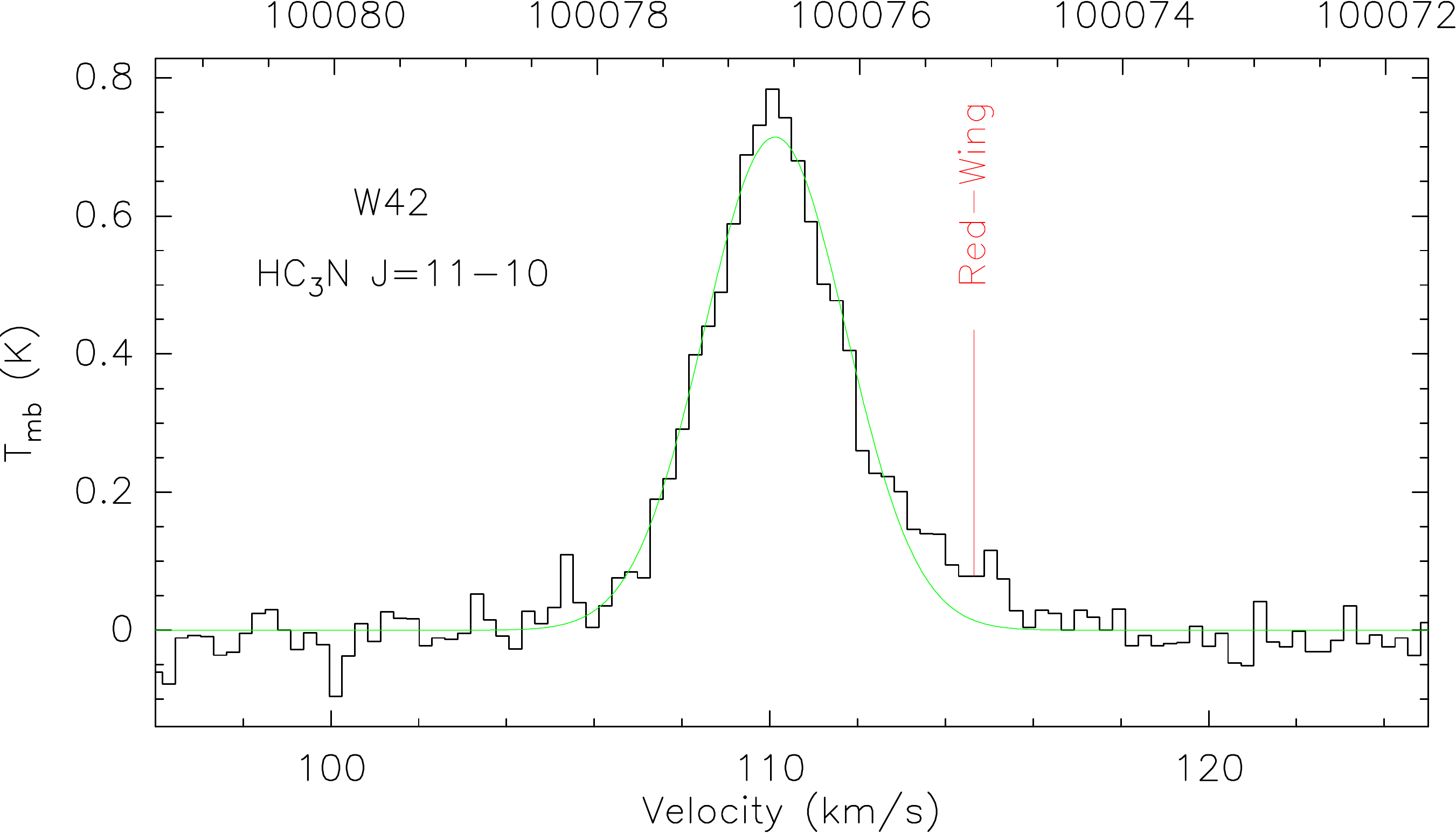}
	\end{minipage}%
	\begin{minipage}[t]{0.495\textwidth}
		\centering
		\includegraphics[width=80mm]{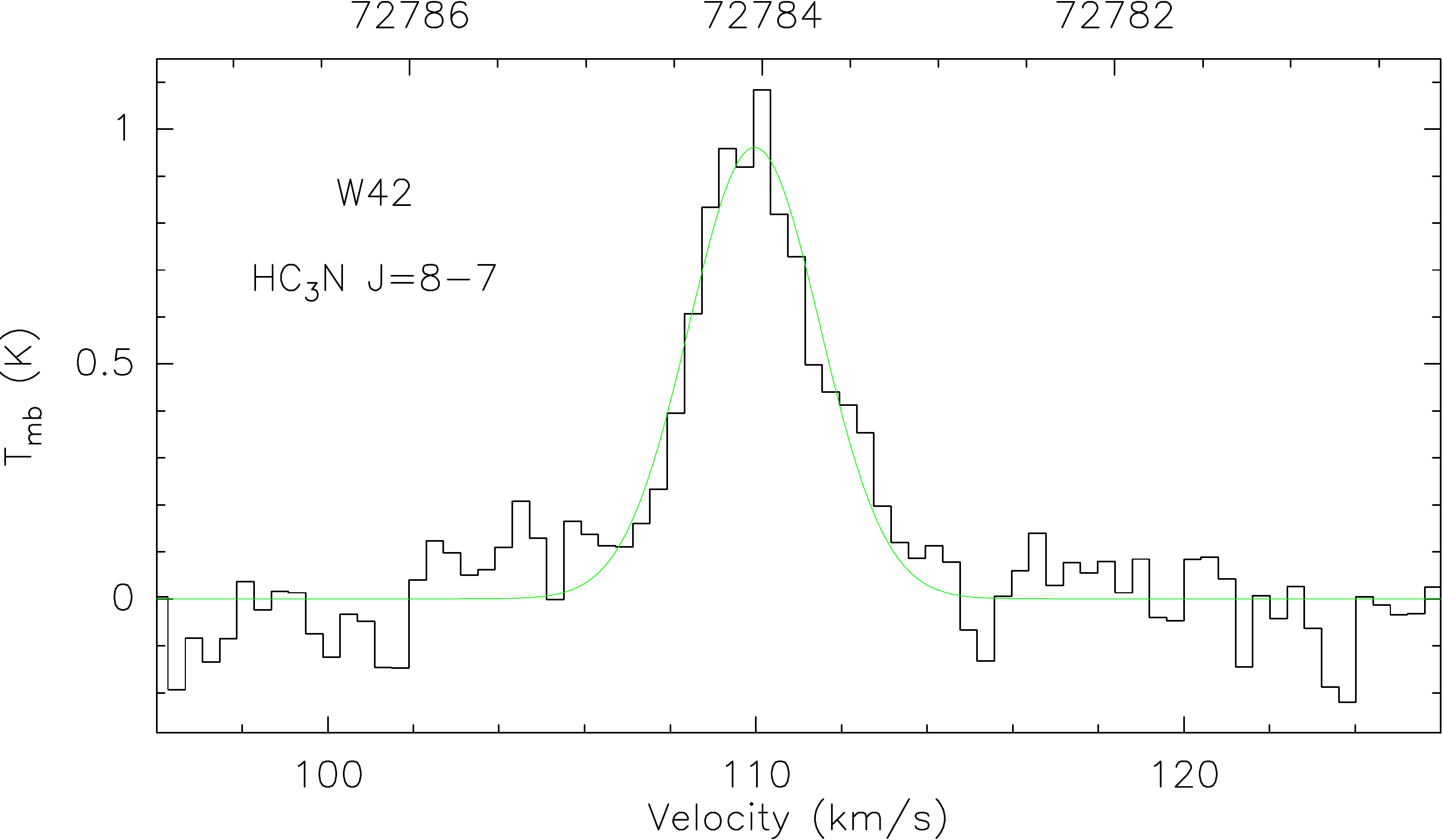}
	\end{minipage}%
	
	\vskip20pt 
	
	\begin{minipage}[t]{0.495\linewidth}
		\centering
		\includegraphics[width=80mm]{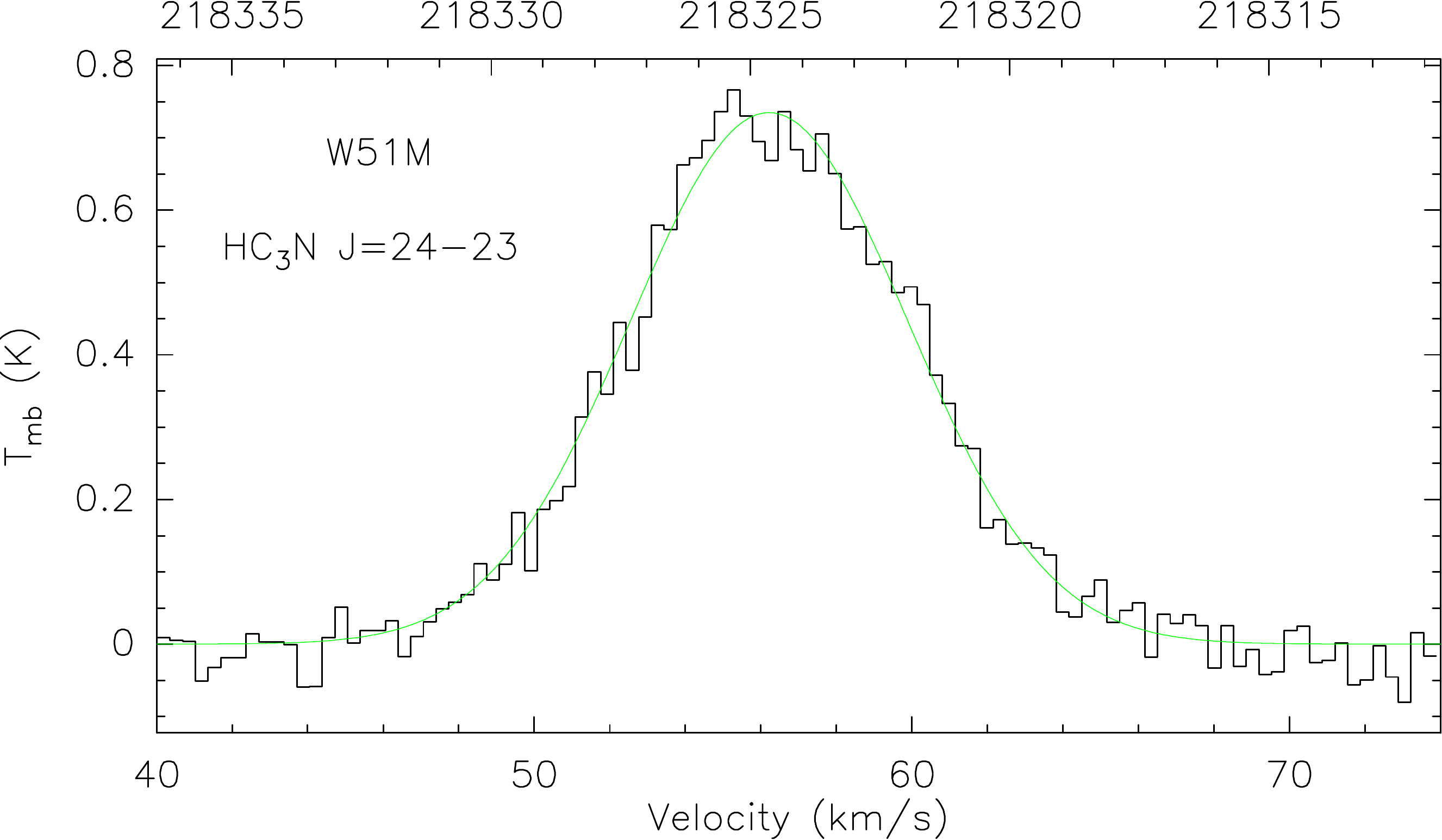}
	\end{minipage}%
	\begin{minipage}[t]{0.495\textwidth}
		\centering
		\includegraphics[width=80mm]{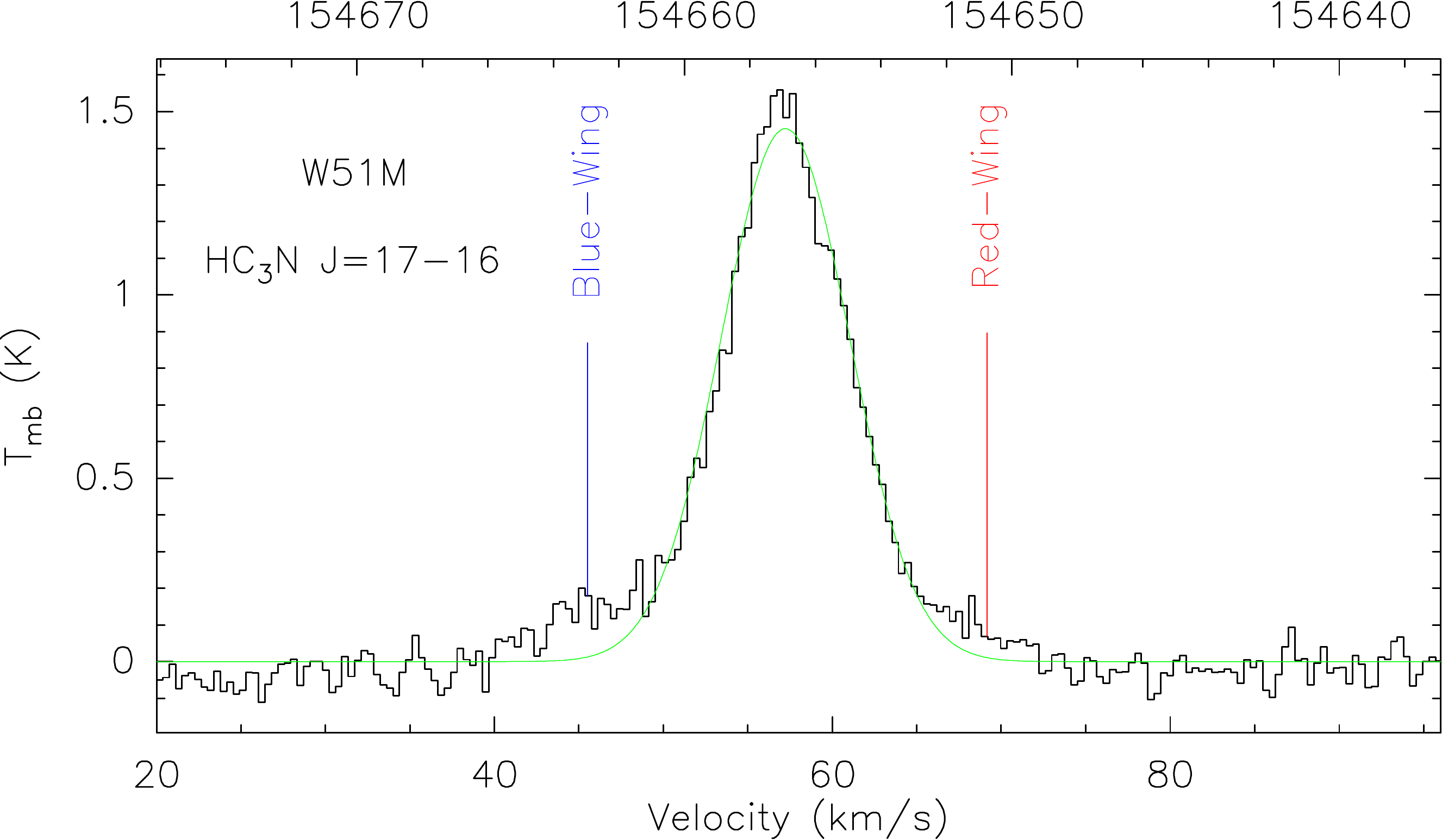}
	\end{minipage}%
	
	\vskip20pt 
	
	\begin{minipage}[t]{0.495\linewidth}
		\centering
		\includegraphics[width=80mm]{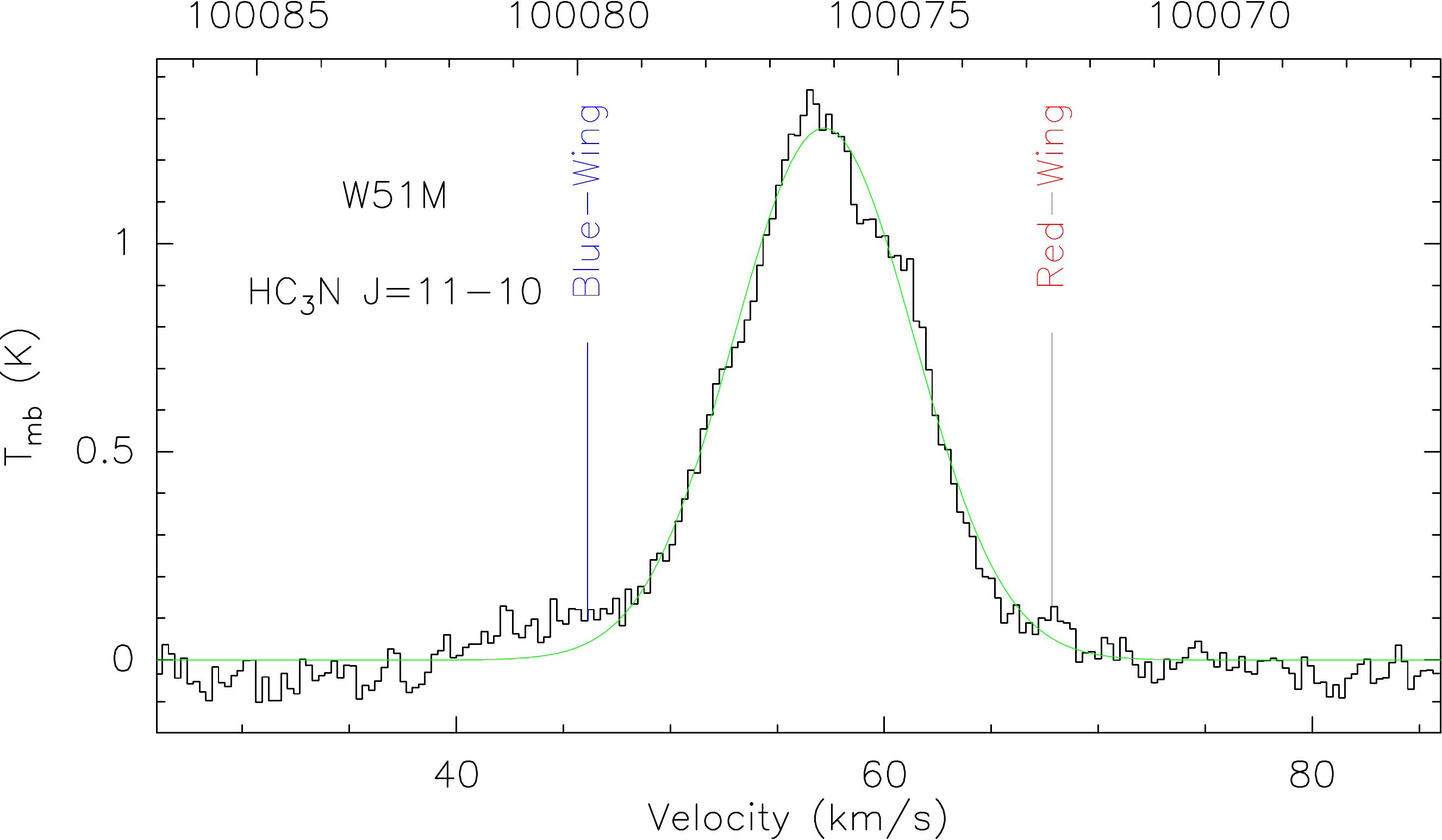}
	\end{minipage}%
	\begin{minipage}[t]{0.495\textwidth}
		\centering
		\includegraphics[width=80mm]{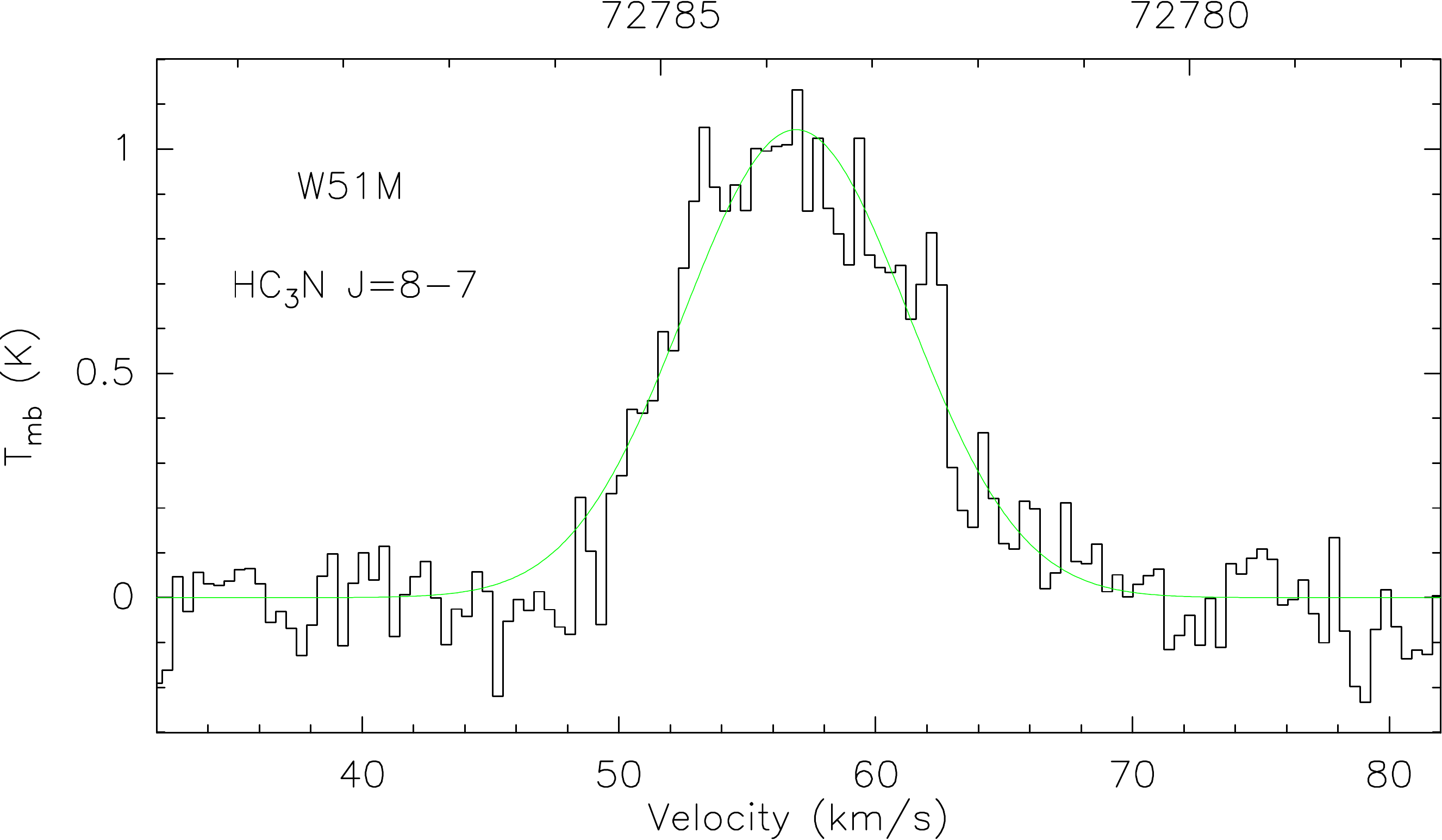}
	\end{minipage}%
	\caption{$-$ Continued.}  
	
\end{figure*}

\clearpage

\addtocounter{figure}{-1}
\begin{figure*}
	\vskip25pt 
	
	\begin{minipage}[t]{0.495\linewidth}
		\centering
		\includegraphics[width=80mm]{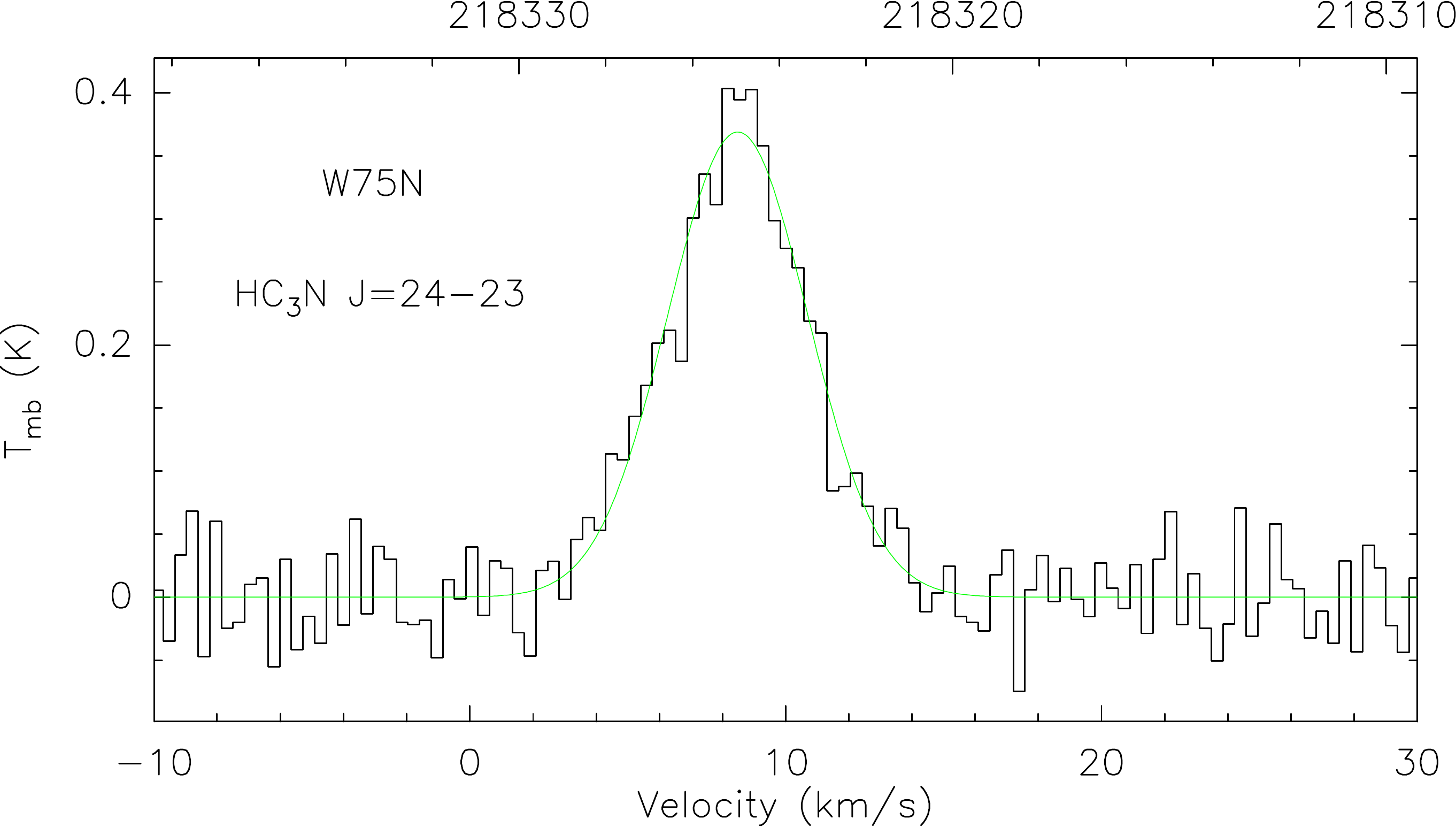}
	\end{minipage}%
	\begin{minipage}[t]{0.495\textwidth}
		\centering
		\includegraphics[width=80mm]{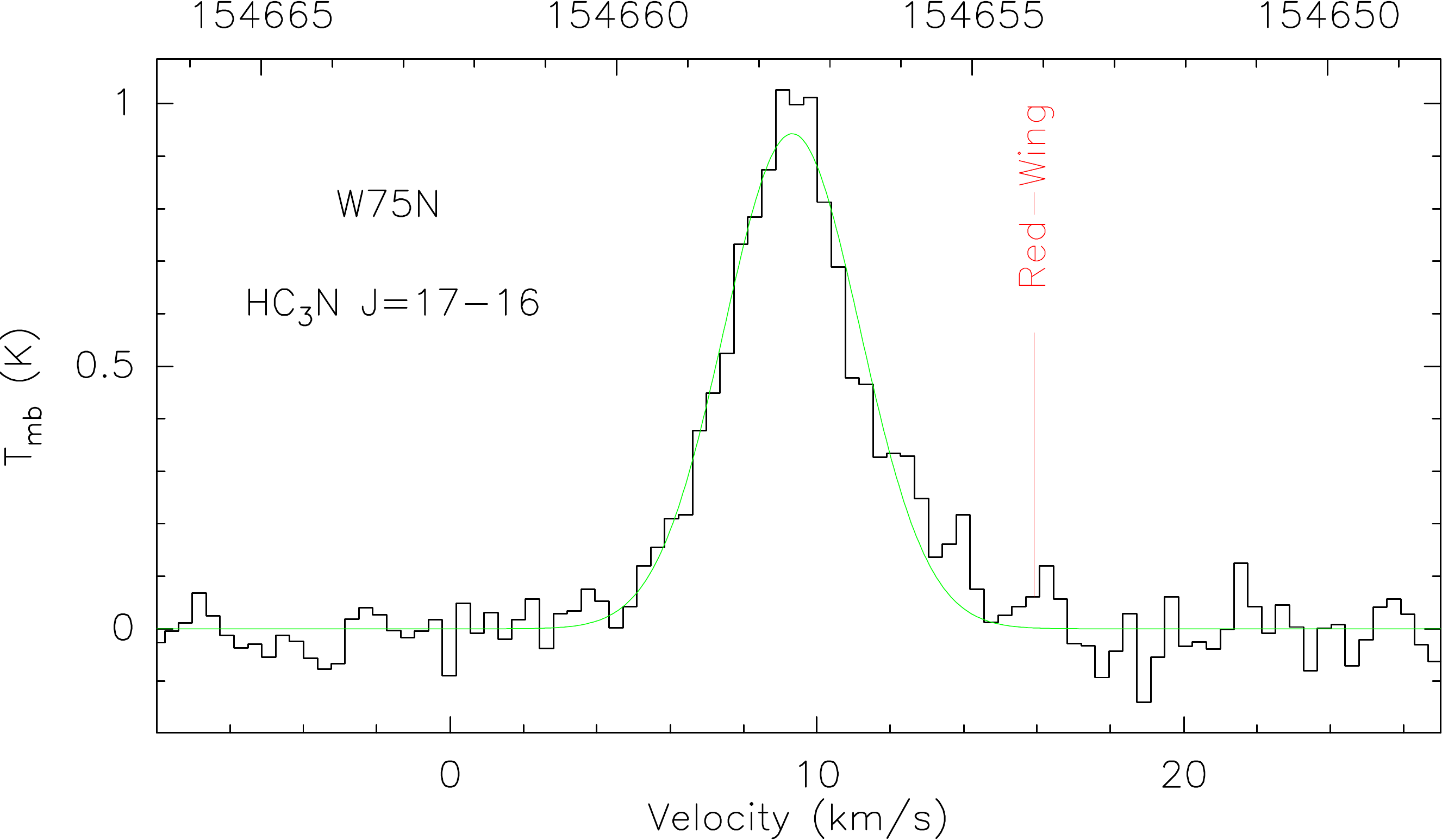}
	\end{minipage}%
	
	\vskip20pt 
	
	\begin{minipage}[t]{0.495\linewidth}
		\centering
		\includegraphics[width=80mm]{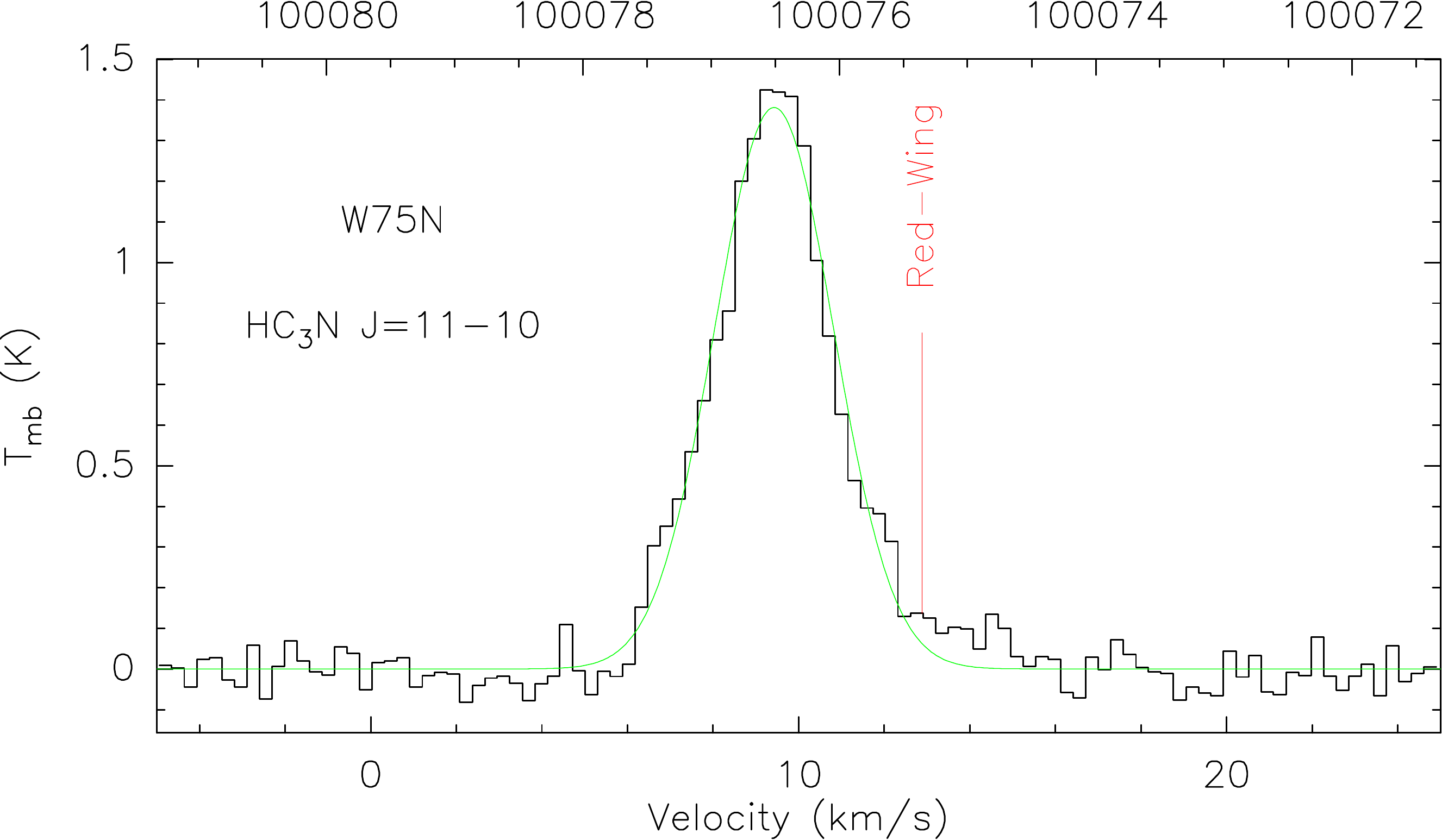}
	\end{minipage}%
	\begin{minipage}[t]{0.495\textwidth}
		\centering
		\includegraphics[width=80mm]{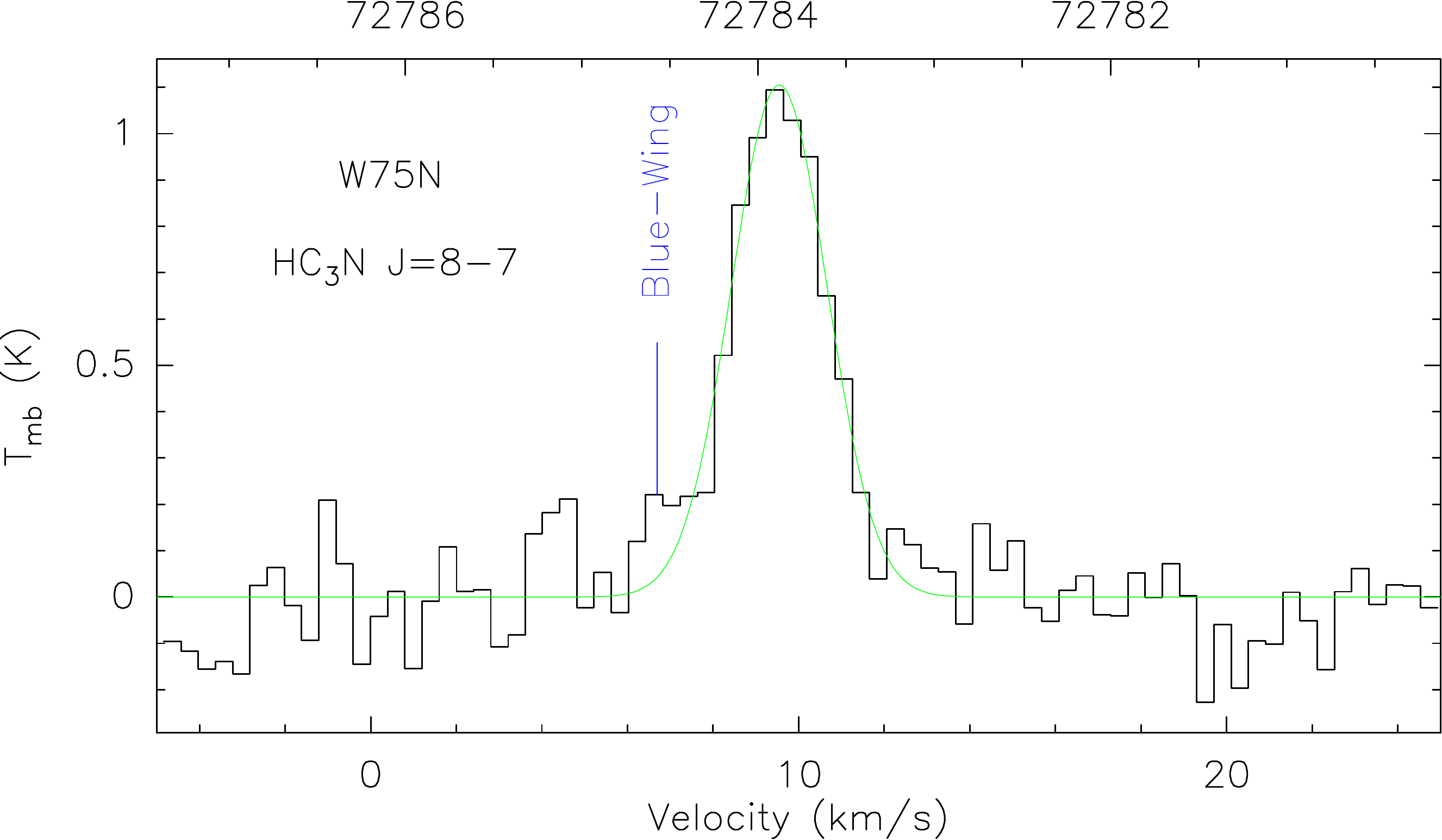}
	\end{minipage}%
	
	\vskip20pt 
	
	\begin{minipage}[t]{0.495\linewidth}
		\centering
		\includegraphics[width=80mm]{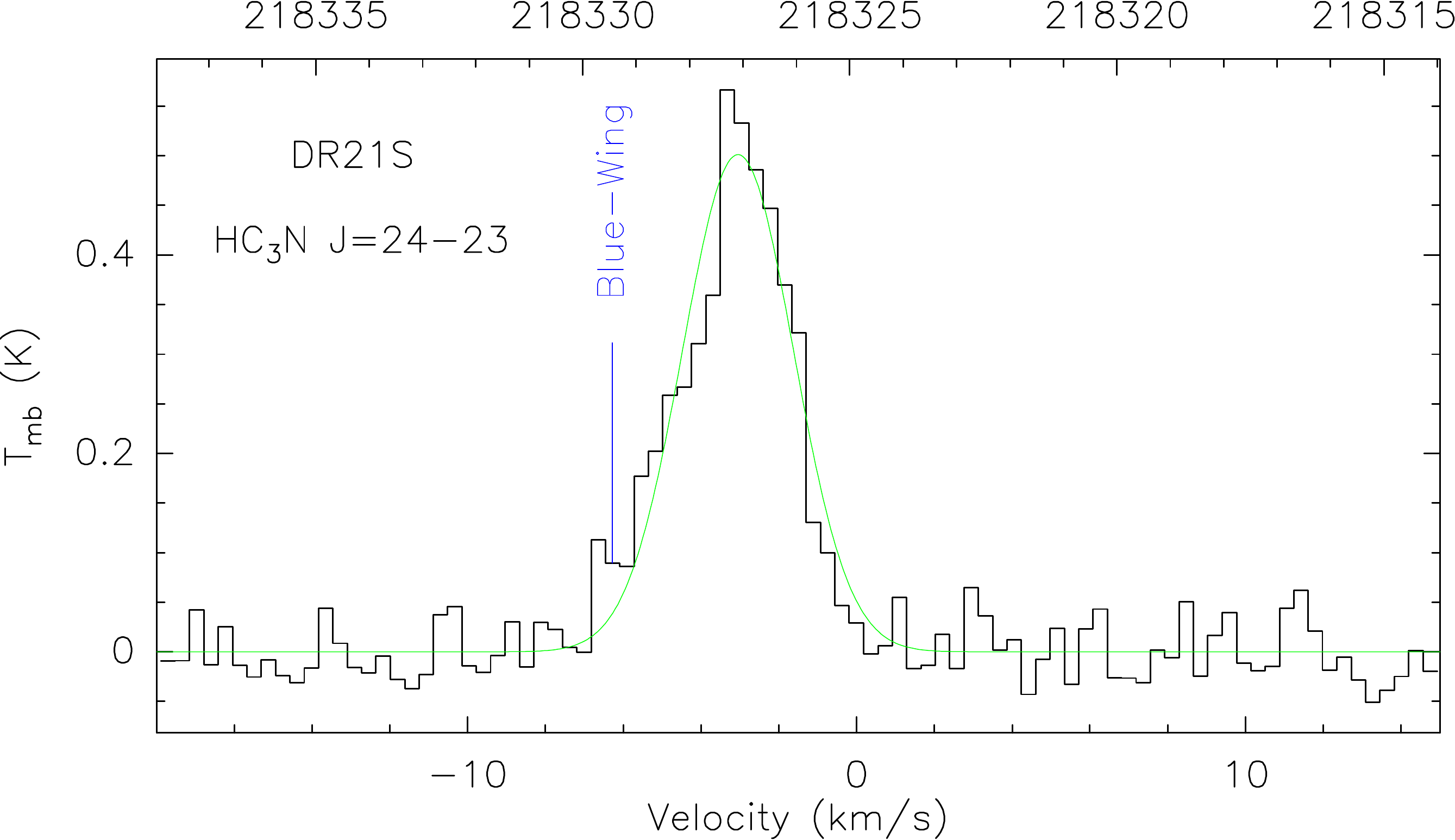}
	\end{minipage}%
	\begin{minipage}[t]{0.495\textwidth}
		\centering
		\includegraphics[width=80mm]{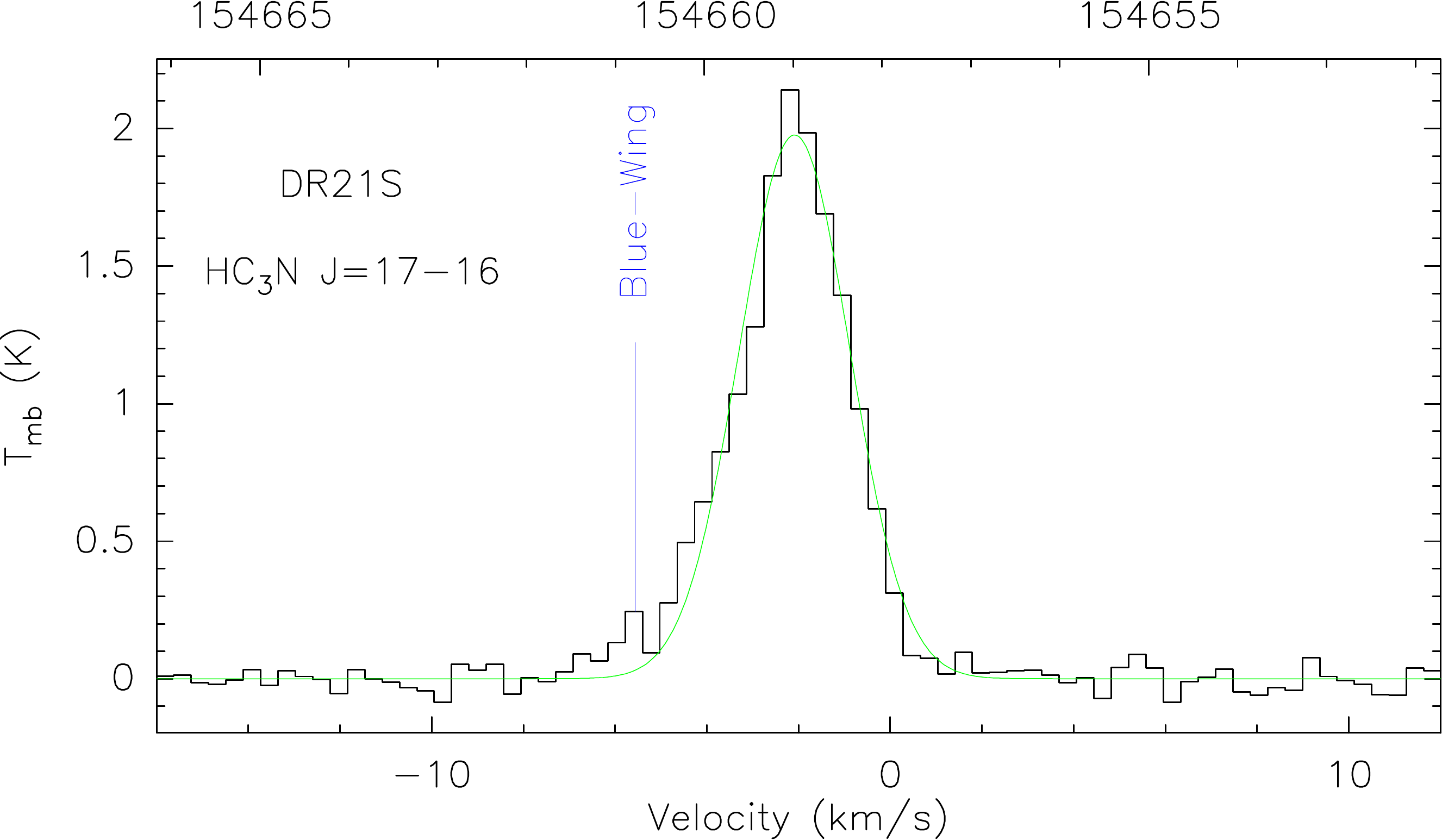}
	\end{minipage}%
	
	\vskip20pt 
	
	\begin{minipage}[t]{0.495\linewidth}
		\centering
		\includegraphics[width=80mm]{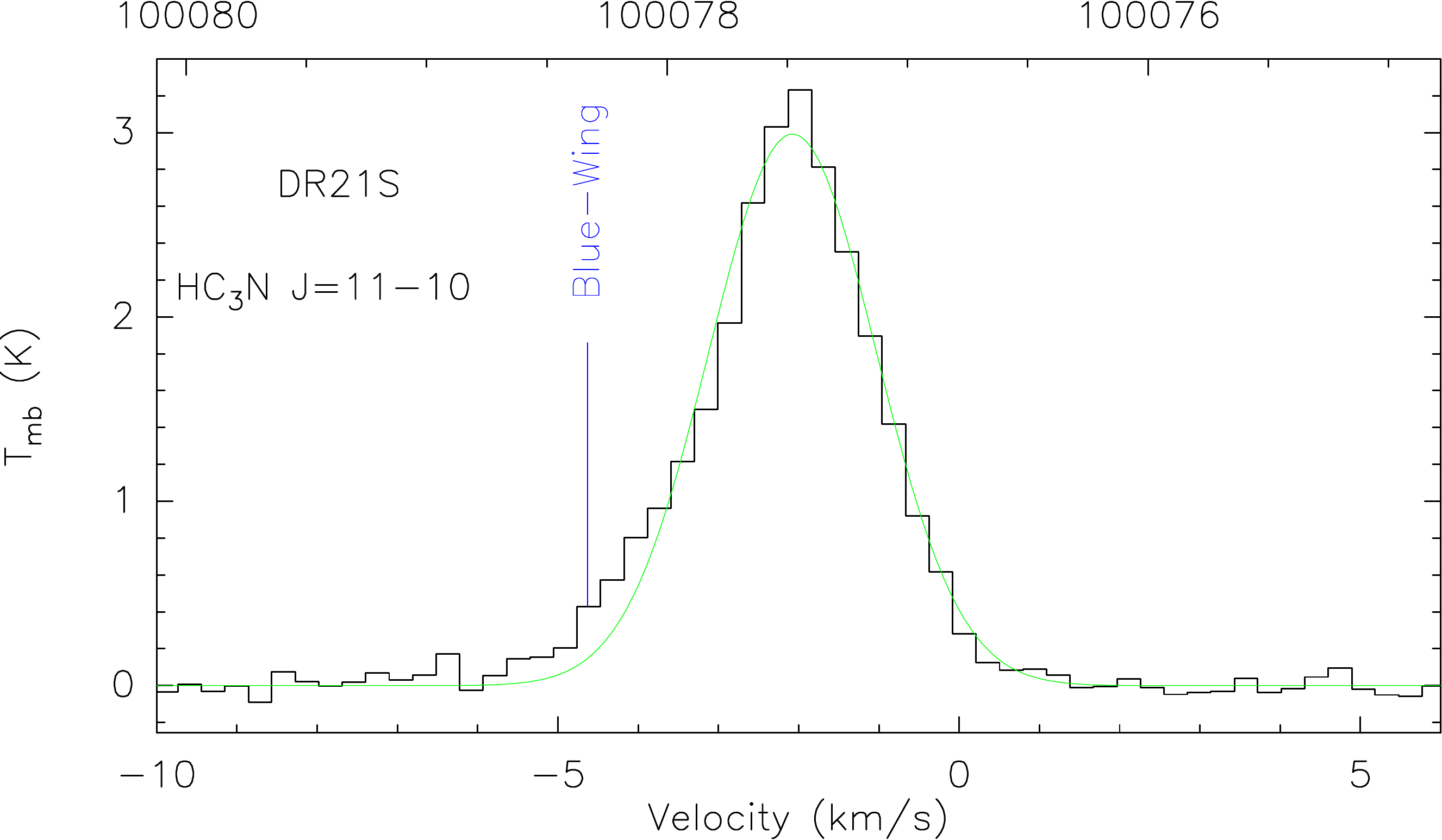}
	\end{minipage}%
	\begin{minipage}[t]{0.495\textwidth}
		\centering
		\includegraphics[width=80mm]{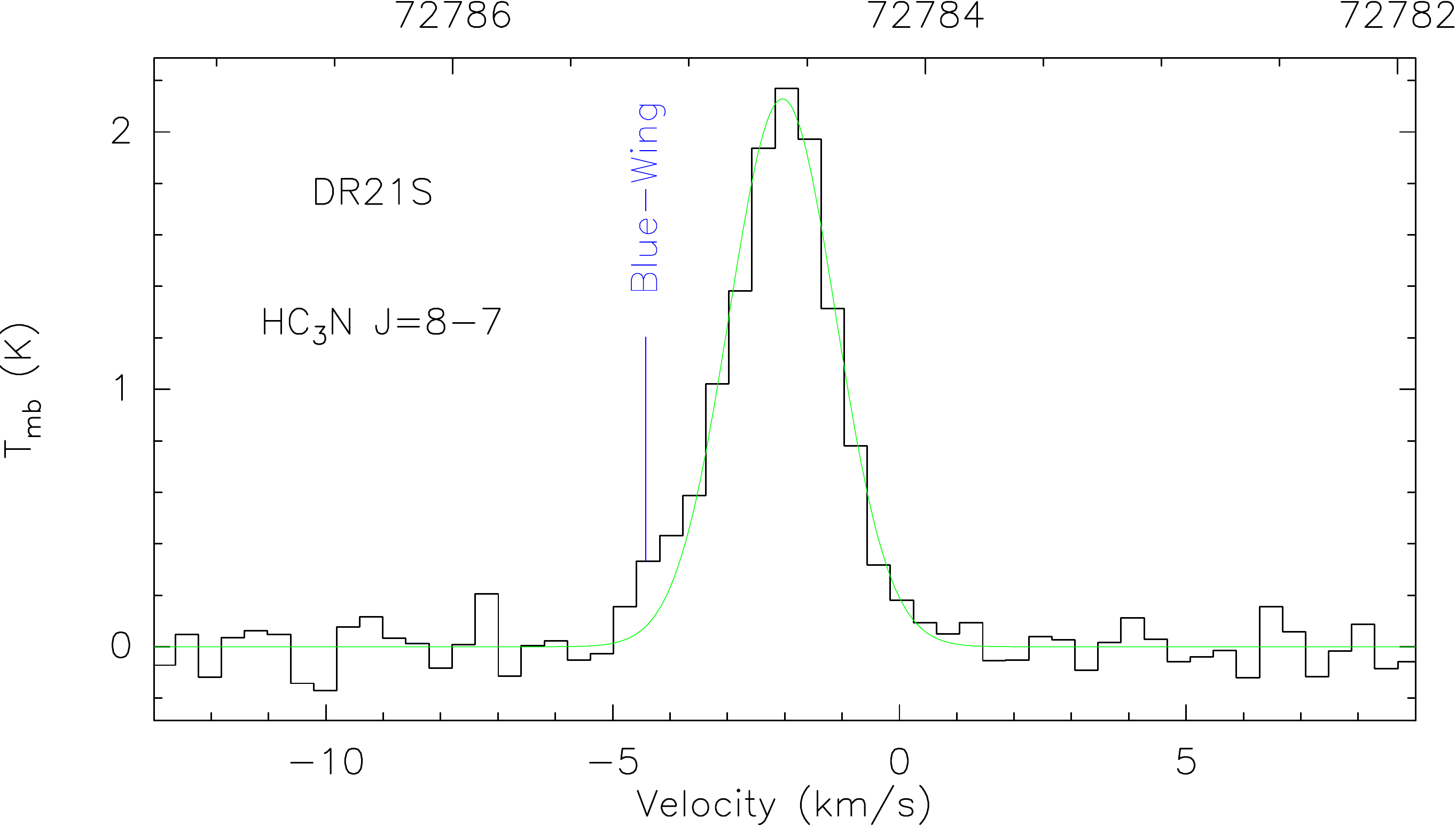}
	\end{minipage}%
	\caption{$-$ Continued.}  
	
\end{figure*}

\clearpage

\addtocounter{figure}{-1}
\begin{figure*}
	\vskip25pt 
	
	\begin{minipage}[t]{0.495\linewidth}
		\centering
		\includegraphics[width=80mm]{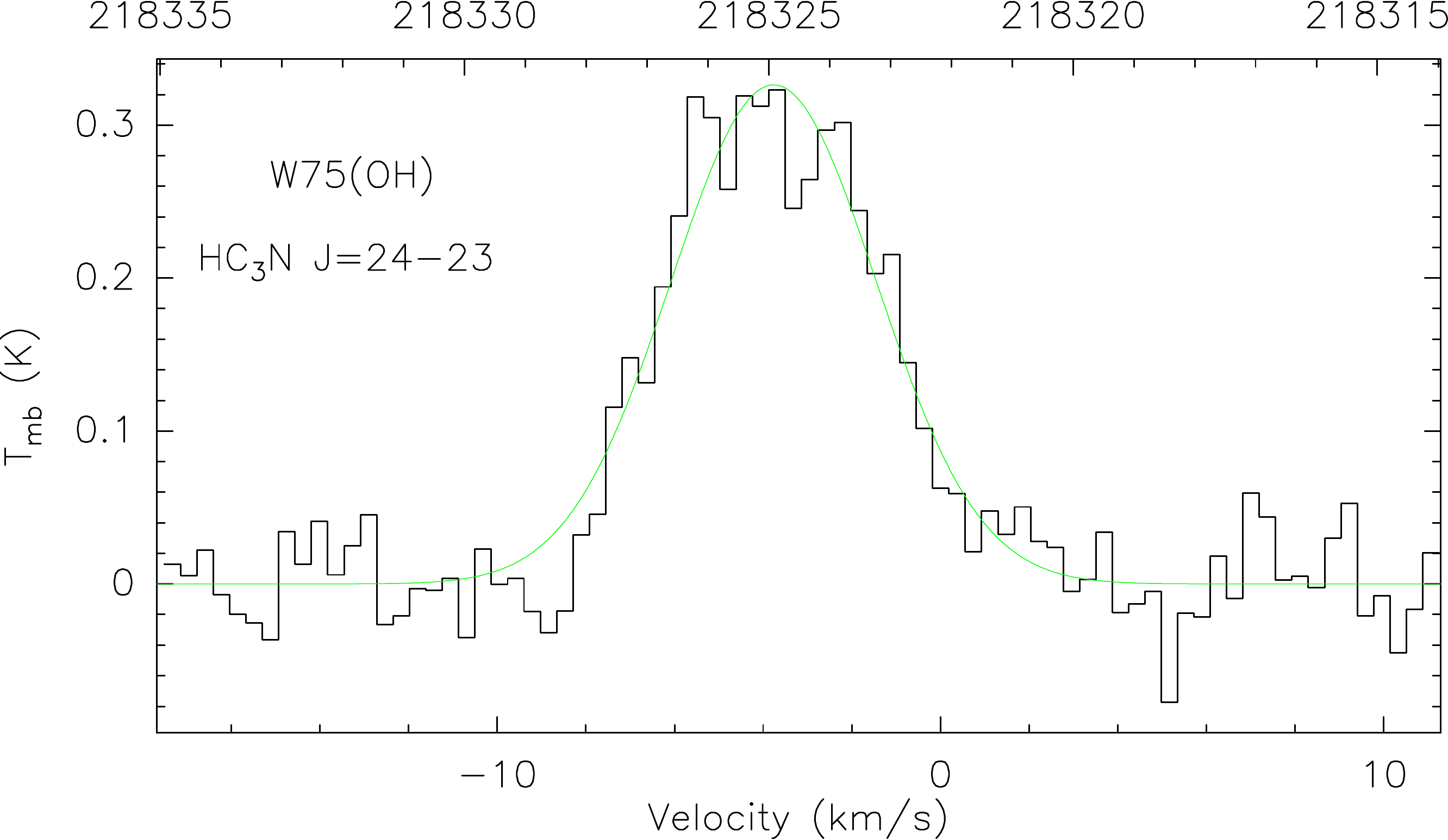}
	\end{minipage}%
	\begin{minipage}[t]{0.495\textwidth}
		\centering
		\includegraphics[width=80mm]{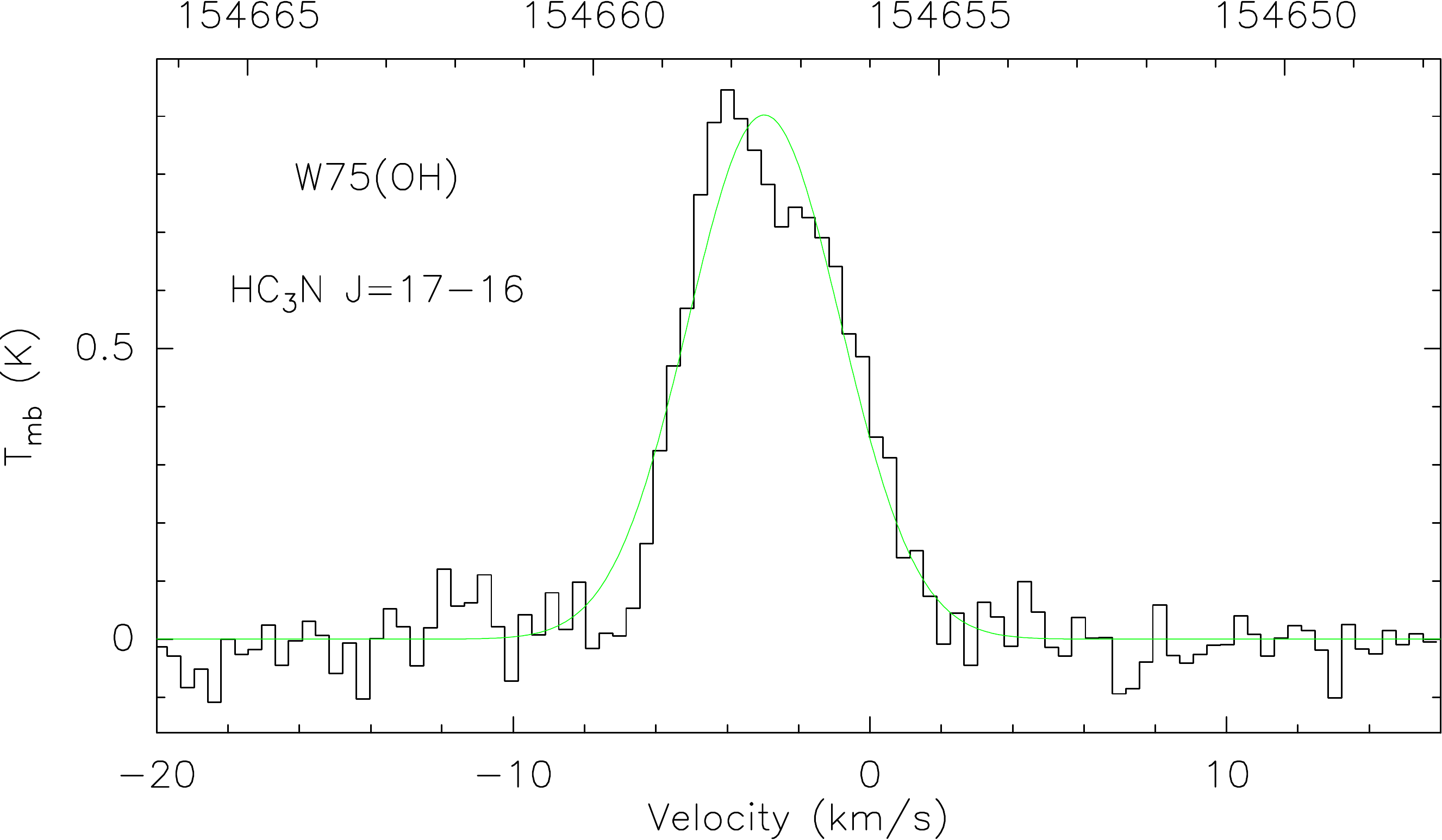}
	\end{minipage}%
	
	\vskip20pt 
	
	\begin{minipage}[t]{0.495\linewidth}
		\centering
		\includegraphics[width=80mm]{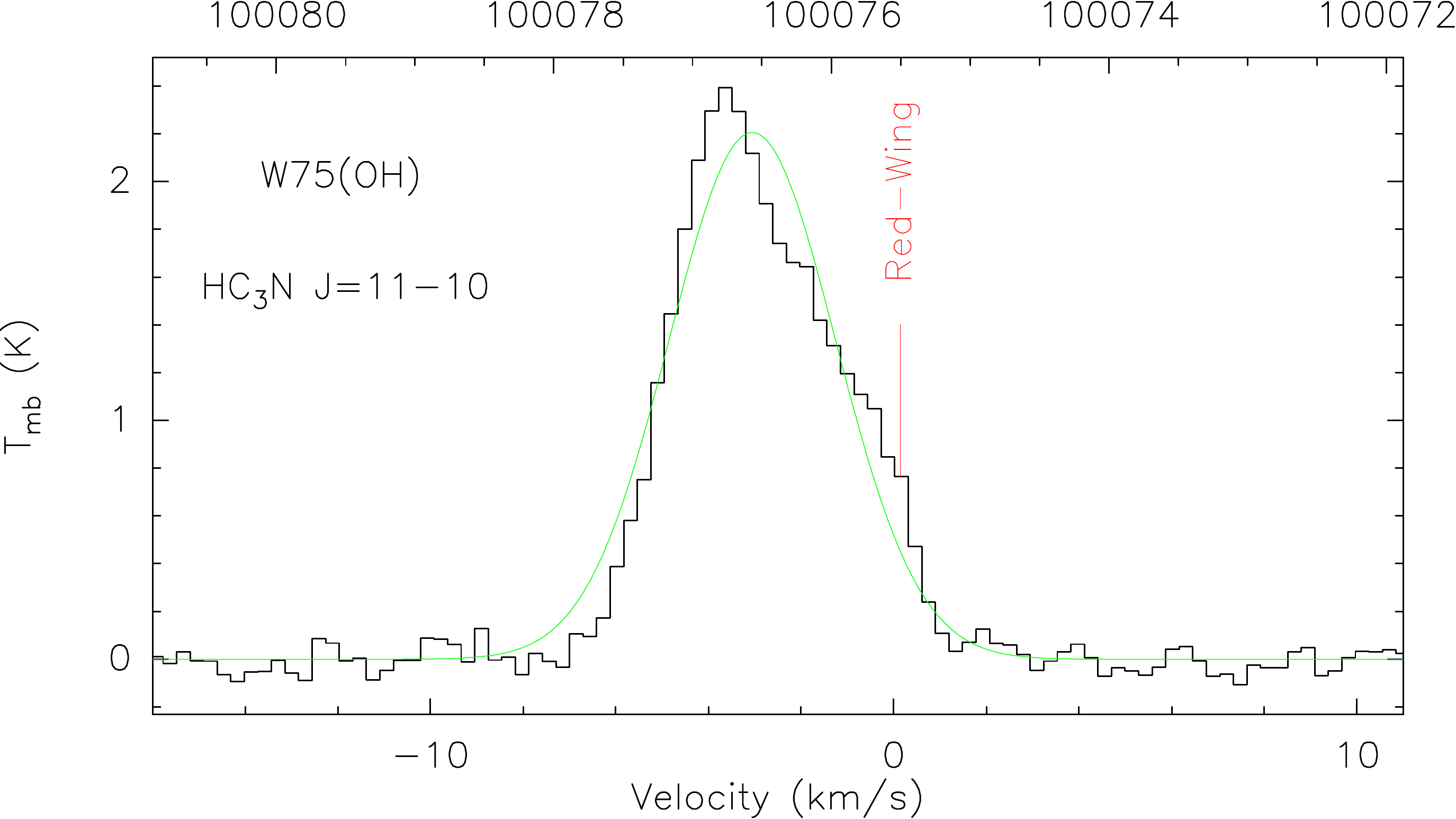}
	\end{minipage}%
	\begin{minipage}[t]{0.495\textwidth}
		\centering
		\includegraphics[width=80mm]{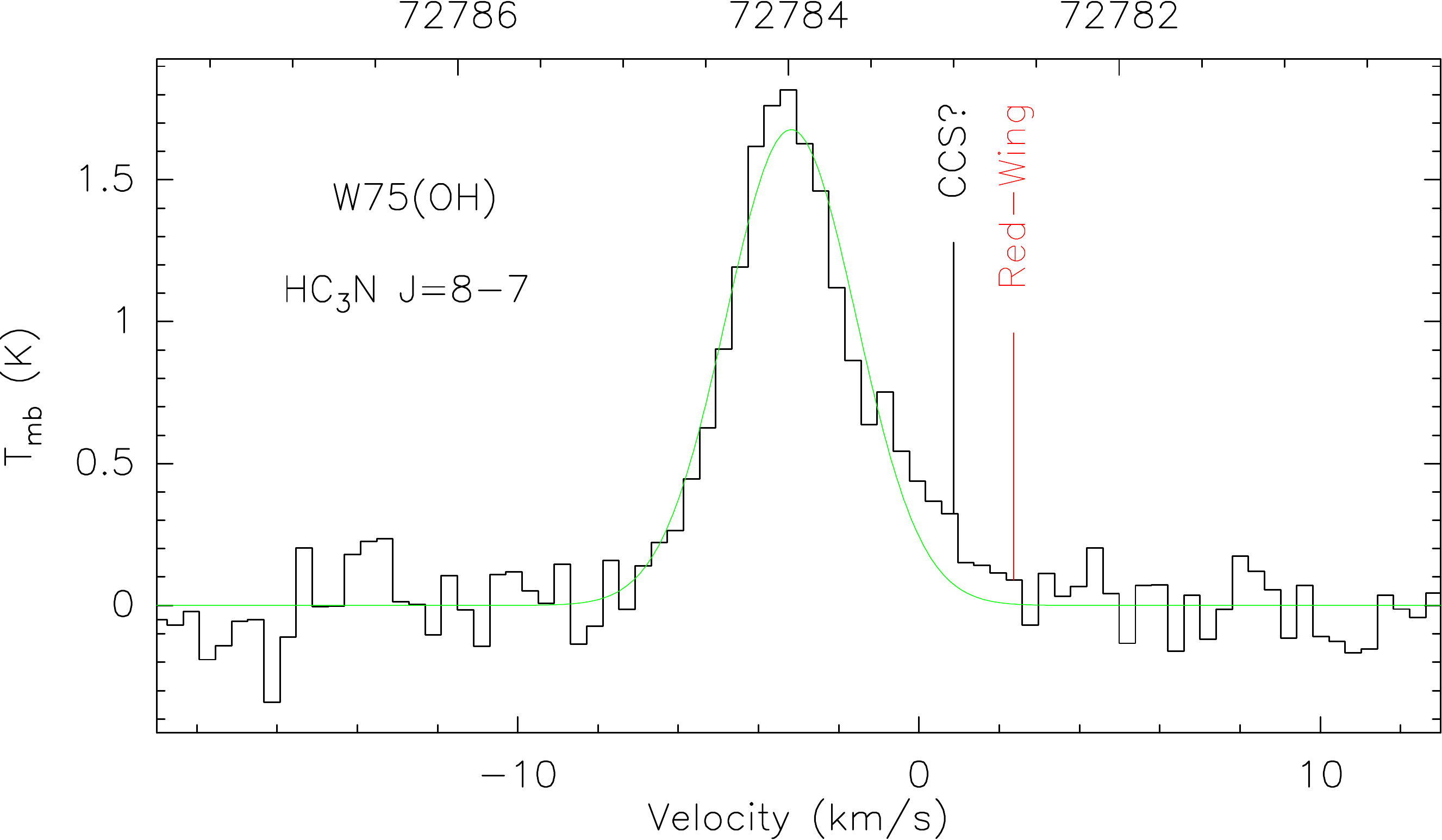}
	\end{minipage}%
	
	\vskip20pt 
	
	\begin{minipage}[t]{0.495\linewidth}
		\centering
		\includegraphics[width=80mm]{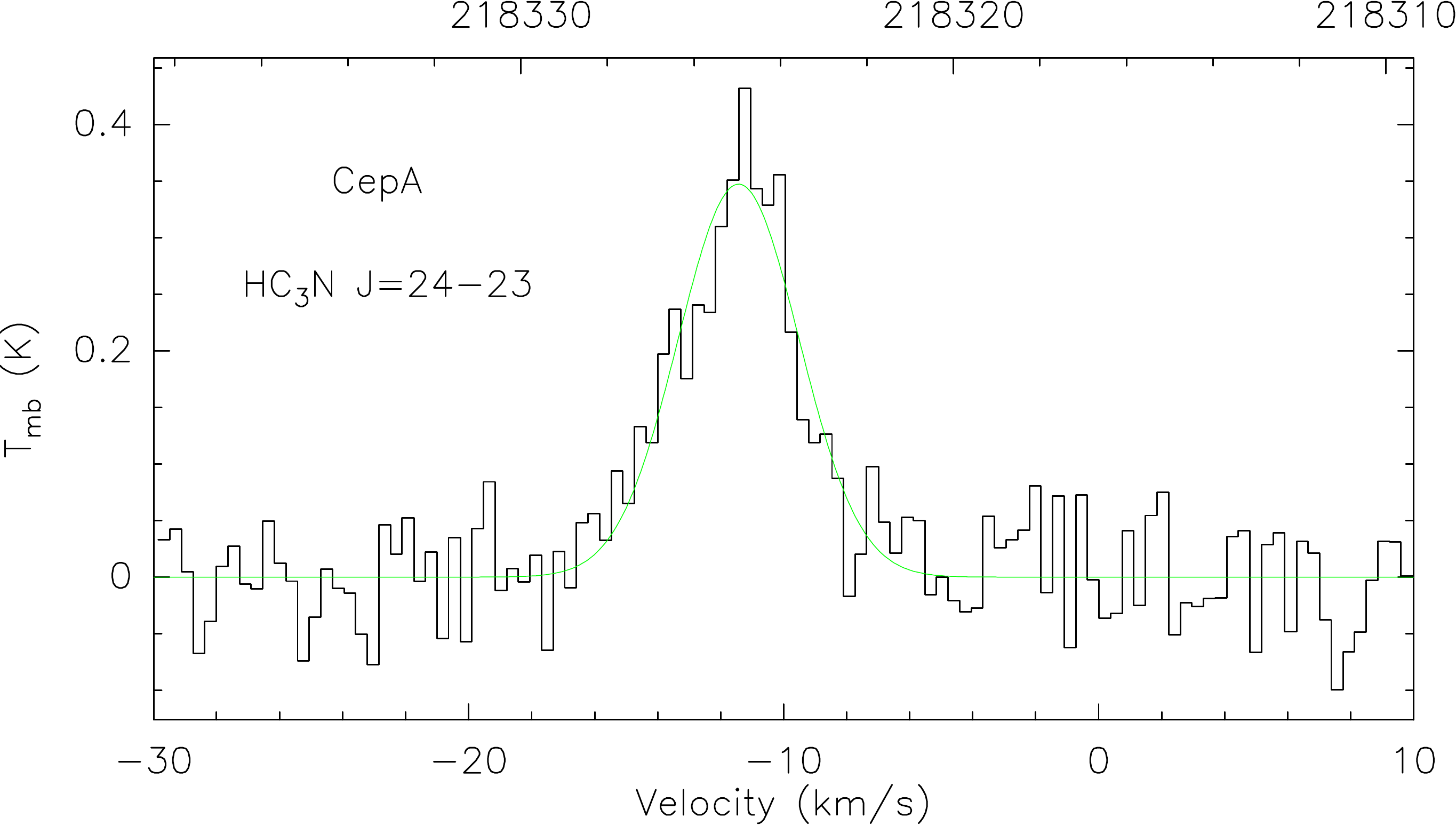}
	\end{minipage}%
	\begin{minipage}[t]{0.495\textwidth}
		\centering
		\includegraphics[width=80mm]{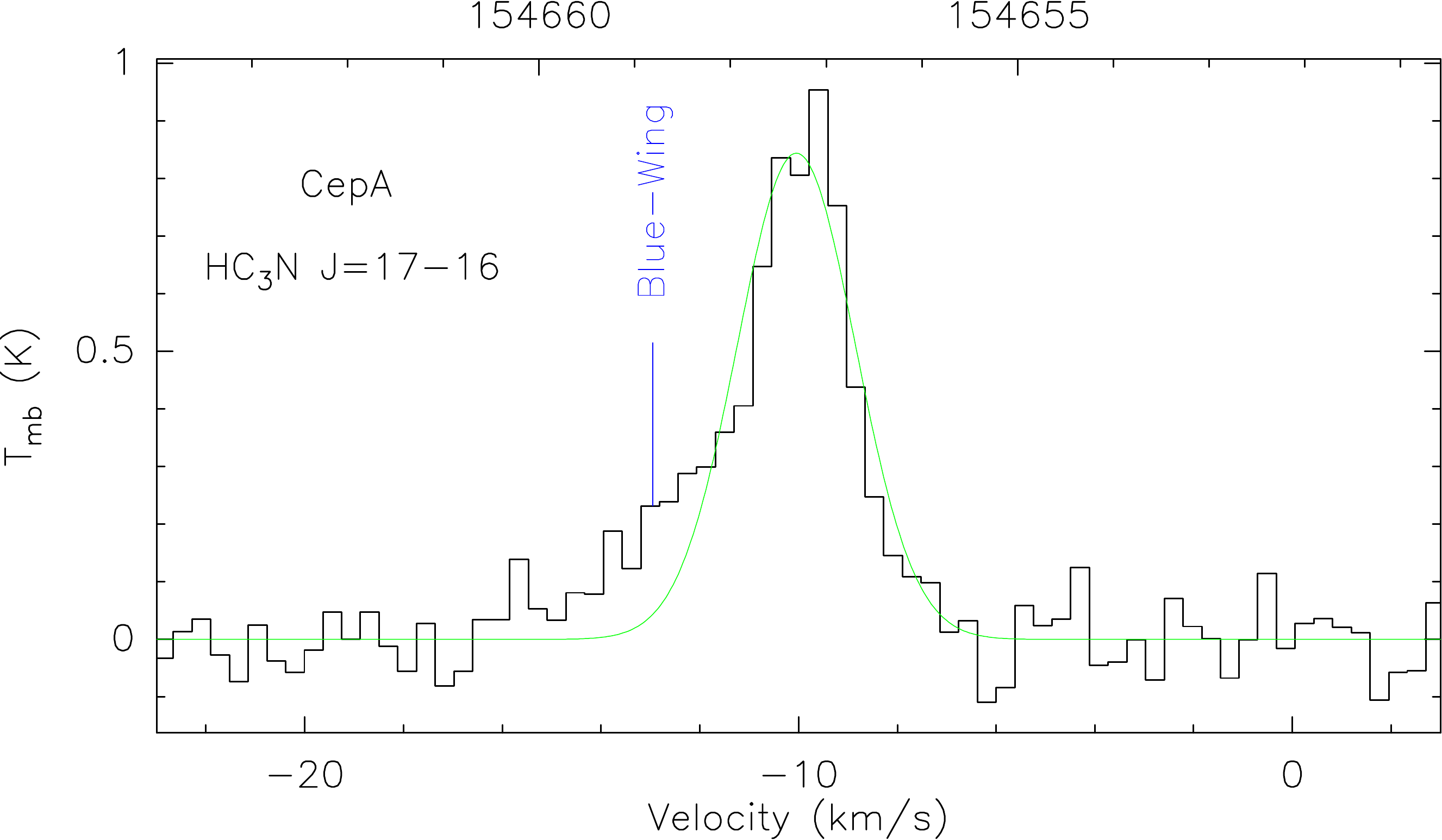}
	\end{minipage}%
	
	\vskip20pt 
	
	\begin{minipage}[t]{0.495\linewidth}
		\centering
		\includegraphics[width=80mm]{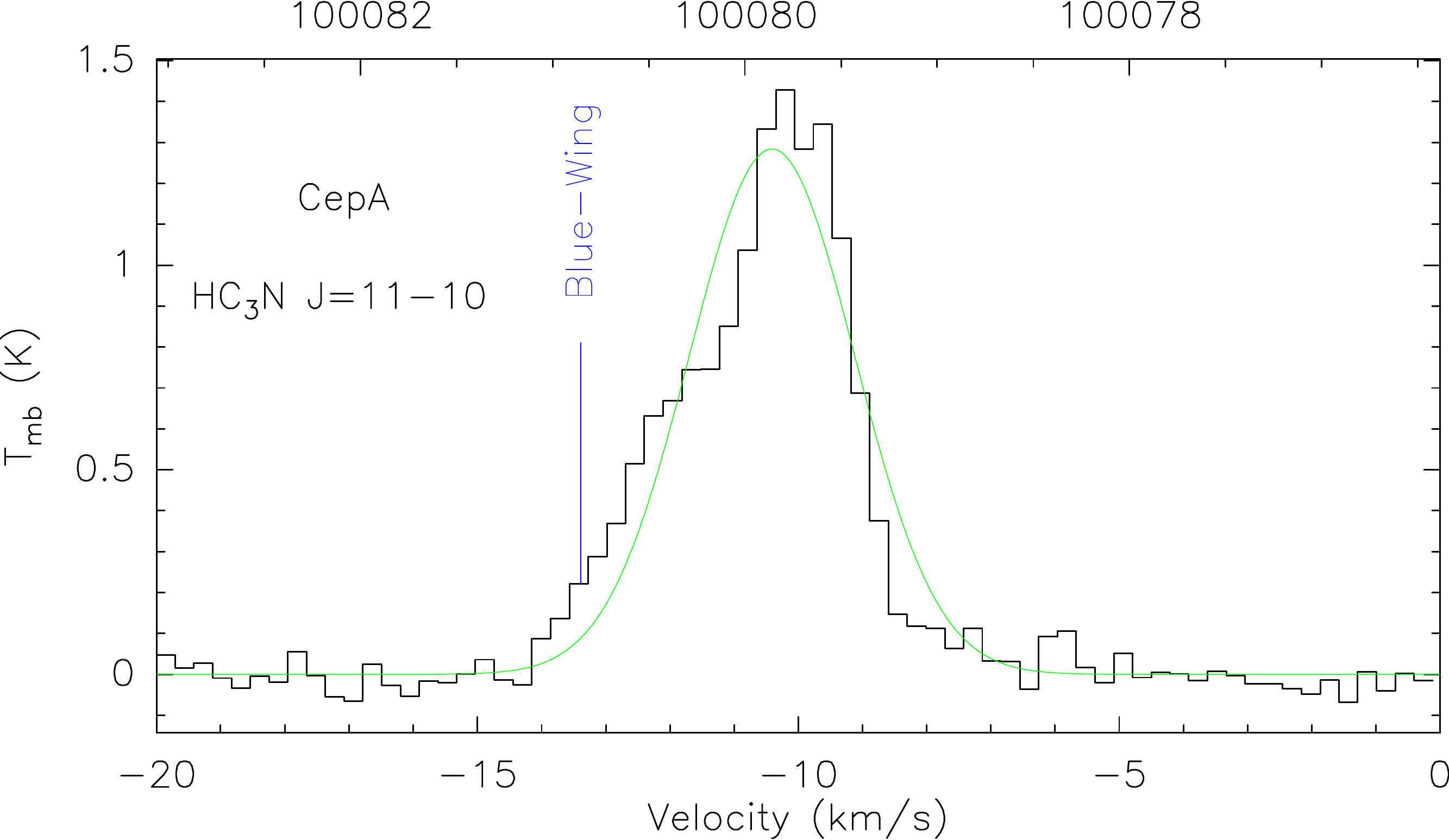}
	\end{minipage}%
	\begin{minipage}[t]{0.495\textwidth}
		\centering
		\includegraphics[width=80mm]{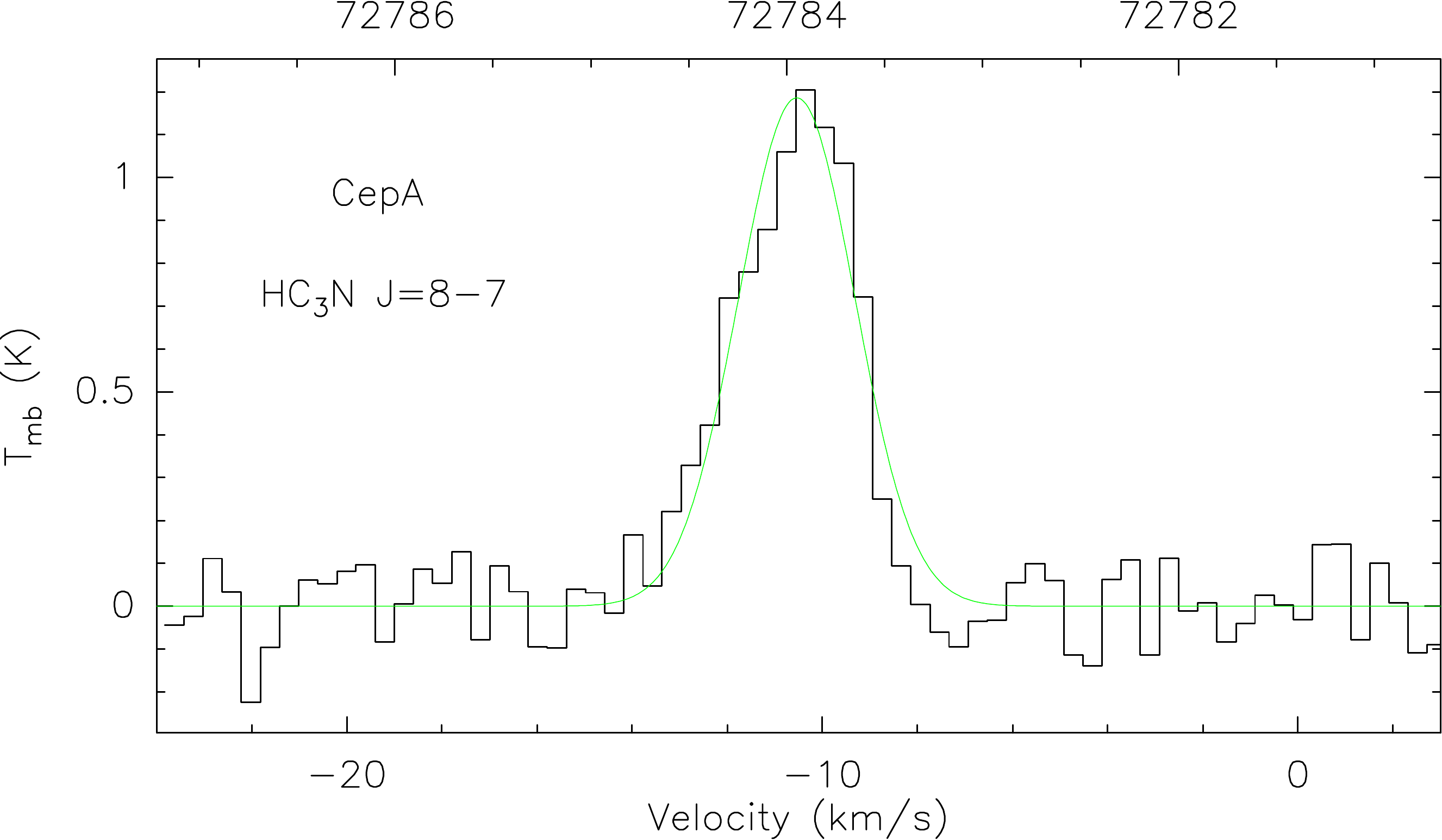}
	\end{minipage}%
	\caption{$-$ Continued.}  
	
\end{figure*}

\clearpage
\appendix
\renewcommand\thefigure{\Alph{section}\arabic{figure}} 
\section{HC$_3$N (24-23) line of sources without line wings}
\label{apx:PvDiagram}
   
\setcounter{figure}{0} 
\begin{figure*}[h]
	\vskip25pt 
	
	\begin{minipage}[t]{0.495\linewidth}
		\centering
		\includegraphics[width=77mm]{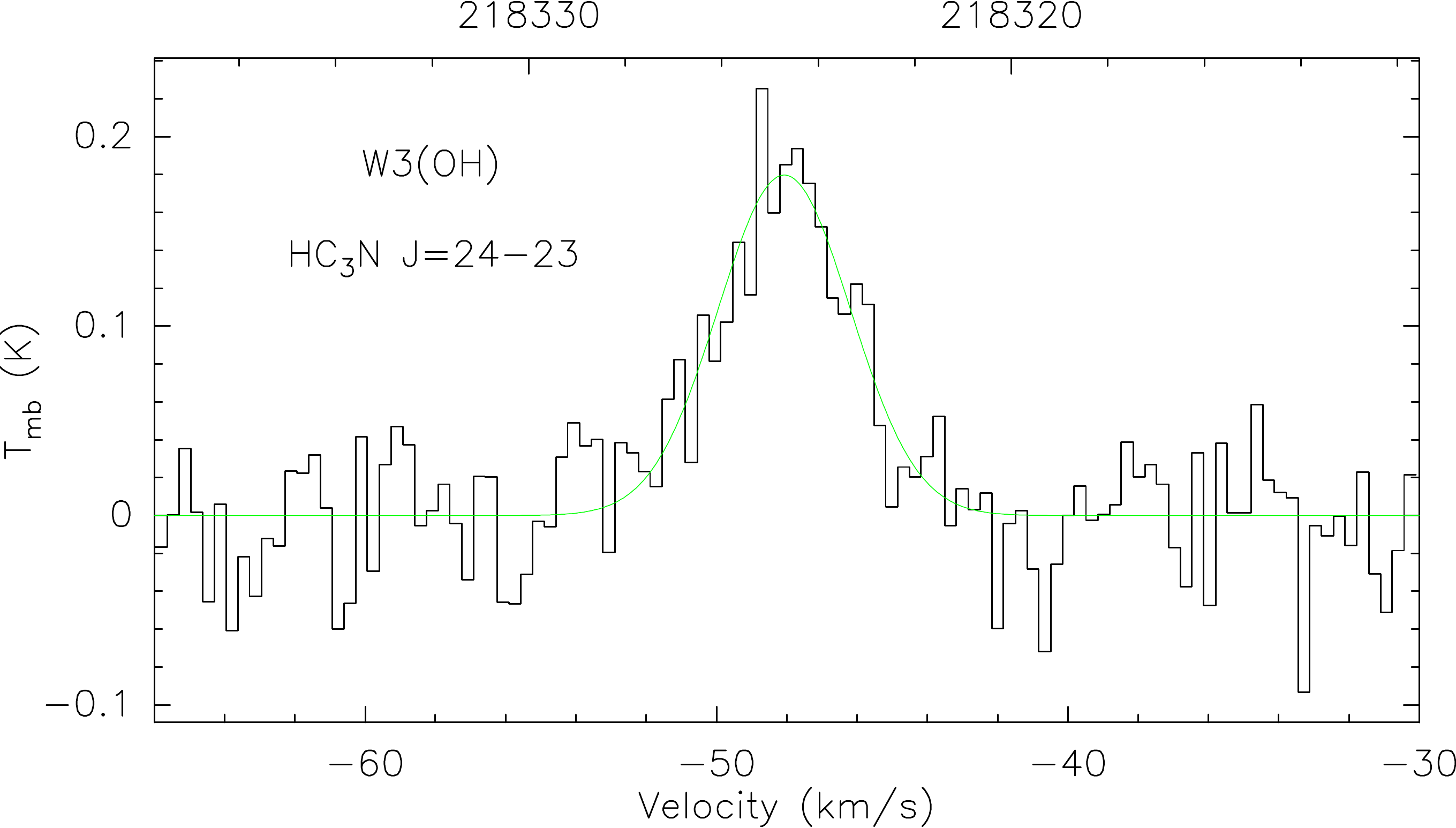}
	\end{minipage}%
	\begin{minipage}[t]{0.495\textwidth}
		\centering
		\includegraphics[width=77mm]{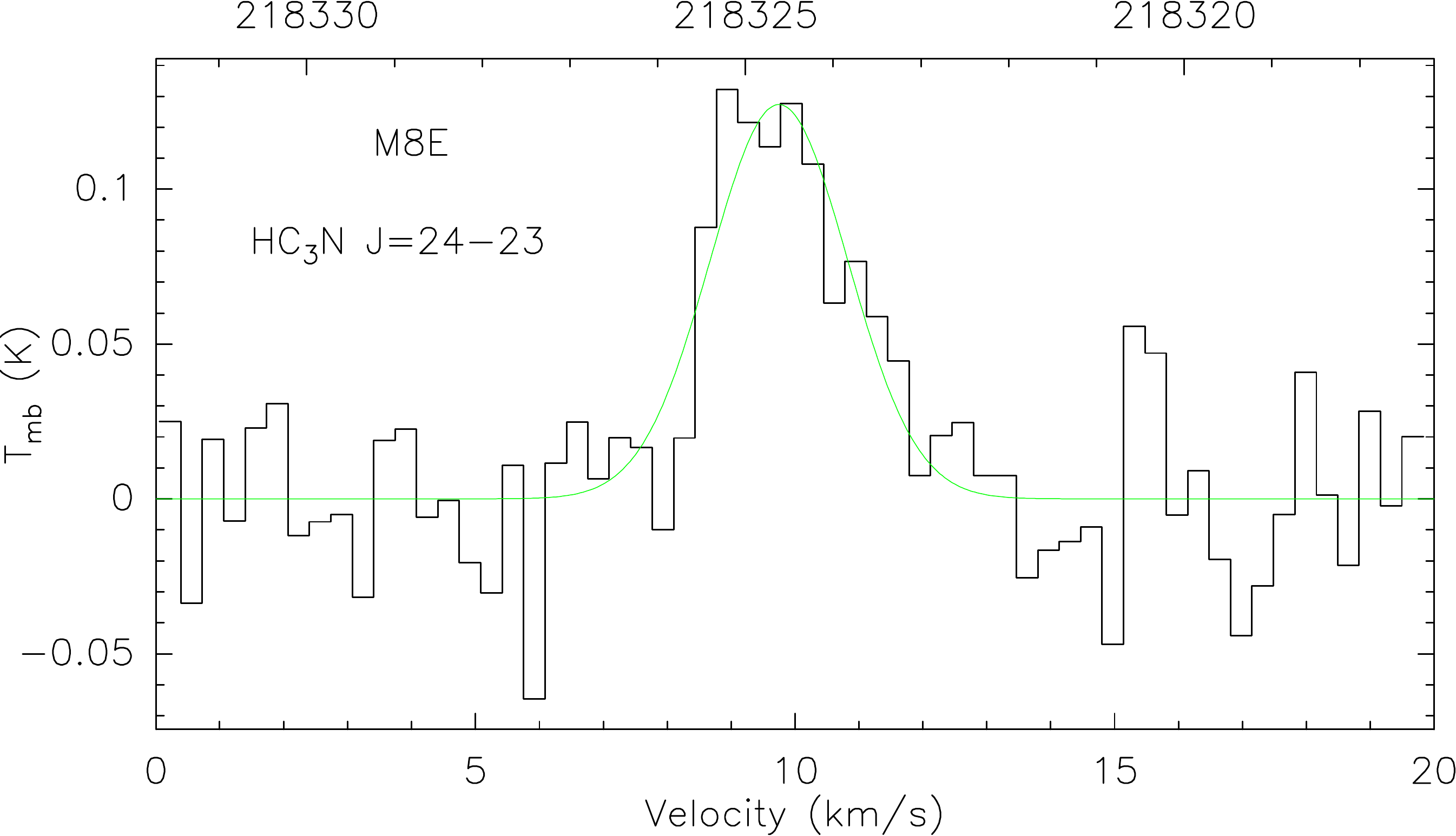}
	\end{minipage}%
	
	\vskip20pt 
	
	\begin{minipage}[t]{0.495\linewidth}
		\centering
		\includegraphics[width=80mm]{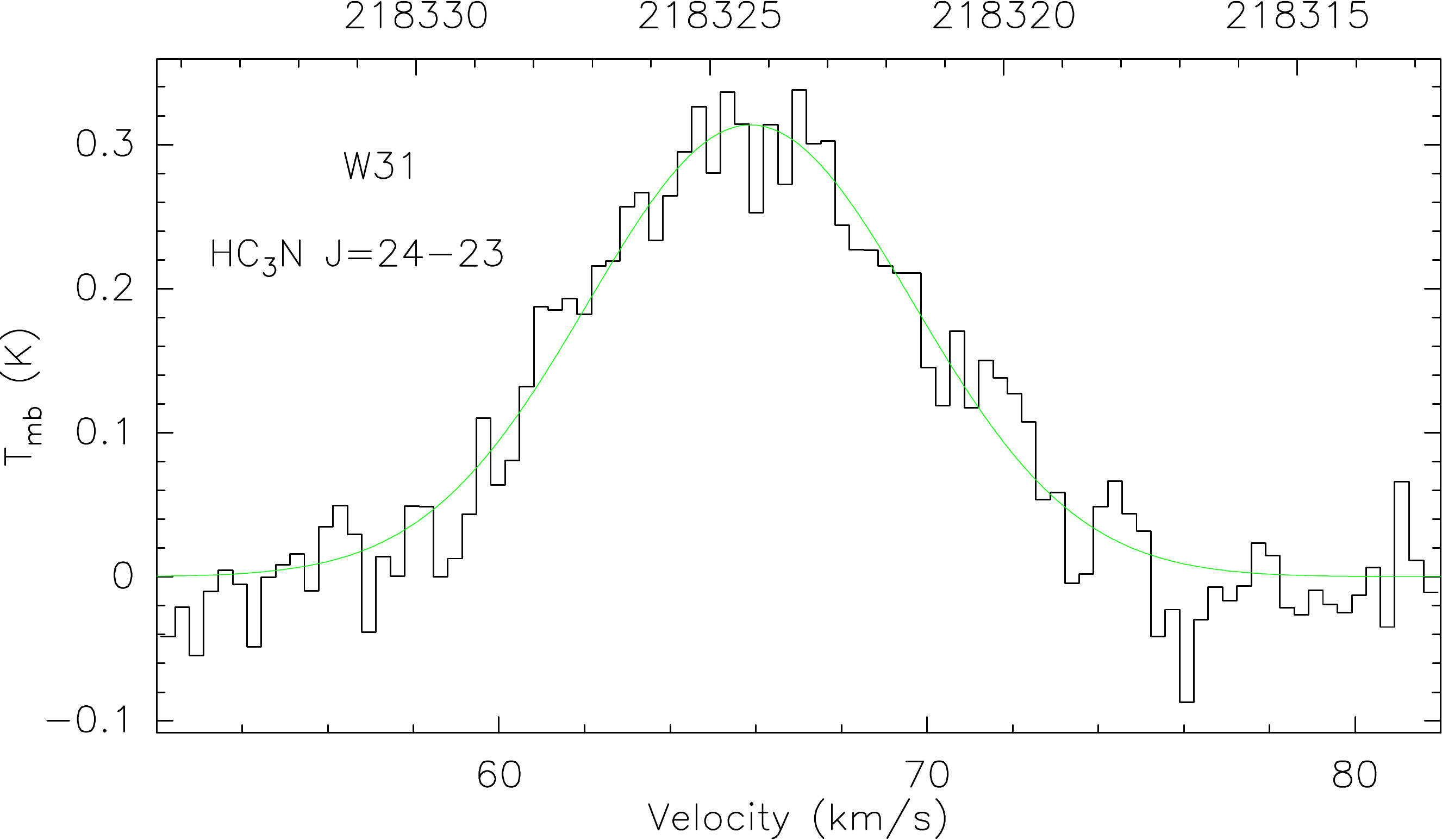}
	\end{minipage}%
	\begin{minipage}[t]{0.495\textwidth}
		\centering
		\includegraphics[width=80mm]{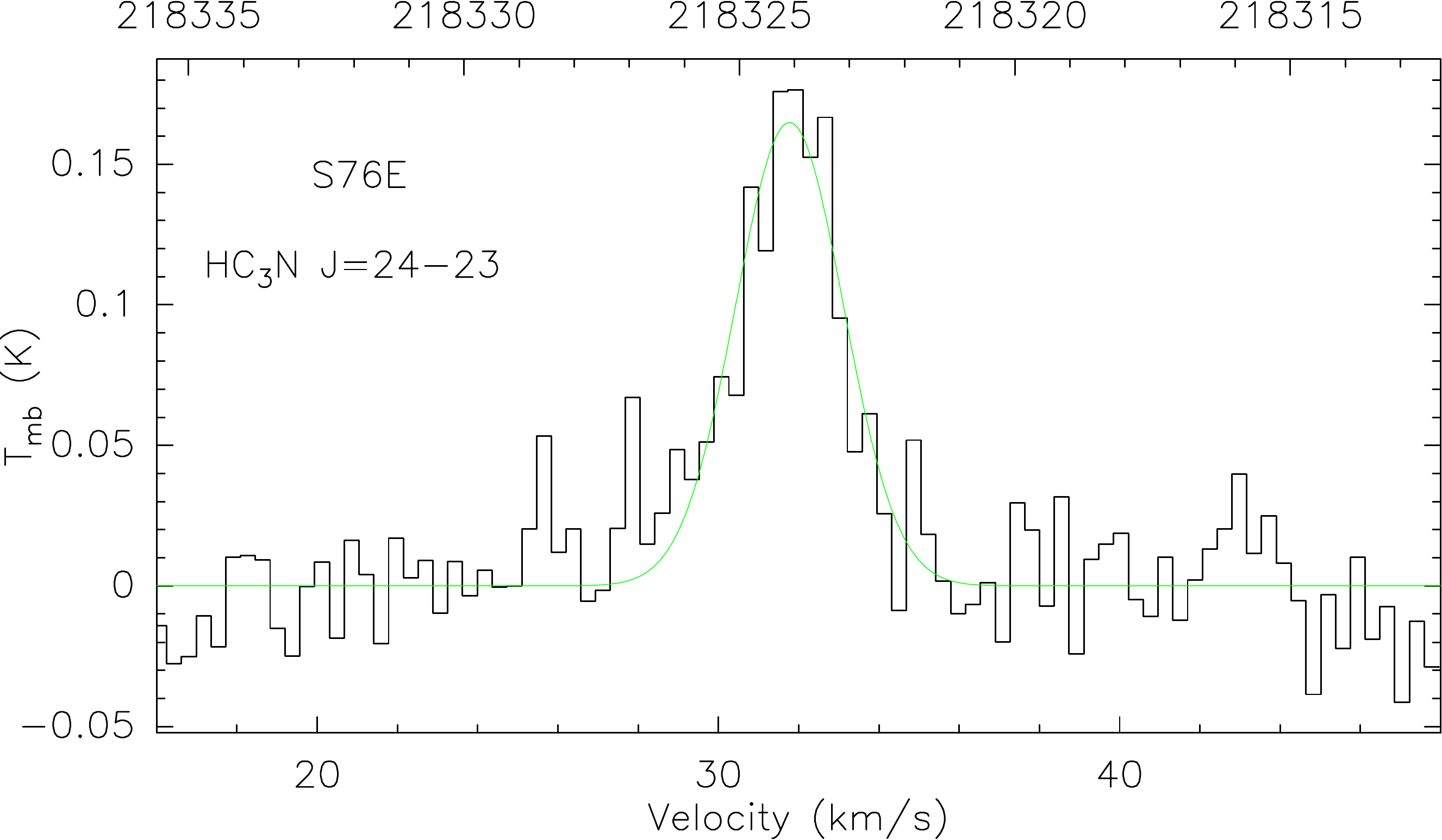}
	\end{minipage}%
	
	\vskip20pt 
	
	\begin{minipage}[t]{0.495\linewidth}
		\centering
		\includegraphics[width=80mm]{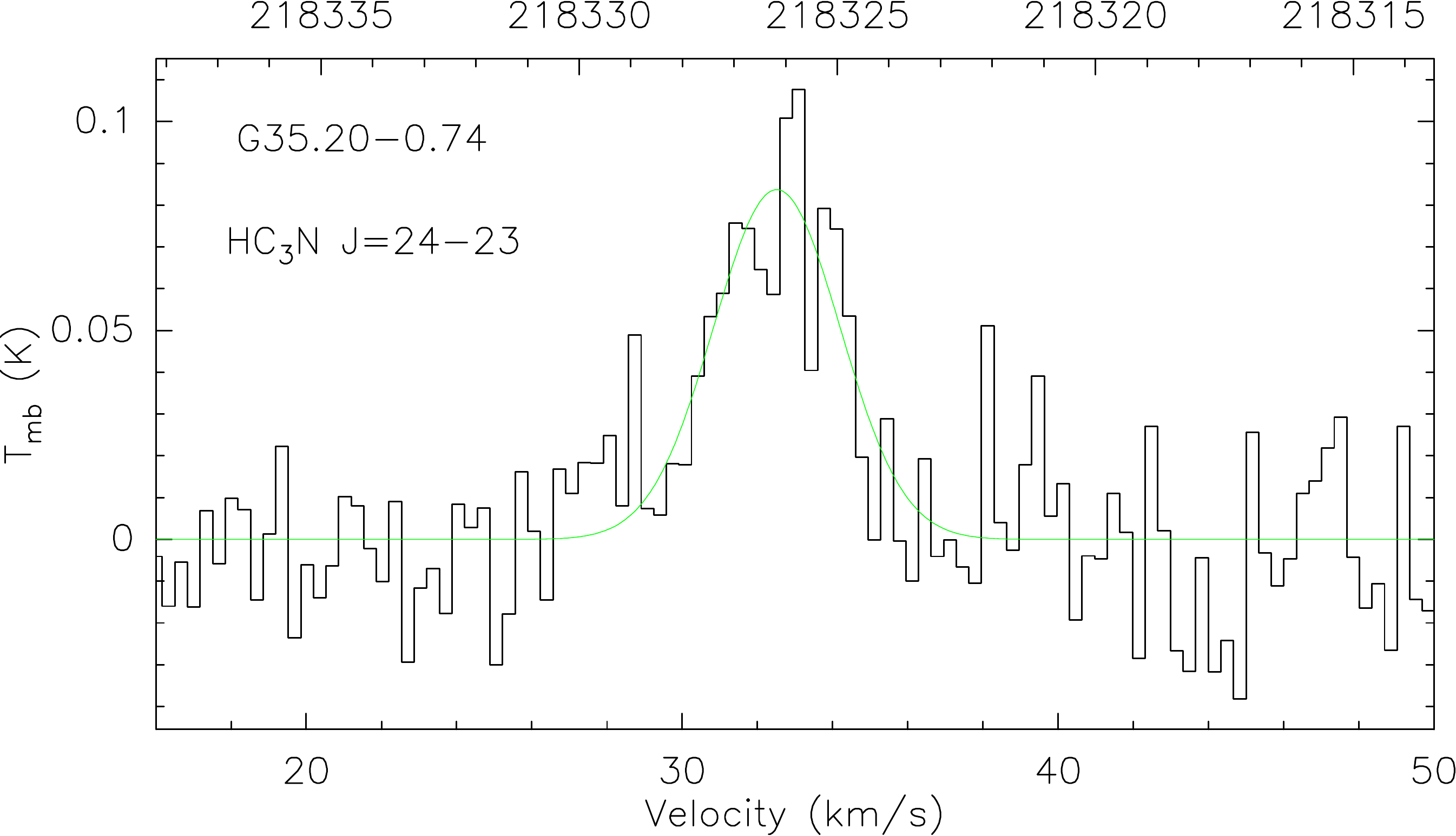}
	\end{minipage}%
	\begin{minipage}[t]{0.495\textwidth}
		\centering
		\includegraphics[width=80mm]{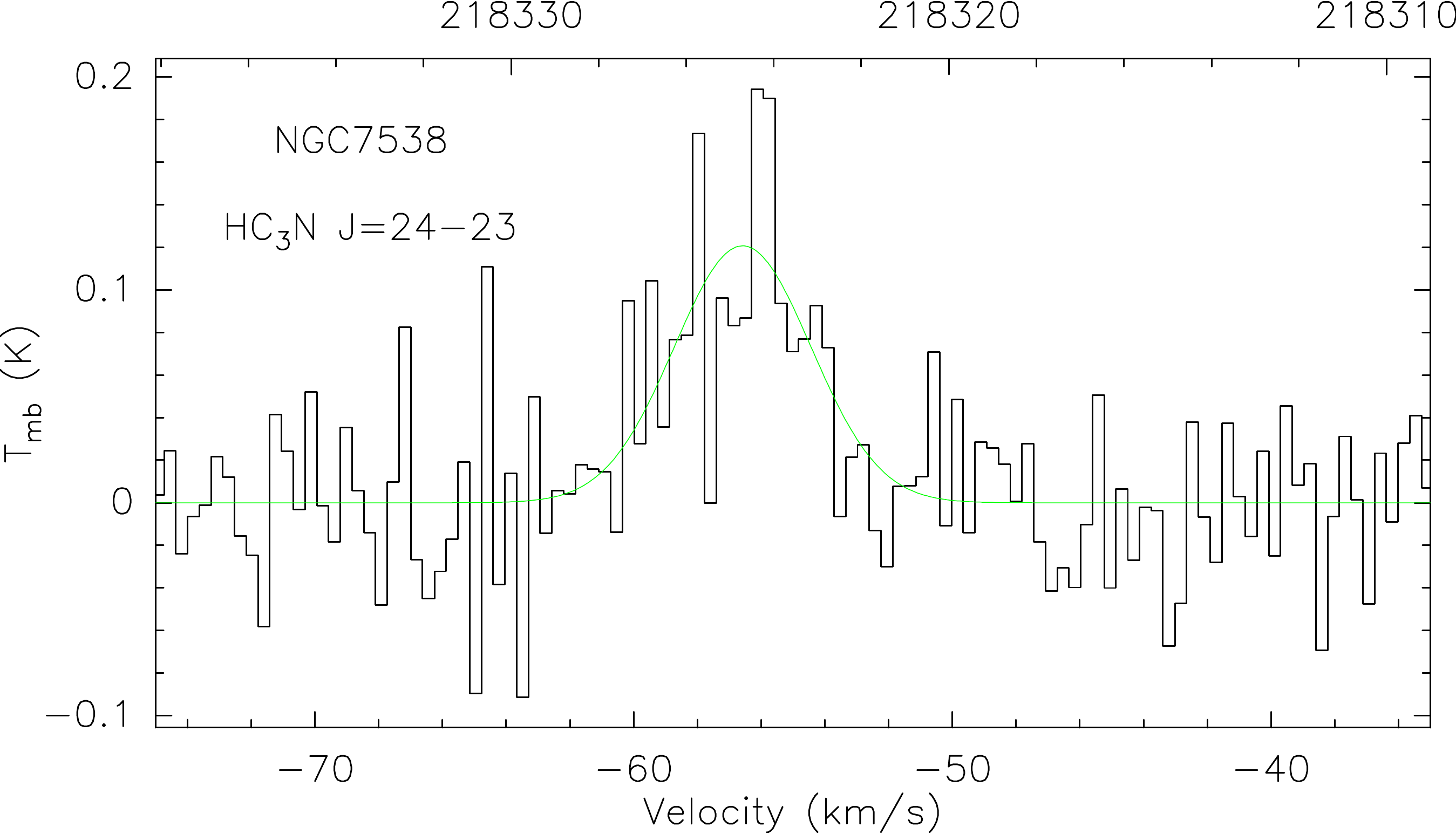}
	\end{minipage}%
	
	\caption{Spectral line of HC$_3$N (24-23) for sources without line wings. The identification of the transition is labeled in the upper left of each figure. Velocity resolution is $\sim$ 0.35\,km s$^{-1}$ overlaid with a Gaussian fit (green line).}
	\label{Figure A1}
\end{figure*}


\end{document}